\RequirePackage[running]{lineno}

\documentclass[aps,prd,twocolumn,nofootinbib,superscriptaddress,letterpaper,amsmath,amssymb]{revtex4}
\pdfoutput=1

\usepackage{ulem}
\usepackage{graphicx}  
\usepackage{dcolumn}   
\usepackage{bbm}        
\usepackage{amsmath}
\usepackage{amsfonts}
\usepackage{amssymb}   
\usepackage{epstopdf}
\usepackage{slashed}
\usepackage{hyperref}
\usepackage{multirow}
\usepackage{color}
\usepackage{float}
\usepackage{subcaption}
\usepackage{soul}
\usepackage{placeins}

\newcommand{\xspec}{{\sc xspec}}
\newcommand{\mrnet}{MR\_Net~}
\newcommand{\exsize}{0.35}

\begin{document}

\title{Deducing Neutron Star Equation of State Parameters \\Directly From Telescope Spectra with Uncertainty-Aware Machine Learning}

\author{Delaney Farrell}
\affiliation{Department of Physics, San Diego State University, San Diego, CA 92115, United States}
\author{Pierre Baldi}
\author{Jordan Ott}
\affiliation{Department of Computer Science, University of California Irvine, Irvine, California 92697, USA}
\author{Aishik Ghosh}
\affiliation{Department of Physics and Astronomy, University of California Irvine, Irvine, California 92697, USA}
\affiliation{Physics Division, Lawrence Berkeley National Laboratory, Berkeley, CA 94720, USA}
\author{Andrew W. Steiner}
\affiliation{Department of Physics and Astronomy, University of Tennessee, Knoxville, Tennessee 37996, USA}
\affiliation{Physics Division, Oak Ridge National Laboratory, Oak Ridge, Tennessee 37831, USA}
\author{Atharva Kavitkar}
\affiliation{Department of Computer Science, TU Kaiserslautern, Germany }
\author{Lee Lindblom}
\affiliation{Center for Astrophysics and Space Sciences,
University of California at San Diego, San Diego, CA 92093, United States}
\author{Daniel Whiteson}
\affiliation{Department of Physics and Astronomy, University of California Irvine, Irvine, California 92697, USA}
\author{Fridolin Weber}
\affiliation{Department of Physics, San Diego State University, San Diego, CA 92115, United States}
\affiliation{Center for Astrophysics and Space Sciences,
University of California at San Diego, San Diego, CA 92093, United States}

\begin{abstract}
Neutron stars provide a unique laboratory for studying matter at extreme pressures and densities. While there is no direct way to explore their interior structure, X-rays emitted from these stars can indirectly provide clues to the equation of state (EOS) of the superdense nuclear matter through the inference of the star’s mass and radius. However, inference of EOS directly from a star’s X-ray spectra is extremely challenging and is complicated by  systematic uncertainties. The current state of the art is to use simulation-based likelihoods in a piece-wise method which relies on certain theoretical assumptions and simplifications about the uncertainties. It first infers the star’s mass and radius to reduce the dimensionality of the problem, and from those quantities infer the EOS. We demonstrate a series of enhancements to the state of the art, in terms of realistic uncertainty quantification and a path towards circumventing the need for theoretical assumptions to infer physical properties with machine learning.  We also demonstrate novel inference of the EOS directly from the high-dimensional spectra of observed stars, avoiding the intermediate mass-radius step. Our network is conditioned on the sources of uncertainty of each star, allowing for natural and complete propagation of uncertainties to the EOS. 
 
\end{abstract}

\maketitle

\vspace{.25in}

\section{INTRODUCTION}

Neutron stars are the densest stellar objects,  providing a unique laboratory for studying matter in physical conditions that cannot be replicated on Earth and are only found in these neutron-packed remnants of massive stars ($\gtrapprox 8 M_\odot$). Insights about the forms of matter which emerge under these extreme conditions can improve our understanding of two of the least well-understood fundamental forces, quantum chromodynamics and gravity.

The neutron-rich matter within the inner regions of a neutron star can reach supranuclear densities of $10^{15} ~\text{g/cm}^3$, potentially leading to transitions to stable non-nucleonic states of strange matter in the form of hyperons~\cite{Tolos:2020aln,li2018competition, spinella2019hyperonic}, deconfined quark matter made of up, down, and strange quarks~\cite{Fukushima:2013rx,orsaria2014quark}, color superconducting phases~\cite{Alford:2007xm,zdunik2013maximum}, or  Bose-Einstein condensates made of negatively charged pions or $K^-$ mesons~\cite{baym1973pion,KAPLAN,ellis1995kaon}. The structure and behavior of matter at such extreme densities are one of the great mysteries in modern science, prompting decades of theoretical and experimental research into the interior composition of a neutron star.

The nature of matter within a neutron star is compactly summarized by its equation of state (EOS),  the relationship between pressure $P$ and energy density $\epsilon$, which show starkly different behaviors under the various states of strange matter hypotheses described above.
This relationship is determined by the microphysical interactions between various particles within the star, relative abundances of different particle species, as well as the star's temperature. Understanding the EOS of supranuclear density matter has been of interest to the nuclear and astrophysics communities for decades, resulting in many proposed phenomenological models for the equation of state of neutron star matter. Variations in models arise from a lack of precise knowledge of nuclear interactions between particles at such extreme conditions, as well as the wide range of densities and isospin asymmetries that are thought to exist within neutron stars \cite{1987ApJ...314..594F}. 

While the internal pressure and density cannot be directly observed, the EOS of a star in static gravitational equilibrium determines stellar properties such as its mass and radius, which in turn determine observables such as the stellar X-ray spectrum.  Conversely, the stellar spectra can be used to infer the masses and radii, which in principle allow inference of the EOS~\cite{Rutledge,Heinke06,Lattimer01,PhysRevD.82.103011,Steiner10te,Lindblom:2013xkra}, though inversion of this second step is numerically very difficult. Additional challenges are due to the small number of neutron star observations, $\mathcal{O}(10)$, and the significant uncertainty of individual measurements.  It is therefore vital that as much information as possible is extracted from each star, and that the uncertainties be propagated accurately, to provide the most complete information possible about the EOS.


\begin{figure*}
    \centering
    \includegraphics[width=0.85\textwidth]{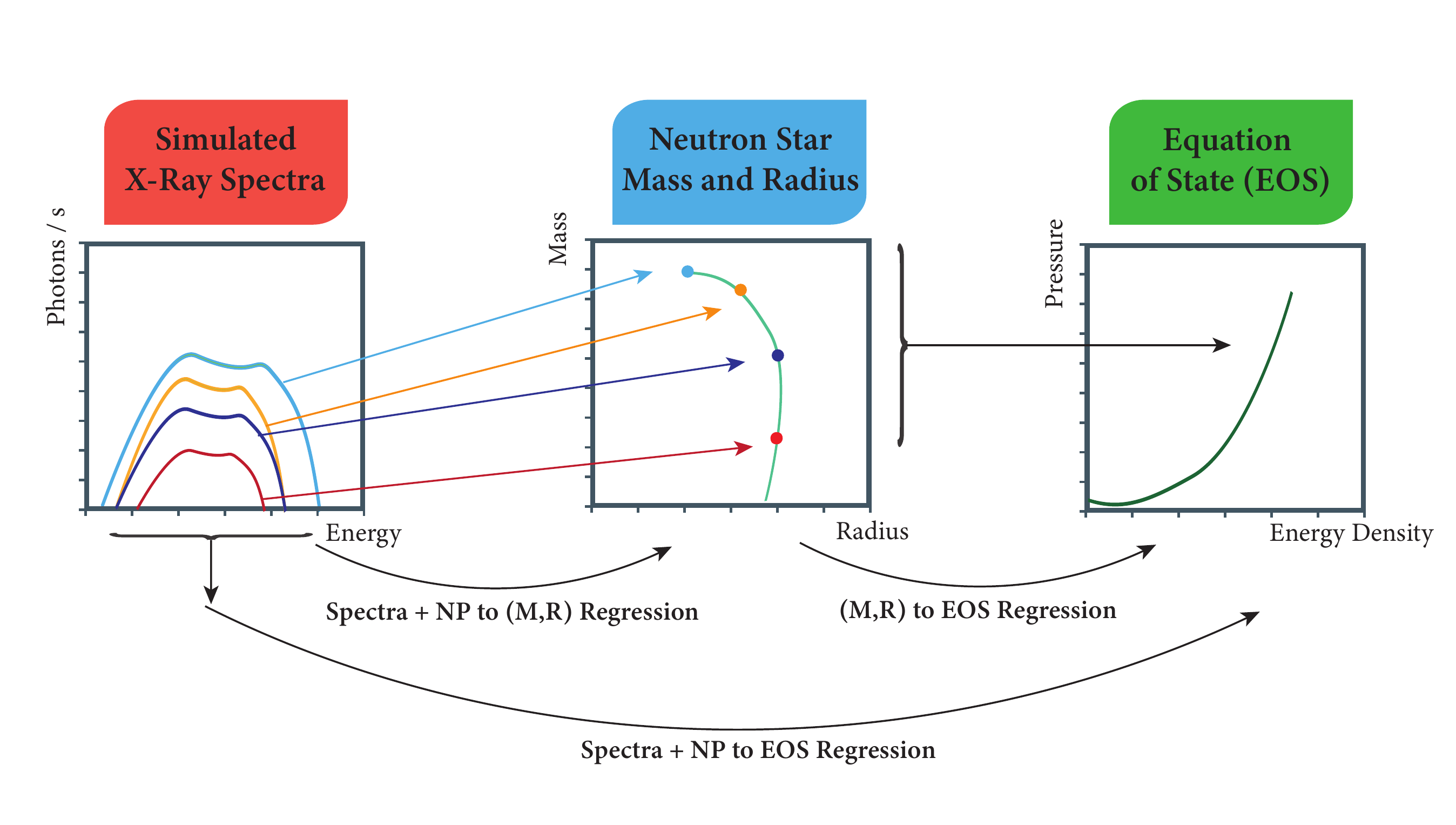}
    \caption{ Overview of the regression task, which involves either inferring stellar summary quantities such as mass and radius, which can then be used to deduce the equation of state as in earlier work~\cite{Fujimoto:2019hxv,Fujimoto:2021zas,Morawski:2020izm,Ferreira:2019bny} or inference of EOS directly from stellar spectra, as is demonstrated in this study.}
    \label{fig:scheme}
\end{figure*}

At the same time, there has been a dramatic burst of progress in artificial intelligence, specifically deep learning~\cite{baldi2021deep}, a modern re-branding of neural networks. This progress has led to breakthroughs not only in traditional areas such as natural language processing and computer vision but also in the natural sciences, including particle physics, often increasing the statistical power of difficult-to-collect data~\cite{baldi2014searching} while allowing robust handling of uncertainties~\cite{Ghosh:2021roe}.  Where earlier neural networks were limited in size, computing progress especially in the form of Graphical Processing Units (GPUs), has enabled the deployment of larger and  deeper networks that can handle more complex and higher-dimensional data~\cite{Baldi:2016fzo,Guest:2016iqz}, allowing direct analysis of data without requiring dimensional reduction, or other preprocessing steps, that can often sacrifice useful information.  The full power of these techniques has not yet been brought to bear on many astrophysical tasks.

In the context of the inference of neutron star EOS, recent work by Fujimoto {\it et al.}~\cite{Fujimoto:2019hxv,Fujimoto:2021zas} demonstrated the ability of deep networks to regress the EOS directly from a set of stellar mass-radius pairs, without the need to extract the functional relationship between mass and radius. Their analysis used a toy model to describe the uncertainties in mass and radius, assuming uncorrelated Gaussian errors randomly drawn from ad-hoc priors. Real measurements, of course, do not often obey these simplifying assumptions, and show complex correlations between mass and radius~\cite{2016ApJ...831..184B}. Related work~\cite{Morawski:2020izm} has demonstrated similar regression, again assuming Gaussian uncertainty on mass and radius values, but with clever efforts to reduce dependence on EOS parameterization.  An alternative approach~\cite{Ferreira:2019bny} uses both neural networks and support vector machines to regress the EOS from stellar radii and tidal deformations. 

More realistic characterization of the uncertainties in the mass-radius plane can be extracted using the state-of-the-art tool \xspec~\cite{xspec}, which assumes a theoretical model for the star and telescope response, allowing for explicit calculation of the likelihood of telescope spectra for various mass and radius values. The likelihood can be used in the standard way to extract best-estimates and uncertainty contours of any shape in the mass-radius plane. However, these complex mass-radius likelihoods cannot be trivially incorporated into the existing EOS inference schemes, motivating the simplifying assumptions of uncorrelation normal distributions which can be described by two width values. An additional concern is that \xspec's contours rely on the simplifying assumptions of the theoretical model. 

What has received less attention in the literature are likelihood-free methods to infer the EOS directly from the telescope spectra, without the intermediate stepping stone of the mass-radius determination and the challenges of its representation. This would allow for the full propagation of realistic uncertainties and the relaxation of assumptions about the theoretical model.

In this paper, we present a technique of EOS inference which allows for the full propagation of the uncertainties in the X-ray spectra, without making simplifying assumptions about the shape of the contours in the mass-radius plane.  We proceed in three steps, beginning from an approach similar to the state of the art but with realistic uncertainty propagation, and moving towards end-to-end infererence. In the first step, our neural network model infers the neutron star EOS from a set of stellar masses and radii extracted from \xspec, but rather than making simplifying assumptions or extracting uncertainty contours from \xspec, we vary the assumed nuisance parameters (NPs) which are the source of the uncertainty to produce new best-estimate mass-radius points. The EOS inference can then be performed on many sets, each corresponding to varied NP values, producing a variation in the inferred  which represents the propagated uncertainty.  In the second step, we investigate a more flexible method of inferring the mass and radius that does not explicitly rely on \xspec's specific theoretical model.  We introduce a network capable of directly analyzing high-dimensional neutron star spectra, performing inference of stellar mass and radii from telescope spectra, a demonstration of the impressive capacity of modern deep networks.  Finally, we perform a first-of-its-kind inference of the EOS parameters directly from a multi-star {\it set} of stellar spectra, without requiring the intermediate step of collapsing the information into mass and radius; see Figure ~\ref{fig:scheme}.  In both cases, we allow for full propagation of uncertainties by conditioning the networks on the stellar nuisance parameters. As this is -- to our knowledge -- the first attempt at full  propagation of these uncertainties for this task, there are no direct benchmarks in prior work. Instead,  we show comparisons between our three methods and visualize the impact on the EOS inference of variation of the nuisance parameters  for a fixed x-ray spectrum.

The paper is organized as follows. In Section~\ref{sec:bg}, we provide background on the physics of the connection between the nuclear equation of state and the stellar observations. Section~\ref{sec:ml} describes the fundamentals of the machine learning concepts on which our studies rely.  Details of the samples of simulated data are given in Section~\ref{sec:data}. Section~\ref{sec:eosmr} demonstrates inference of the EOS parameters from mass and radius, while Section~\ref{sec:mr} describes how mass and radius parameters can be inferred directly from stellar spectra, and Section~\ref{sec:e2e} shows end-to-end inference of EOS parameters from a set of neutron star spectra.

\section{Background}
\label{sec:bg}

\subsection{Equation of State for Dense Matter}

Neutron star interiors present a unique opportunity to study matter under conditions beyond the reach of terrestrial laboratories: matter that is extremely high in density, relatively cold in temperature, and isospin-asymmetric \cite{lattimer}. Perhaps the closest experimental constraints have come from ultrarelativistic heavy-ion collisions (as conducted at the RHIC~\cite{muller2006results}, the LHC~\cite{aad2008atlas}, and FAIR~\cite{spiller2006fair}), which probe the nature of hot, 
symmetric nuclear matter Extremely neutron-rich matter is more recently probed in studies of neutron skin or giant monopole and dipole resonances, but these studies are limited in nature \cite{nskin}. Unfortunately, these experiments currently lack the temperature and density constraints present within the core of a neutron star - meaning the dense matter encountered within a neutron star cannot yet be replicated by experiment. 

While properties of superdense matter cannot be derived directly from experiments, certain general principles from general relativity and quantum chromodynamics (QCD) guide the theoretical investigation into neutron star matter. Neutron star structure is controlled by the long-range gravitational force~\cite{glendenning2012compact,RevModPhys.88.021001} which holds the star together and short-range strong interactions between nucleons and nuclei which provide the pressure that prevents the star from collapsing. At low nuclear densities (below nuclear saturation), effective field theories based on QCD provide a systematic basis for nuclear forces, which offers good constraints on two-nucleon interactions \cite{Machleidt:2011zz,RevModPhys.88.021001,qcd}. At higher densities, the QCD framework predicts that baryonic matter (where quarks are confined within hadrons) will experience a phase transition to quark-gluon plasma (QGP), where quarks and gluons are freed from hadronic boundaries \cite{qcd}. Other stable states of non-nucleonic matter may also occur, like the formation of hyperons, color superconducting phases of quark matter, or Bose-Einstein condensates of different mesons \cite{baym1973pion, ellis1995kaon}. Theoretical uncertainties have resulted in a wide range of proposed phenomenological models for the EOS of neutron star matter, which can then be tested by experiment or observation.

The EOS of neutron star matter is intrinsically linked to macroscopic characteristics like gravitational mass $M$ and radius $R$ through the general relativistic stellar structure equation known as the Tolman-Oppenheimer-Volkoff (TOV) equation \cite{PhysRev.55.374,PhysRev.55.364}. This equation assumes the object is spherically symmetric, non-rotating, and non-magnetic. The TOV equation is given by (assuming geometrized units where $G=c=1$): 
\begin{equation} \label{eq:1}
    \frac{dP}{dr} = -\frac{(\epsilon+P)(m+4\pi r^3 P)}{r^2\left(1-\frac{2m}{r}\right)}
\end{equation}
where $m$ is the gravitational mass enclosed within a sphere of radius $r$. The mass of the star can be solved for as:
\begin{equation} \label{eq:2}
    \frac{dm}{dr} = 4\pi r^2 \epsilon
\end{equation}
where the total gravitational mass $M$ of a star with radius $R$ is given by $M \equiv m(R) = 4\pi \int_0^R dr ~r^2\epsilon$. Given an EOS, numerically solving the TOV equation for $M$ and $R$ is straightforward. These equations create a one-to-one map from the EOS to the $M-R$ relation \cite{Lindblom:2013xkra}; the inverse form of this map can therefore provide constraints on the EOS from observable properties. To mathematically invert the TOV equation, at least two stars' mass and radius must be known exactly, a feat is not possible with current observational technology. Solving the inverse problem is therefore much more complicated, potentially even intractable without making significant numerical assumptions.

\subsection{X-Ray Spectroscopy for Neutron Stars}

Many reliable observations of neutron stars come from X-ray emission, either from electromagnetic radiation from pulsars or thermal emission in quiescent low-mass X-ray binaries (qLMXBs). qLMXBs are particularly desirable to place strong constraints on neutron star structure as they are likely to have low magnetic fields (10$^{8-9}$ G), resulting in minimal effects on the radiation transport or temperature distribution on the star's surface \cite{2016ApJ...831..184B, campana1998neutron, potekhin2014atmospheres}. Additionally, these binaries are identified in globular clusters where distances, ages, and reddening are well-constrained \cite{Heinke_2003}. The distinctive soft thermal spectra from these sources come from a long-lived thermal glow resulting from heat stored in the deep crust of the neutron stars within the binary system during accretion, which is then re-radiated from the whole surface when accretion stops \cite{brown1998crustal}. For the context of this work, the inference of EOS will come from simulated thermal spectra from qLMXBs.

Observation of neutron star emission, whether X-ray or gravitational wave, has long served as a way to constrain mass and radius for neutron stars (eg. \cite{hebeler2013equation, Steiner18ct}), but uncertainties arise in the inference of these properties for a variety of reasons. In the case of X-ray radiation from qLMXBs, constraints on mass and radius are determined by fitting the emitted spectrum with an appropriate atmosphere model (where the surface composition is known or can be determined by the X-ray spectrum) and combining the spectroscopic measurements with the distance of the source. Models for thermal X-ray radiation are based on a light-element atmosphere, as the lightest element that is present in the atmosphere floats to the top due to rapid gravitational settling on neutron star surfaces \cite{2016ApJ...831..184B}. Atmospheric models used on X-ray spectra from qLMXBs gave the first broad constraints on neutron star radius, and more modern analyses of X-ray spectra have provided tighter constraints on both radius and EOS.

The high-resolution imaging and spectroscopy of NASA’s \textit{Chandra X-ray Observatory} have provided powerful insight into neutron star properties like cooling \cite{cooling, wijnands2017cooling}, mass and radius \cite{2016ApJ...831..184B}, and binary mergers of exotic stars \cite{mag}. \textit{Chandra}'s telescope contains a system of four pairs of mirrors that focus incoming X-ray photons to the Advanced CCD Imaging Spectrometer (ACIS), which measures the energy of each incoming X-ray. The observed spectrum, along with a corresponding instrument response, is then fit to a well-motivated parameterized model. Many such models for spectral fitting exist in \xspec~\cite{xspec}, an X-ray spectral fitting package distributed and maintained by the aegis of the GSFC High Energy Astrophysics Science Archival Research Center (HEASARC). These parameterized models differ for different types of X-ray sources, as well as assumptions about the source's atmosphere, magnetic field, temperature (a full list of models can be found in the \xspec\ manual \cite{xspec}). \xspec\ has been used numerous times in the past to analyze data from \textit{Chandra} as well as other spectrometers like \textit{NICER}, \textit{Nustar}, and \textit{XMM-Newton}, making it a valuable resource for inference of neutron star properties.

\section{Machine Learning}
\label{sec:ml}

Machine learning methods, in particular deep learning, aim to extract useful knowledge from data automatically and are rapidly being applied across many data-rich fields of science~\cite{baldi2021deep}. In regression tasks such as EOS inference, one is interested in constructing a function $f$ whose inputs are the observed data and whose outputs are an estimate of some parameter of interest.  The optimal function $f$ is not known initially, but an approximation can be learned from a set of input-output example pairs.

In order to approximate $f$, machine learning methods first begin with a suitable class $C$ of parameterized functions (e.g. polynomials of a certain degree, neural networks of a certain architecture) and then seek to find the best approximation to $f$ within
the class $C$. This is typically done through a stochastic gradient descent procedure that seeks to iteratively minimize the approximation error on the training set.

The well-known technique of linear regression is the most elementary form of regression, and can be viewed as a form of shallow learning (no hidden layers). Deep learning generalizes linear regression by using multi-layer neural networks as the class $C$ and thus enabling the construction of sophisticated and flexible non-linear approximations. 
With sufficient training data and computing power, deep learning methods can handle large-scale problems with high dimensional data and avoid heuristic simplifications that lose information. It is not uncommon to deal with problems with input sizes in the range of up to $10^9$ examples, each with dimensions of $10^{3-4}$, with neural networks that can have up to $10^{11}$ free parameters. Training sets can range in size from $10^1$ to $10^{10}$ or more.  Unlike shallow learning and linear regression,  deep learning does not require that the number of parameters be equal to the number of training examples~\cite{neyshabur2017exploring}. More recent, attention-based architectures, such as transformers~\cite{transformers,quarks2022baldi}, allow networks to take advantage of structures and symmetries in the data, and are applied in sections below.

When the interpretation of data depends on external unmeasured or poorly-known parameters, such as neutron star temperature or distance, it can be useful to apply  {\it parameterized networks}~\cite{Baldi:2016fzo}. Such networks learn a task as a function of the external parameter, allowing for evaluation of a fixed dataset under varying assumptions about the parameter~\cite{Ghosh:2021roe}.

\section{TRAINING SAMPLES}
\label{sec:data}

Samples of simulated neutron stars, used to train networks and evaluate their performance, are described below. 

Each simulated star is described by two high-level summary quantities, mass and radius, which are drawn from the mass-radius relation determined by the EOS, as well as three nuisance parameters that are independent of the EOS and can vary from star to star.  These five parameters are sufficient to determine the expected simulated Chandra telescope spectrum in the chosen NS theoretical model. In the case of EOS inference, sets of stars with consistent EOS are grouped to form training and testing sets. Details of each step of the generation are provided below.

\subsection{Equation of State}

The equation of state of the hadronic matter within the core is modeled with the relativistic non-linear mean field model GM1L~\cite{Typel}. The version used here only accounts for protons and neutrons but can be extended to include hyperons and $\Delta$ baryons~\cite{Spinella:2020WSBook}. The corresponding saturation properties of symmetric nuclear matter for the GM1L parametrization are shown in Table \ref{tab:gm1l} \cite{Spinella:2020WSBook,Malfatti19}. These properties include the nuclear saturation density $n_0$, energy per nucleon $E_0$, nuclear compressibility $K_0$, effective nucleon mass $m^*_N/m_N$, asymmetry energy $J$, asymmetry energy slope $L_0$, and the value of the nucleon potential $U_N$. The value of $L_0$ listed in Table \ref{tab:gm1l} is in agreement with the value of the slope of the symmetry energy deduced from nuclear experiments and astrophysical observations \cite{Oertel}.

The most commonly used constraints on $K_0$ come from  experimental values of the giant monopole resonance, which lie in the range of 220 to 260 MeV \cite{Shlomo2006,Garg_2018}. The analysis of \cite{Stone2014}, however, suggests a higher range of 250 to 315 MeV. The value of $K_0=300$~MeV considered in our paper falls into the latter category, but this will not dramatically impact the neutron-rich equation of state appropriate for neutron star interiors. The GM1L equation of state for the core is paired with two models for the crust. For the outer crust, which falls in the density range $10^4-10^{11} ~\text{g/cm}^3$, we use the Baym-Pethick-Sutherland (BPS) model \cite{1971ApJ...170..299B}. For the inner crust, with densities in the range $10^{11}-10^{14} ~\text{g/cm}^3$, we use the Baym-Bethe-Pethick (BBP) model \cite{BAYM1971225}.

\begin{table}[]
    \caption{Parameters of the model used to select example equations of state for the generation of simulated data samples. Shown are properties of the  symmetric nuclear matter at saturation density for the GM1L parametrization of neutron star interiors\cite{Typel}; see text for details.}
    \label{tab:gm1l}
    \centering
    \begin{tabular}{ccc}
             \hline \hline
         Saturation Property ~~~~& Value ~~~~&  Units \\
                  \hline 
        $n_0$ & 0.153 & fm$^{-3}$ \\
        $E_0$ & $-16.3$ & MeV \\
        $K_0$ & 300.0 & MeV \\
        $m^*_N/m_N$ & 0.70 & - \\
        $J$ & 32.5 & MeV \\
        $L_0$ & 55.0 & MeV \\
        $U_N$ & $-65.5$ & MeV \\
         \hline \hline
    \end{tabular}
\end{table}

To limit the number of parameters the networks must learn, the essential features of the high-density portion of the EOS needed to be represented efficiently by just a few values. This can be done accurately by constructing parametric representations based on spectral fits, formed as generalizations of the Fourier series used to represent periodic functions \cite{PhysRevD.82.103011}. An EOS, defined as $P = P(\epsilon)$ or $\epsilon(P) = \epsilon$, can be represented as a linear combination of basis functions $\epsilon_k(\phi)$: 
\begin{equation} \label{eq:ep}
    \epsilon(p) = \sum_{k}\epsilon_k \Phi_k(p)
\end{equation}

\noindent where $\phi_k(p)$ can be any complete set of functions. The EOS is therefore determined by the spectral coefficients $\epsilon_k$, making $ \epsilon_k = \epsilon(p,\epsilon_k)$. There are two important conditions that a physical EOS must satisfy to ensure microscopic stability. The first is that the EOS must be non-negative, or $p(\epsilon) \geq  0$, and the second is that pressure must be monotonically increasing with density \cite{PhysRevD.82.103011}. Because these conditions are not naturally respected by arbitrary basis functions in a spectral representation, representing an EOS with a straightforward spectral expansion will likely produce data that violates microscopic stability and is therefore erroneous. To ensure these two conditions are met, we instead turn to a faithful construction of spectral representations of the EOS; the process for constructing these is outlined in detail in \cite{PhysRevD.97.123019} and \cite{PhysRevD.82.103011}. 

The spectral representation of GM1L is formed from representing the EOS in terms of the relativistic enthalpy, $h$, where the EOS can be rewritten as a pair of equations $P = P(h)$ and $\epsilon = \epsilon(h)$. The enthalpy can be defined as
\begin{equation} \label{eq:hp}
    h(P) = \int_{0}^{P} \frac{dP'}{\epsilon(P')c^2+P'}
\end{equation}
where $c$ is the speed of light \cite{PhysRevD.97.123019}. Inverting Eq.~\ref{eq:hp} obtains the equation $P = P(h)$, which can recover the EOS $\epsilon(P)$ as $\epsilon(h) = \epsilon[P(h)]$. The pair of equations $P = P(h)$ and $\epsilon = \epsilon(h)$ can be expressed from a reduction to quadrature:
\begin{equation}
    P(h) = P_0 + (\epsilon_0c^2+P_0) \int_{h_0}^{h} \mu(h')dh',
\end{equation}
\begin{equation}
    \epsilon(h) = -P(h)c^{-2}+(\epsilon_0+p_0c^{-2}) \mu(h).
\end{equation}

The function $\mu(h)$ is defined as
\begin{equation}
    \mu(h) = \exp \left \{ \int_{h_0}^{h} [2+\Gamma(h')]dh'  \right \}
\end{equation}
which is dependent on the sound speed or velocity function $\Gamma(h)$ and constants $P_0 = P(h_0)$ and $\epsilon_0 = \epsilon(h_0)$ \cite{PhysRevD.97.123019}. Similar to Eq.~\ref{eq:ep}, the velocity function can be represented as a spectral expansion:
\begin{equation}
    \Gamma(h,\nu_k) = \exp \left[  \sum_{k} \nu_k\Phi_k(h) \right]
\end{equation}
where $\Phi_k(h)$ is any complete set of basis functions on the domain [$h_0,h_{max}$]. 

\begin{figure}[hbt!]
    \centering
        \includegraphics[width=0.5\textwidth]{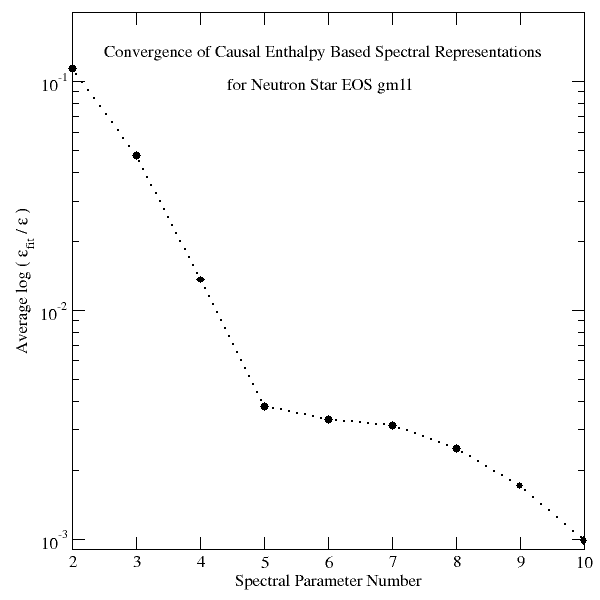}
    \caption{ Relative error in spectral parameterization of the equation of state, shown as a function of the number of parameters used.}
    \label{fig:sp_err}
\end{figure}

We constructed enthalpy-based causal spectral fits for the GM1L EOS with up to 10 parameters. Figure \ref{fig:sp_err} shows that just two spectral parameters produce parameterizations with a mean relative error of only 10\%, and additional parameters can reduce the error to 5\% or lower.  A small number of parameters is preferred due to the increased complexity of learning multiple parameters, and the danger of  Runge's phenomenon when applying our networks to current neutron star observations with accurate readings of mass and radius, which are still relatively few. Runge's phenomenon arises when attempting to fit equispaced data points with polynomials of high degree; increasing the order of the polynomial interpolation can result in issues with convergence or divergence rates for certain functions \cite{runge_phen}. When applied to neutron star observation, attempting to fit a small number of data points (with varying accuracy in observation method) with a model having many parameters may result in fits that accurately fit the data, but are very poor representations of the actual physics. Based on the reasons listed above, we chose to use two spectral parameters to represent the EOS, hereafter referred to as $\lambda_1$ and $\lambda_2$.

\begin{figure}[hbt!]
    \centering
        \includegraphics[width=0.42\textwidth]{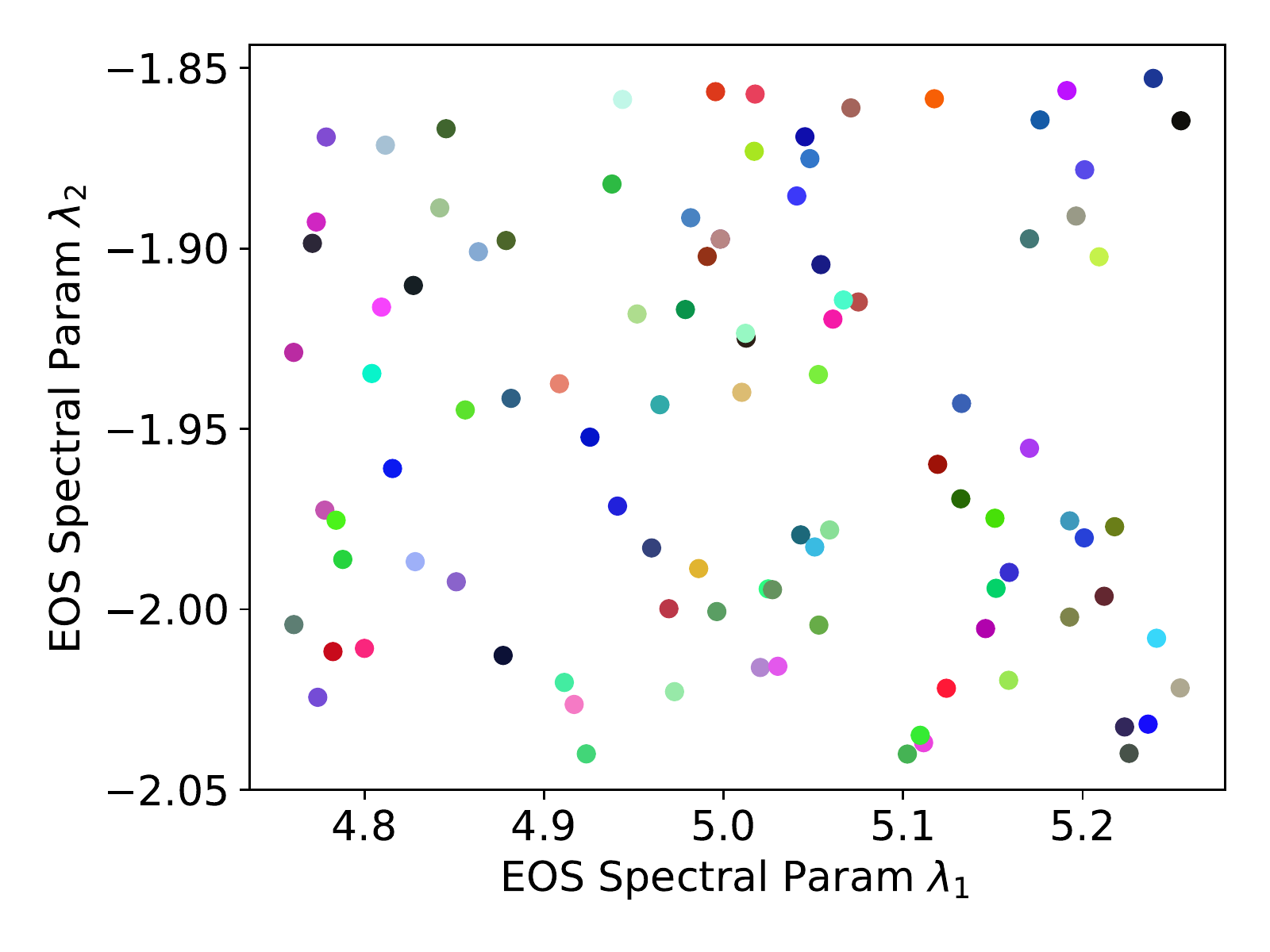}
    \includegraphics[width=0.42\textwidth]{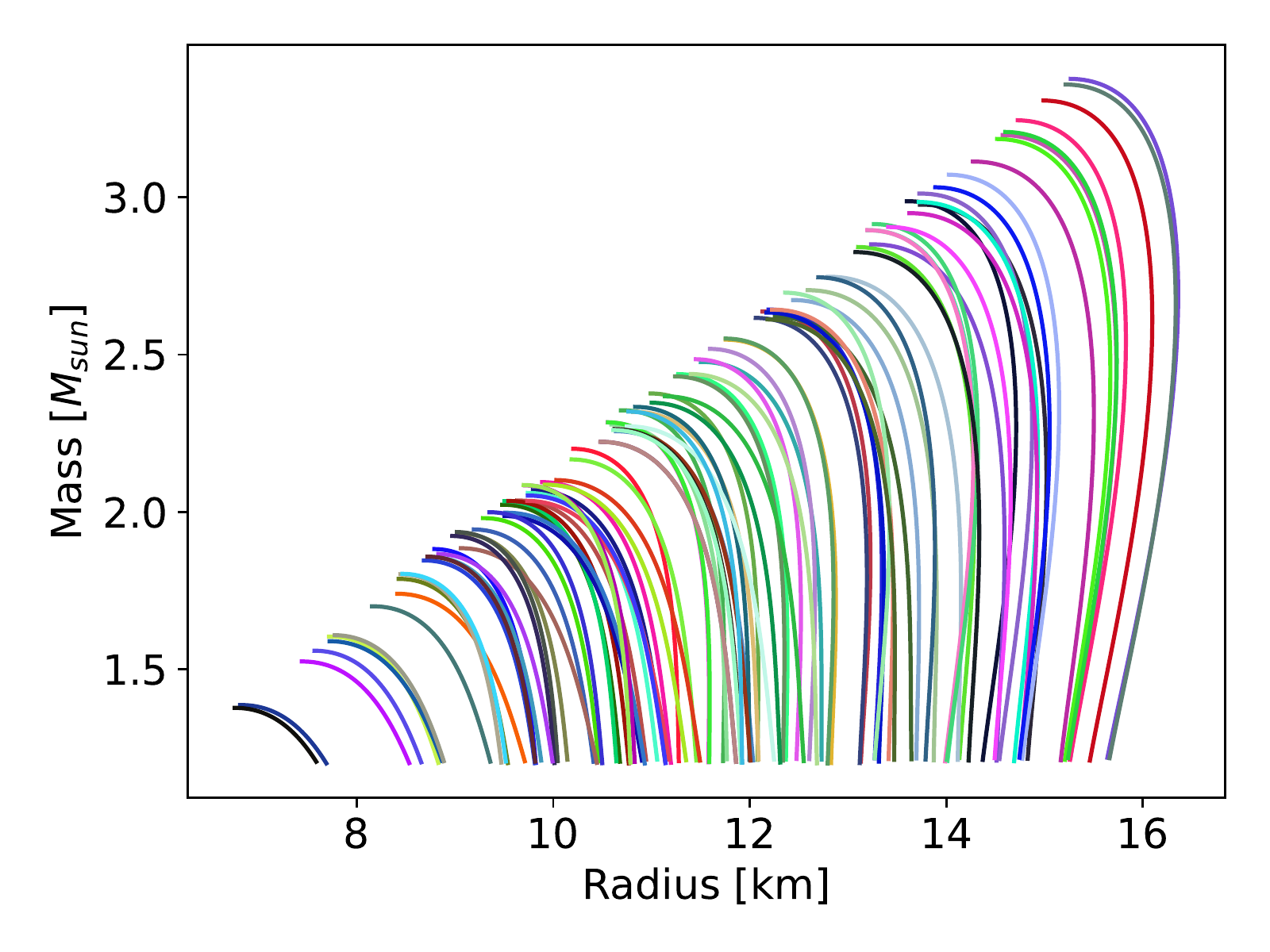}
    \caption{Examples of training data. Top, 100  samples in EOS spectral parameter space $(\lambda_1,\lambda_2)$ randomly selected from the full set of 10,000 EOS spectral pairs. Bottom, neutron star mass-radius curves determined by the selected EOS parameters.}
    \label{fig:tdata_mr}
\end{figure}

To create many samples needed for training and testing, spectral parameters were then constructed from the expression:
\[\lambda_{\text{generated}} = \lambda_{\text{true}} \cdot(1+2 \cdot\mathit{scale} (-0.5+\mathit{ran2}))\]
where $\lambda_{\text{generated}}$ represents the newly constructed spectral parameter, $\lambda_{\text{true}}$ is the best fit (true) spectral parameter of GM1L, and $\mathit{scale}$ is a scaling parameter set to 0.05.  $\mathit{ran2}$ are uniformly distributed random numbers in the range 0 to 1 generated by the $\mathit{ran2}$ function given in \cite{Press2007}. This process was repeated to create $10^4$ different EOS variations. Each EOS variation was used to generate a coinciding $M-R$ relation using equations \ref{eq:1} \ref{eq:2}, examples of which are seen in Figure ~\ref{fig:tdata_mr}, from which 100 $(M,R)$ pairs are selected, each representing stellar parameters consistent with that EOS. Due to the random component of our EOS generation, some models have a mass peak below the current observed mass limit, 2.1 $M_\odot$. All models have a minimum mass of at least 1 $M_\odot$. The physicality of predicted results will be discussed in further detail \ref{sec:Discussion}.

\subsection{Modeling X-ray Spectra}

The relation between stellar parameters $(M,R)$ is determined by the EOS, and samples from the allowed curve are used as input to generate simulated X-ray Chandra spectra, such as the Chandra observation of the quiescent low-mass X-ray binary (qLMXB) X7 in the globular cluster 47 Tuc~\cite{2016ApJ...831..184B}. 

The \xspec\ program~\cite{xspec}, which can be used for spectral fitting, is also capable of generation of simulated spectra, via the  \texttt{fakeit} command when a NS model and telescope response matrix are provided.


The NS theoretical model \texttt{NSATMOS}~\cite{Heinke_2006} selected includes a hydrogen atmosphere model with electron conduction and self-irradiation. The Chandra telescope response specified in Ref.~\cite{Heinke_2006} was also used to describe the instrument response and telescope effective area.

\begin{figure}[hbt!]
    \centering
    \includegraphics[width=0.29\textwidth]{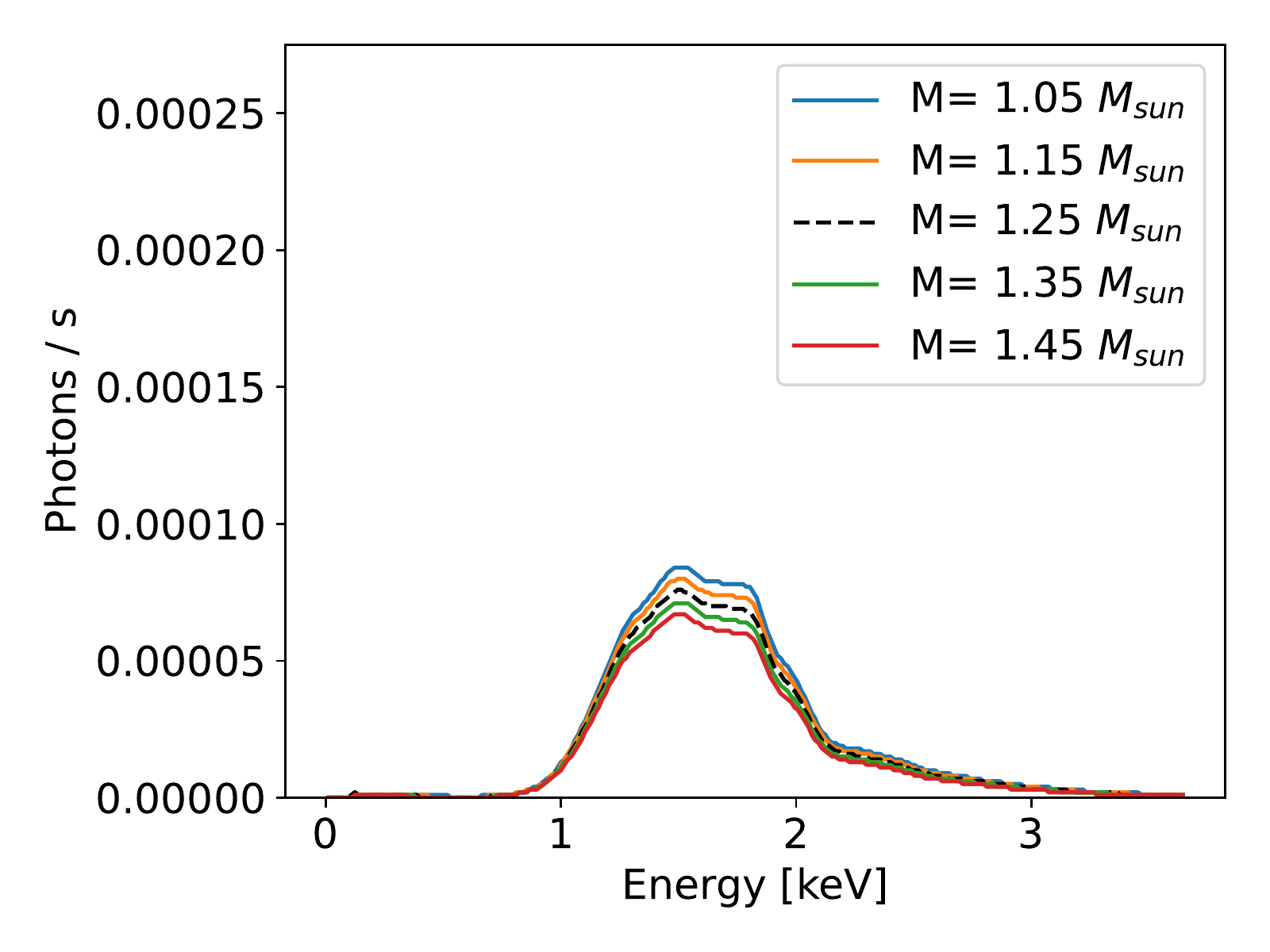}
    \includegraphics[width=0.29\textwidth]{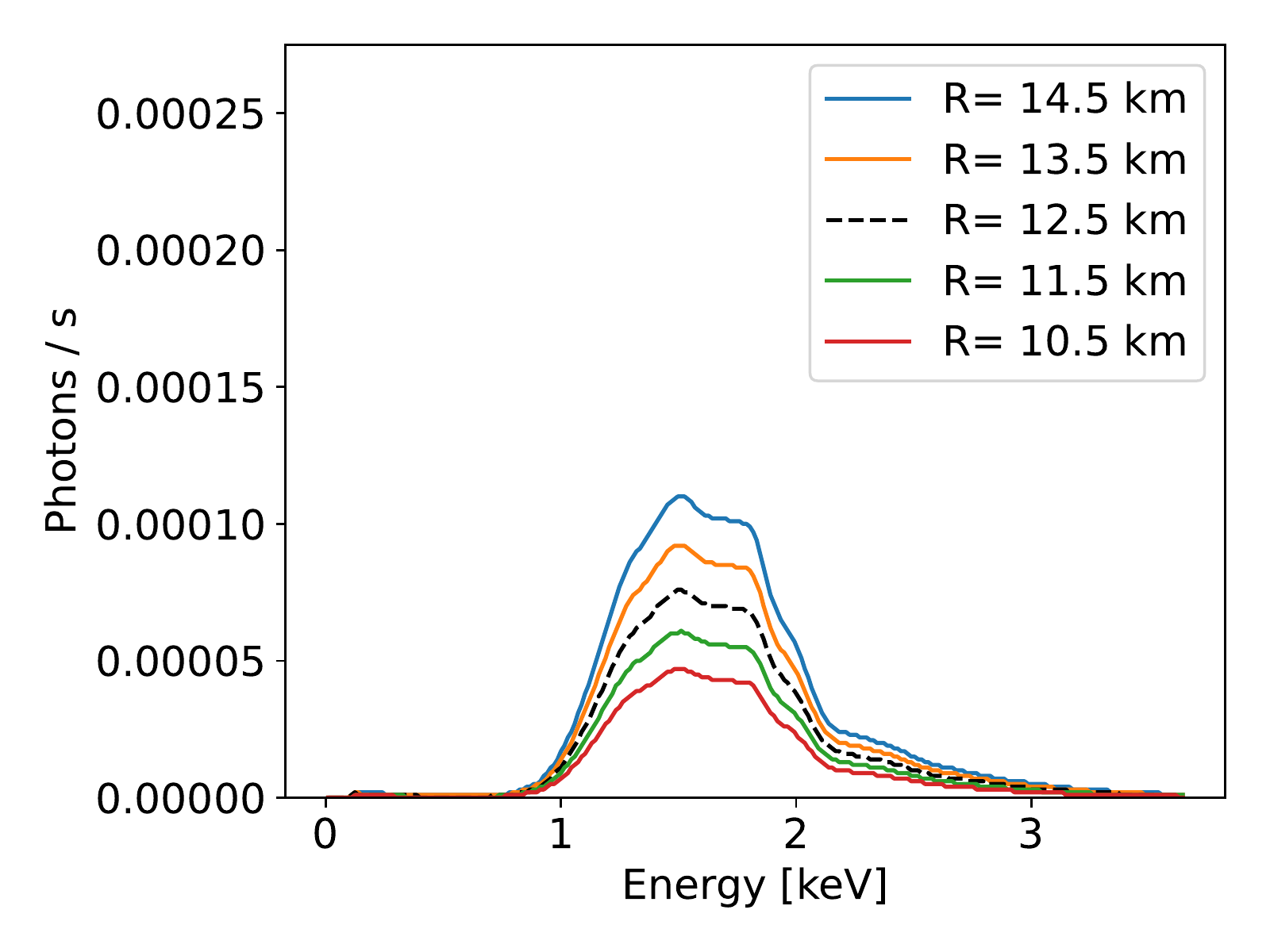}
    \includegraphics[width=0.29\textwidth]{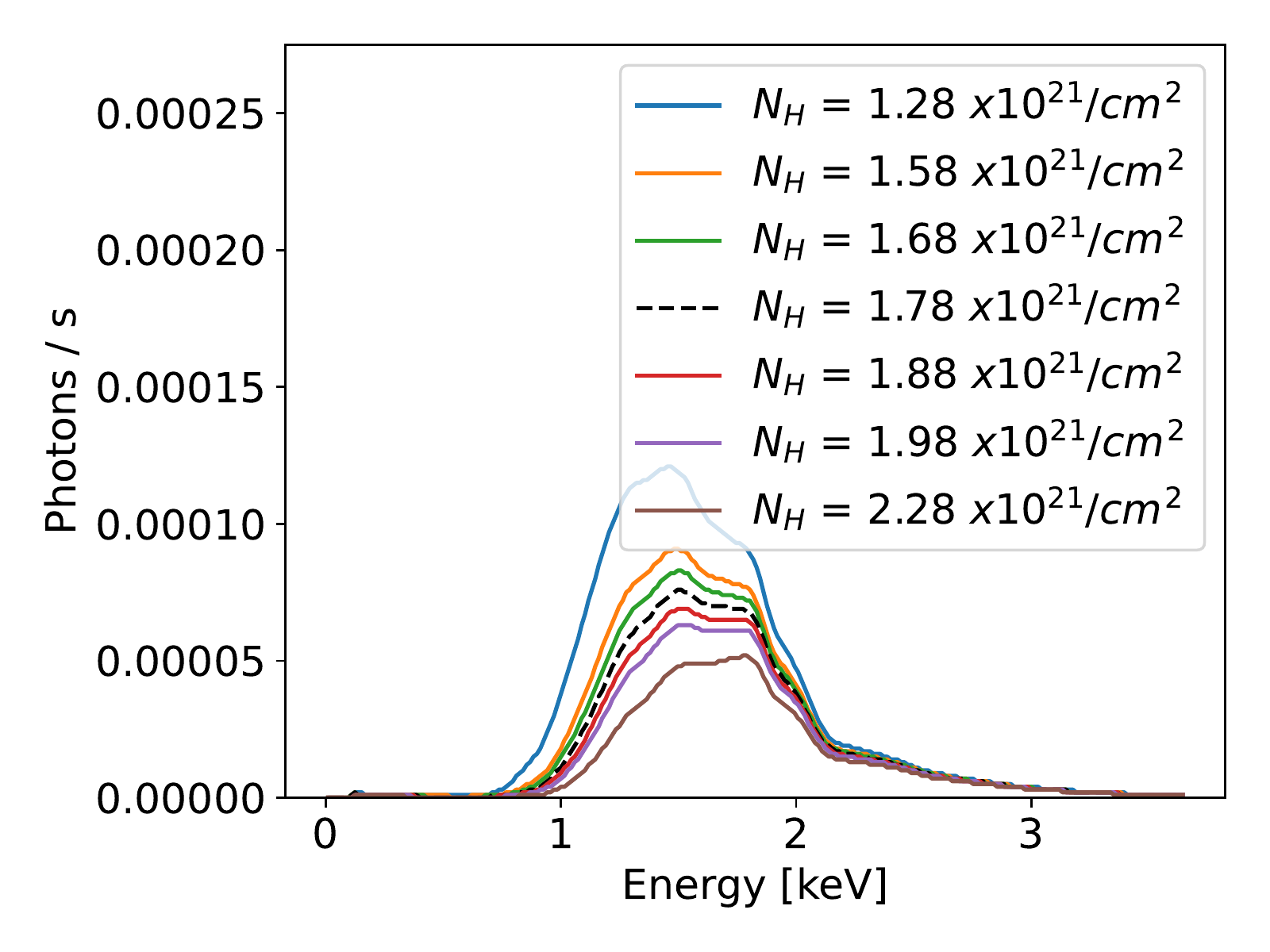}
    \includegraphics[width=0.29\textwidth]{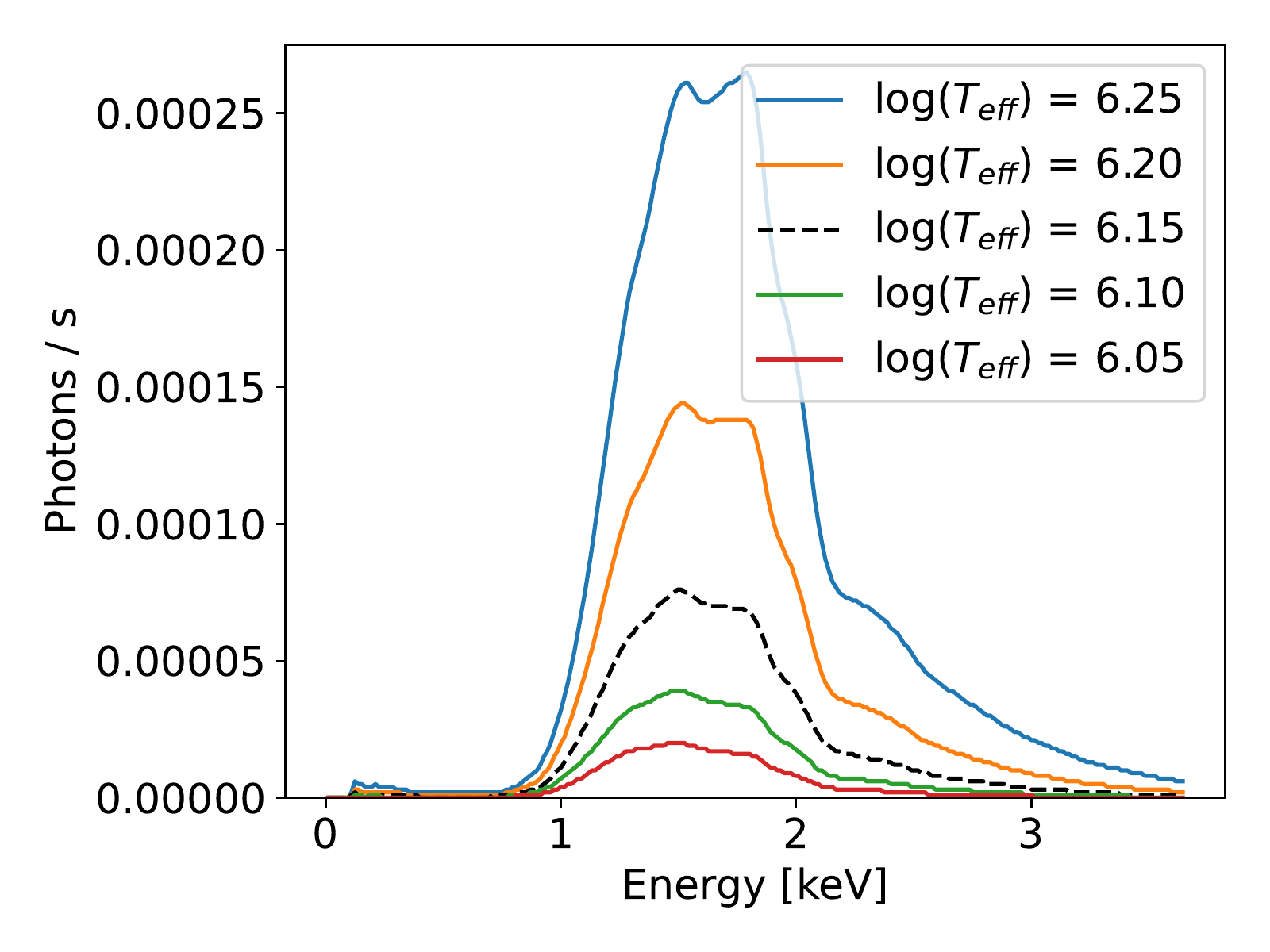}
    \includegraphics[width=0.29\textwidth]{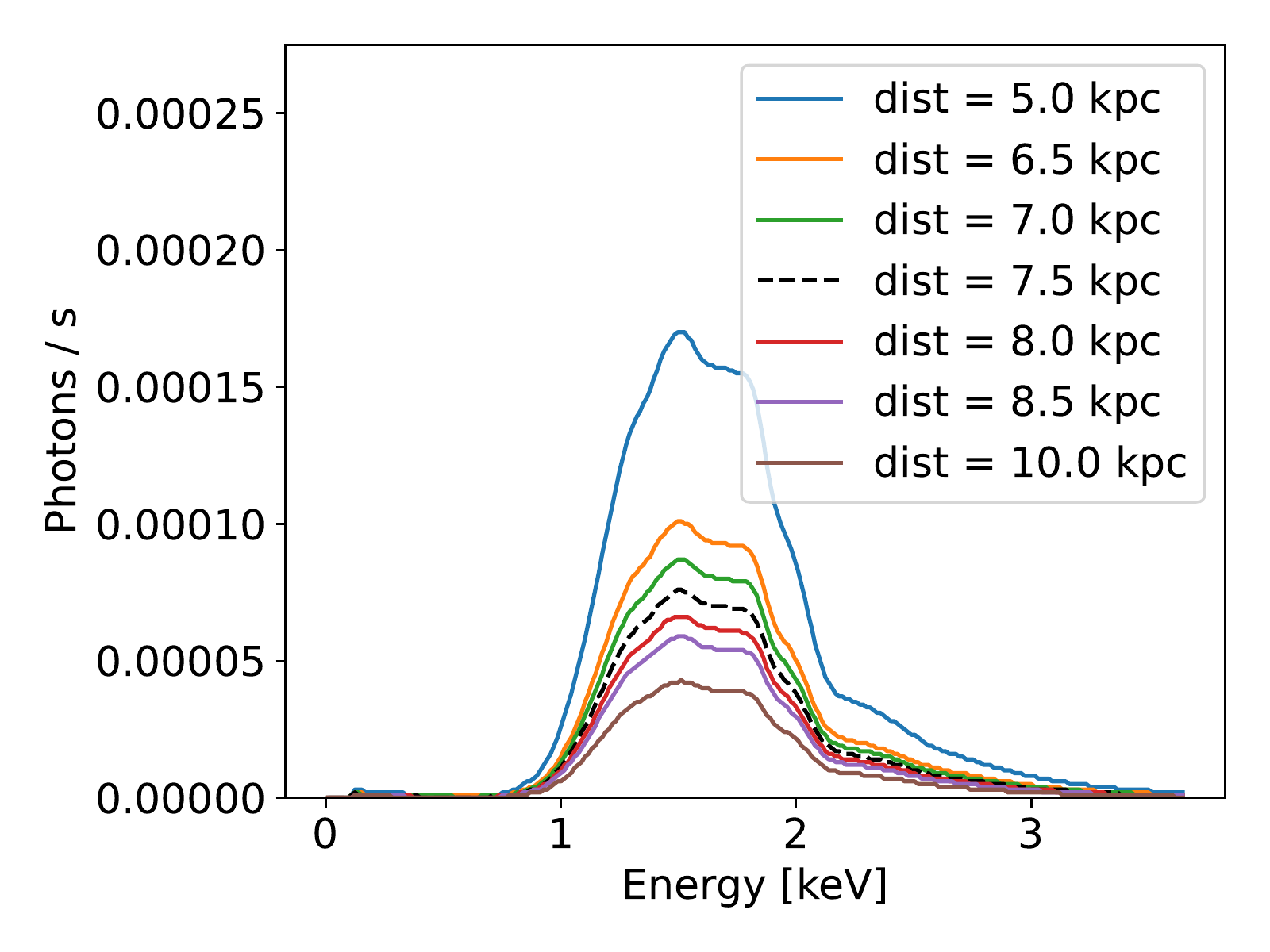}
    \caption{ Examples of simulated stellar spectra expected for several values of stellar parameters.  Each pane shows the expected rate of photons in Chandra per energy bin, for variations of the parameters of interest (mass M, radius R) as well as for variations of the nuisance parameters ($N_H$, log($T_{\textrm{eff}}$), distance). The dashed black line has the same parameters in each pane.}
    \label{fig:tdata_spectra}
\end{figure}

\subsection{Nuisance Parameters}

The \texttt{NSATMOS} model has five parameters to describe each star:
 gravitational mass $M$ in units of $M\odot$, radius $R$ in units of km, and three additional parameters related to observation. For the context of $M-R$ and subsequent EOS inference, only $M$ and $R$ are parameters of interest, whose values come from those generated by the GM1L EOS and so provide information relevant to the physical question. The remaining three nuisance parameters are the effective temperature of the surface, $T_{\mathrm{eff}}$, the distance to the star, $d$, and the hydrogen column, $N_H$ which parameterizes the reddening of the spectrum by the interstellar medium.  These parameters influence the observed spectrum of a given neutron star. Lack of knowledge of these values is a leading source of uncertainty in the inference of mass and radius, and hence EOS.
 
 Using Table 1 in Ref.~\cite{Steiner18ct} as a guide, we find that distances typically range between 2 and 10 kpc, and hydrogen columns lie between 0.2 and $5 \times 10^{21}~\mathrm{cm}^{-2}$. While neutron stars with larger distances and larger hydrogen columns exist, they are sufficiently distant as to be difficult to obtain spectral information. From Table 3 in Ref.~\cite{Lattimer14ns}, effective temperatures at the surface typically lie between 50 and 200 eV, or from $6 \times 10^5$ and $2.4 \times 10^6$ K. Note that core temperatures are typically a few orders of magnitude larger. Again colder neutron stars most certainly exist but are more difficult to observe.
    
Examples of generated spectra for varying stellar parameters are shown in Figure ~\ref{fig:tdata_spectra}. The generated spectra are very sensitive to the effective surface temperature, with lesser sensitivity to other parameters. The dependence of the curves in Figure ~\ref{fig:tdata_spectra} to the changing nuisance parameters is not surprising: roughly proportional to radius and distance squared, but higher power in temperature.

The networks detailed below provide estimates of either the neutron star mass and radius or the EOS parameters, conditioned on NP values. Uncertainty in regressed parameters of interest due to uncertainties in the NP can then be fully propagated via variation of the NPs used during regression.  To demonstrate the impact of NP uncertainties, we define three example scenarios of uncertainties, dubbed ``true", ``tight", and ``loose", which describe the quality of prior information on the NP values for each star. 
     
     In the ``true" scenario, the NPs are set to the true value used to generate the spectra, such that the NP prior is essentially a delta function. In the ``tight" scenario, the uncertainty is described as a narrow  Gaussian for each NP, with distance having a width of 5\%, hydrogen column having a width of 30\%, and $\log(T_\textrm{eff})$ having a width of 0.1.
  In the ``loose" scenario, the uncertainties are described by a wider Gaussian, with distance having a width of 20\%, hydrogen column having a width of 50\%, and $\log(T_\textrm{eff})$ having a width of 0.2. These ranges are shown in Table \ref{tab:nps}. The sensitivity to NP values is reflected in the performance of the networks below.

\begin{table}[]
    \caption{Description of ``true", ``tight", and ``loose" nuisance parameter (NP) scenarios. Shown are the width of each Gaussian distribution representing the prior knowledge of each NP. For distance and $N_H$, width is relative; for log($T_{\text{eff}}$), it is absolute. See text for details and references.}
    \label{tab:nps}
    \centering
    \begin{tabular}{lcrrr}
             \hline \hline
         Nuis. Param. & True & Tight & Loose \\
                  \hline 
        Distance & exact  & 5\% & 20\% \\
        Hydrogen Column $N_H$ &exact & 30\% & 50\% \\
        log($T_{\text{eff}}) $ &exact    & $\pm$0.1  & $\pm$0.2  \\ 
         \hline \hline
    \end{tabular}
\end{table}


\section{Inference of EOS from Mass and Radius}
\label{sec:eosmr}

Previous applications of machine learning to the task of inferring the equation of state have begun from the stellar mass and radii~\cite{Fujimoto:2019hxv,Fujimoto:2021zas,Morawski:2020izm}, or equivalent parameters~\cite{Ferreira:2019bny}, though with simple ad-hoc descriptions of the uncertainty on stellar mass and radius values, often modeled as two-dimensional uncorrelated Gaussians rather than fully propagating the underlying uncertainties. In this section, we tackle the same problem, but where the stellar data are more realistic and the underlying uncertainties are fully propagated to the EOS estimation.  Specifically, the best estimates of stellar mass and radius are derived using state-of-the-art tools that extract them from realistic stellar spectra, which include the impact of stellar nuisance parameters and limited observation time. In addition, this mass-radius estimation is conditioned on the nuisance parameters, such that variations in those nuisance parameters lead to variations in the mass and radius estimates.  This connects directly to the neural network regression of EOS parameters from mass-radius values parameterized in the nuisance parameters, allowing for the direct propagation of the underlying uncertainties to give a measure of the resulting uncertainty on the regressed EOS parameters.

Below, we describe the extraction of realistic mass and radius values with \xspec\, and their subsequent use in NN regression of the EOS parameters and the estimation of the uncertainty.  For comparison, we also provide a demonstration of the regression of EOS parameters using polynomial regression. In subsequent sections, we consider an alternative extraction of mass and radius using a NN, as well as end-to-end regression of EOS directly from stellar spectra. 

\subsection{Mass and Radius inference by XSPEC}

Sample stellar spectra are generated as described above, including Poisson noise corresponding to an observation time of 100 ks, and nuisance parameters variations as specified in Table ~\ref{tab:nps}.

Given a sample observed X-ray spectrum, the \xspec\  code scans the mass and radius parameter space, searching for values that best describe it. For each mass-radius pair, the expected spectrum is calculated using the chosen model and telescope response function,  identical to those used to generate the sample spectra being fit.  The fitted values are those which minimize a bin-wise $\chi^2$, and reported errors are those which generate a fixed increase in the $\chi^2$ metric.

\begin{figure}
    \centering
    \includegraphics[width=0.35\textwidth]{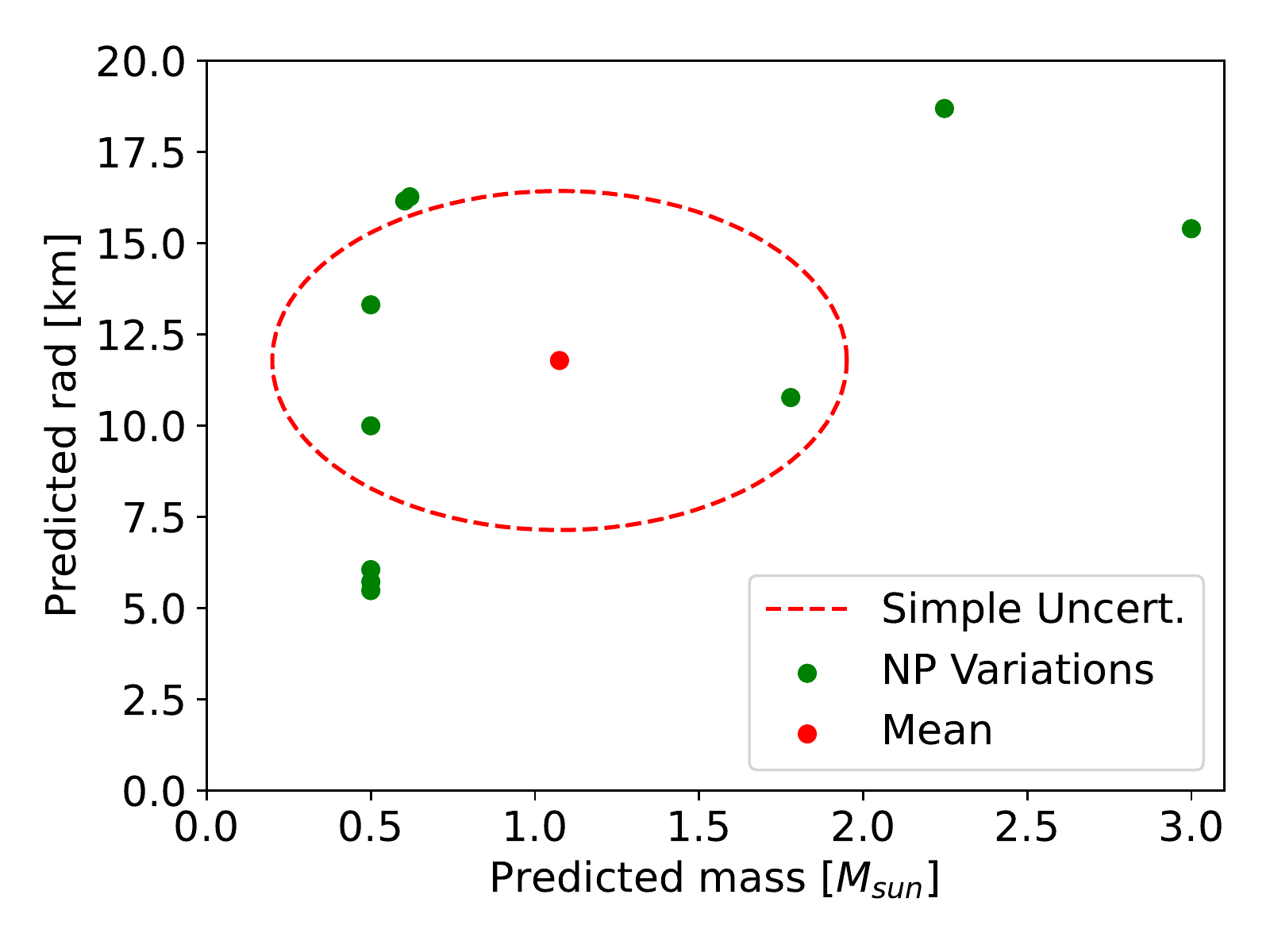}
        \includegraphics[width=0.35\textwidth]{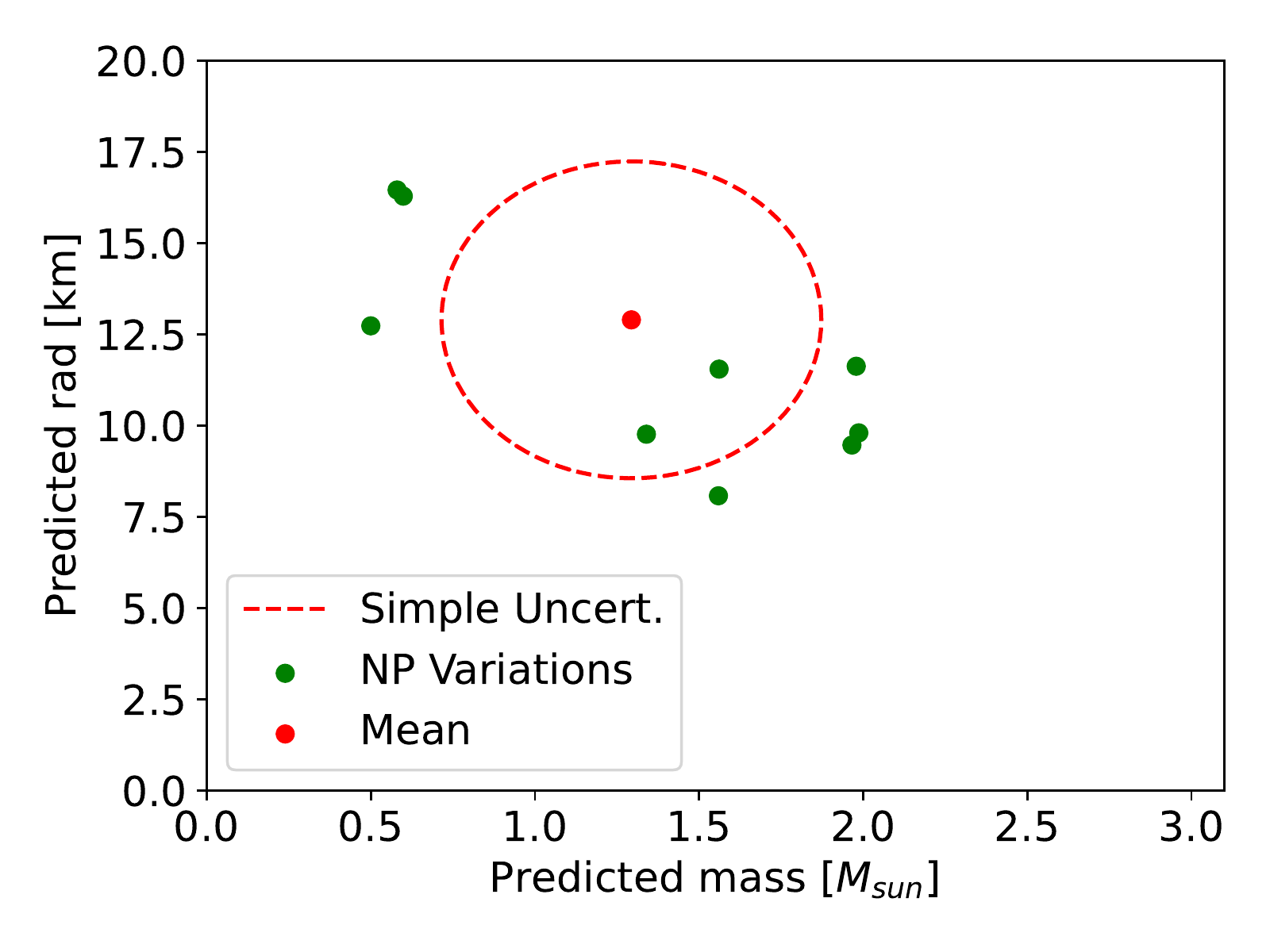}
 \includegraphics[width=0.35\textwidth]{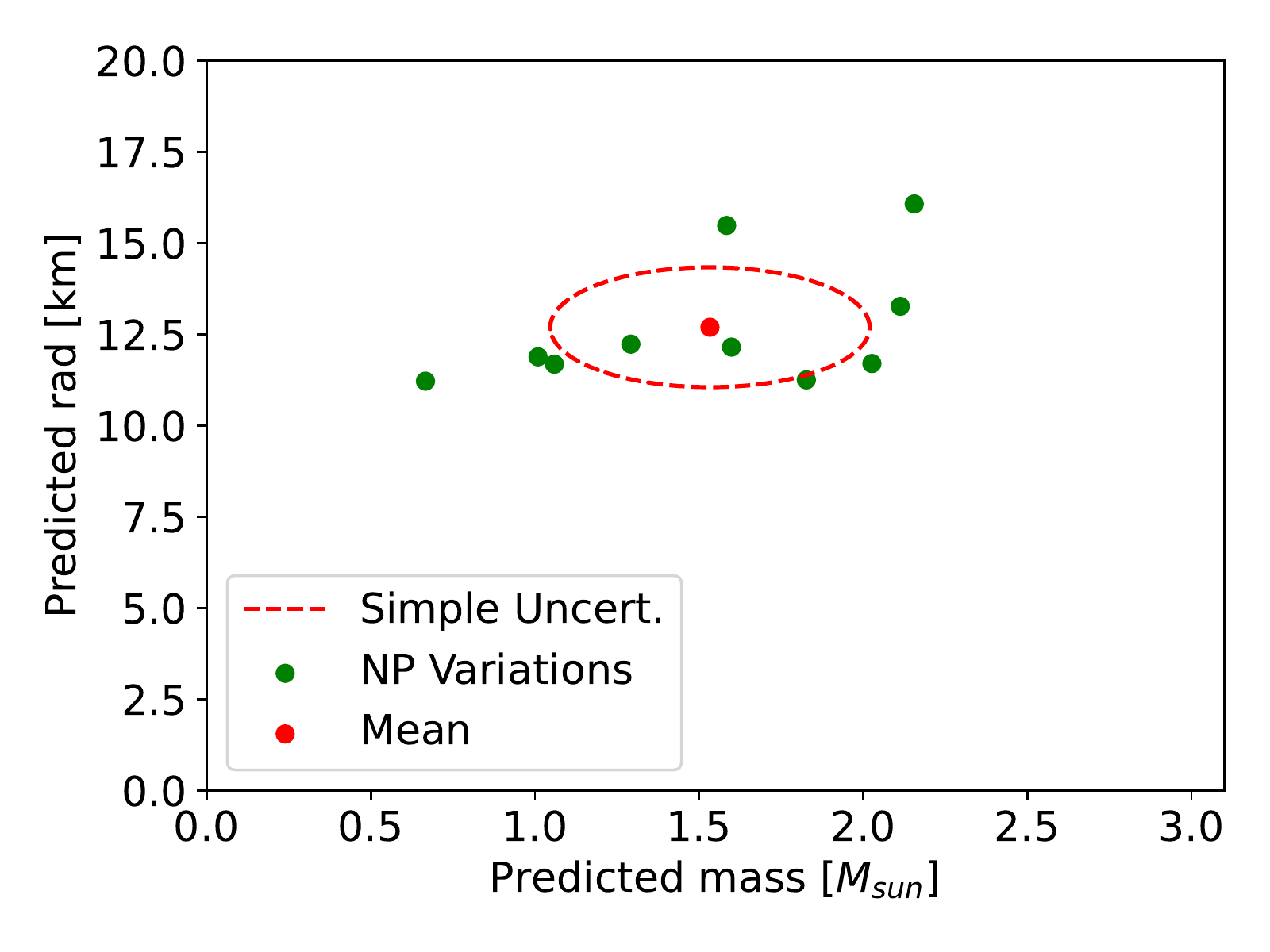}
        \includegraphics[width=0.35\textwidth]{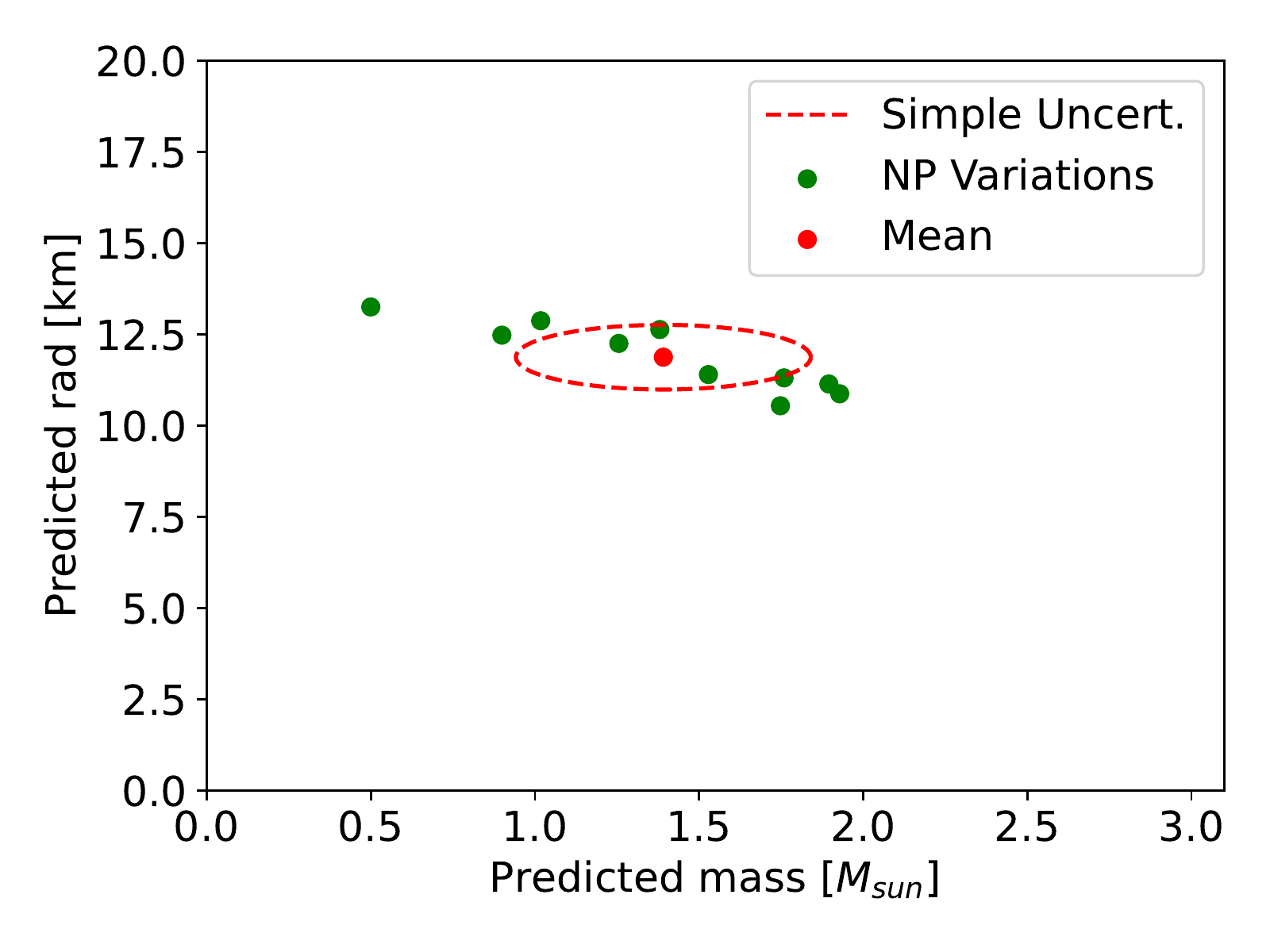}
    \caption{ Estimation of the mass and radius of a neutron star from the underlying stellar spectra, by \xspec. Each pane represents one star, and shown (green) are estimates for several independent values of the nuisance parameters drawn from the associated priors, and the mean value (red). Top two cases have loose priors, bottom two have tight. The dashed ellipse, whose widths are set to the standard deviation of the mass and radius estimates, is a demonstration of the inadequacy of a simple uncertainty model.}
    \label{fig:xspec_mr_np}
\end{figure}

To propagate the uncertainty due to the lack of knowledge of the NP values, the fit on a given spectrum is performed several times with varying assumed values of the NPs drawn from the appropriate prior. The variation in the resulting fitted values then describes the impact of uncertainty on the NPs.  For this reason, during each single \xspec\ fit, the NP values are not allowed to vary, but are frozen.  Figure~\ref{fig:xspec_mr_np} shows examples for individual stars, demonstrating the variation of the stellar parameter estimates with varying NP values. For illustrative comparison to ad-hoc models of uncertainty, the standard deviation in mass and radius are used to define the widths of a 2d error ellipse, though it is clear that this fails to capture the complex nature of the impacts of the underlying uncertainties; these simple error models are not used in our analysis.

\xspec\  is also capable of {\it floating} the nuisance parameters, varying their values to improve an individual fit, and reporting an uncertainty envelope in the mass-radius plane. This can be helpful in the case where the mass and radius and their envelope are the final targets. However, to propagate the uncertainty downstream requires that we have the full posterior in the mass-radius plane or samples from it. An estimate and envelope do not provide that capacity, though they can allow for ad-hoc parameterizations of the prior as have been performed previously. We condition on the nuisance parameters to allow full propagation of the NP uncertainty through to EOS estimation, as we do below. 

\begin{figure}
    \centering
   \includegraphics[scale=0.4]{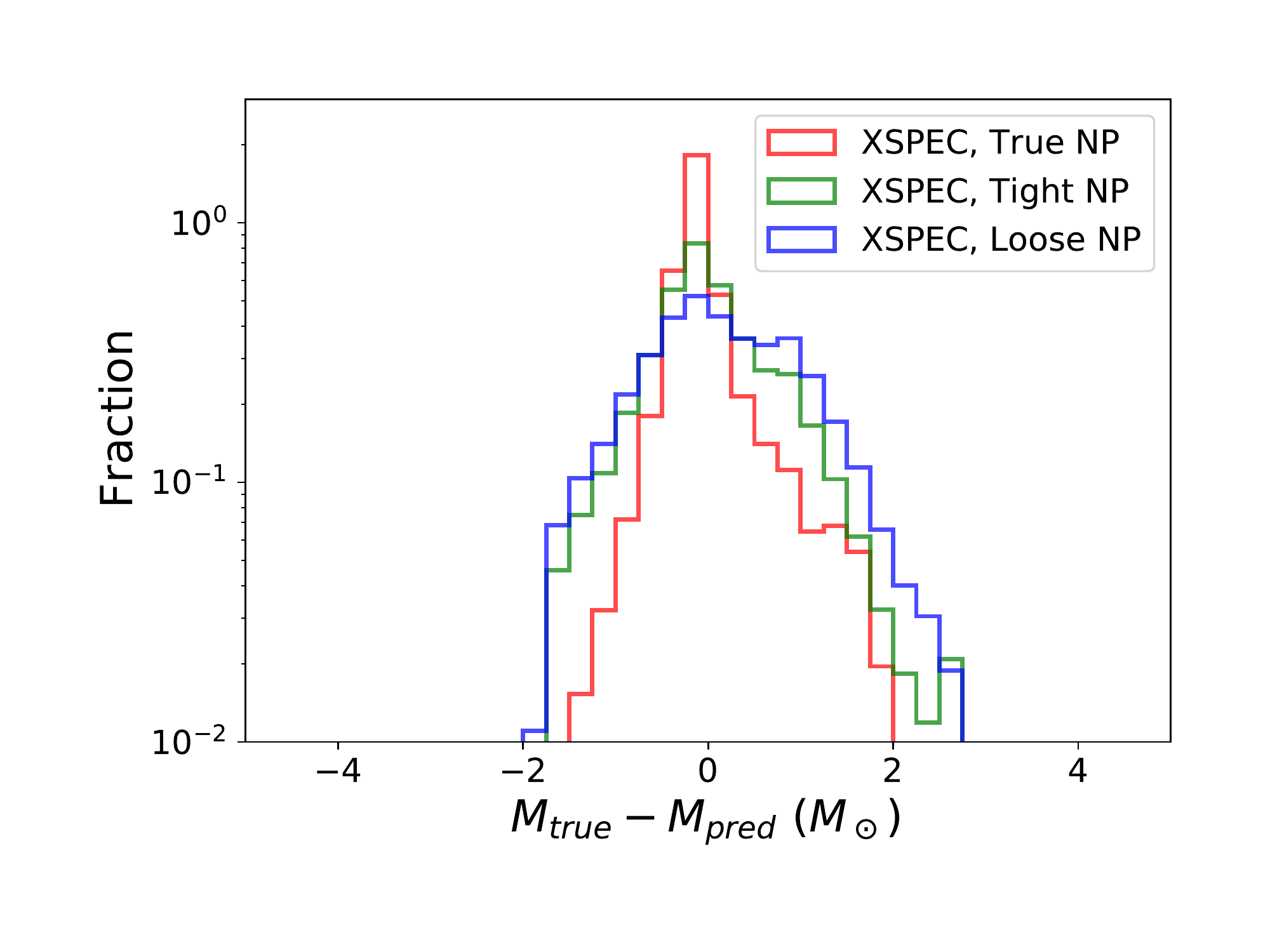}
   \includegraphics[scale=0.4]{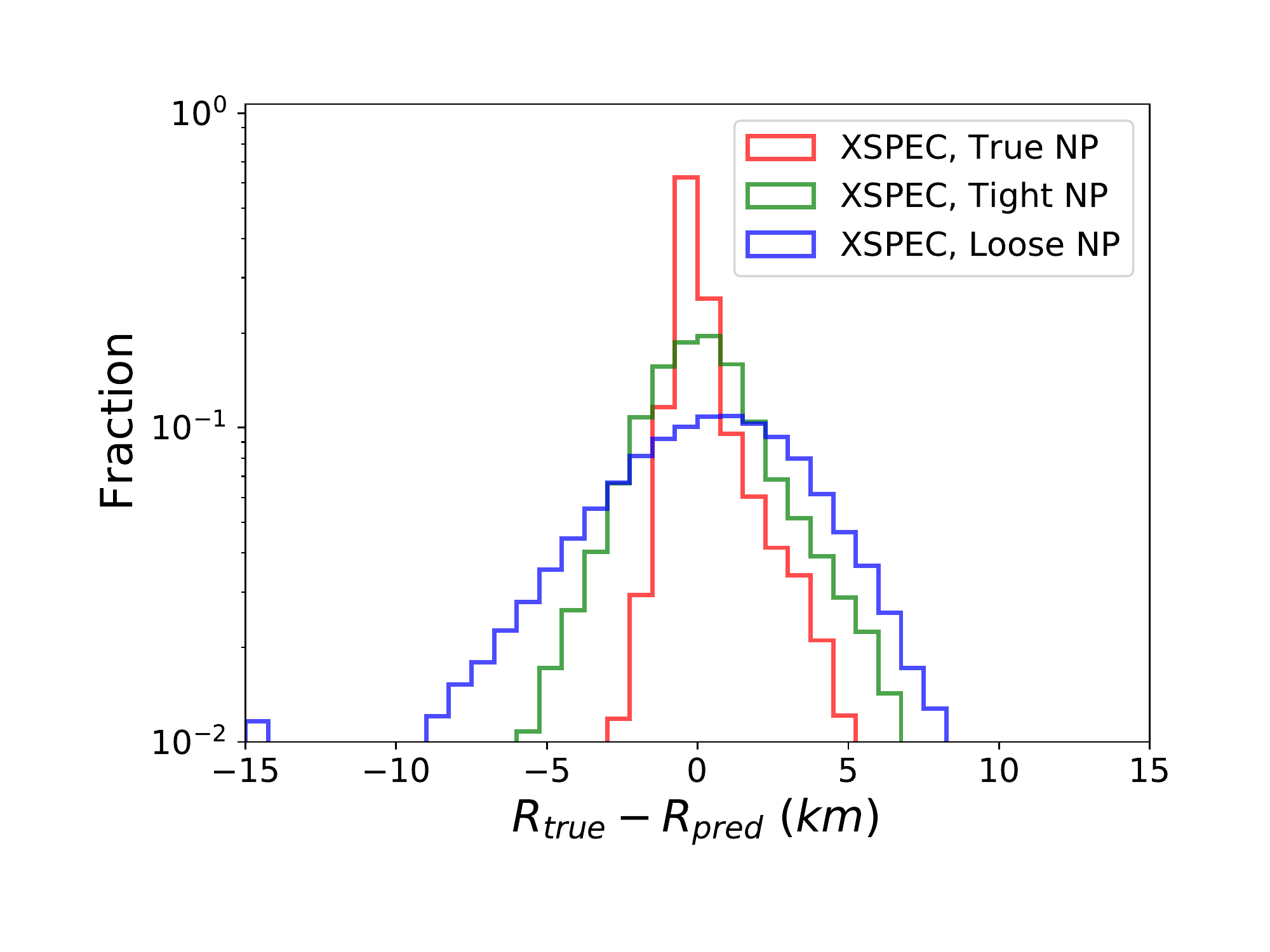}
    \caption{ Performance of \xspec\ inference of neutron star mass and radius, as measured by residuals between the fitted ("pred") and true values under three treatments of the nuisance parameters (NPs). In the ``true" case, the NPs are fixed to their true values; in the ``tight" and ``loose" cases, they are drawn from narrow or wide priors, respectively; see text for details. Cases in which \xspec\  fails to converge are shown as large negative residuals.}
    \label{fig:xspec_mr}
\end{figure}

Performance of \xspec\  regression of mass and radius is shown in Figure ~\ref{fig:xspec_mr}, where the residuals increase as expected with wider priors on the nuisance parameters.  In addition, note that in the case of the ``loose'' priors, there is a small fraction of cases where \xspec\ fails to converge on an estimate, as the nuisance parameters are fixed to a value far from the value used to generate the spectrum. 

One important note regarding \xspec's performance is the same theoretical model, \texttt{NSATMOS}, is used in both the data generation and regression. Because of this, the regression models discussed in the next section can, at best, match \xspec's performance for the evaluation dataset. Nonetheless, it is an important step in demonstrating the capacity of these methods to perform such inference without explicitly relying on a single theoretical model.


\subsection{Inference of EOS from mass and radius}

In this section, we demonstrate the inference of EOS parameters from the stellar mass and radius data, using neural network regression parameterized in the NPs to allow propagation of the uncertainty. In addition, we build polynomial regression models which serve as a benchmark, following Ref.~\cite{Fujimoto:2019hxv}.


\subsubsection{Neural Network regression}
Deep feed-forward neural networks are trained to provide the EOS parameters given a collection of ten stars, each represented by their mass and radius. Each network has two outputs, $\lambda_1$ and $\lambda_2$.

We train three networks, one for each of the true, tight, and loose NP scenarios.
All networks have identical architecture, 10 hidden layers with 32 nodes each followed by an output layer with 2 nodes. Rectified linear units are used as activation functions for the hidden layers while linear activations are used for the output layers. They were trained up to 1000 epochs with a mean squared error (MSE) loss and an Adam optimizer~\cite{adam}, and the performance is evaluated on independent validation data. The performance was not found to be highly sensitive to hyper-parameter tuning, so permutation symmetry preserving architectures were not explored. The networks were implemented using \texttt{Tensorflow 2.7.0} on a single \texttt{NVIDIA RTX A5000} GPU.

\subsubsection{Polynomial regression}

Following the example of Ref \cite{Fujimoto:2019hxv}, as a performance benchmark we also construct a polynomial regression model to regress EOS parameters $\lambda_1$ and $\lambda_2$ from stellar mass and radius information. The input for each model is a 20 x 1 vector containing the following information from each of ten stars: mass $M$, radius $R$.



Each 20 x 1 input to the polynomial regression network represents a set of stars, all of which are chosen from the same EOS. We construct two multivariate polynomial regression models of degree two, following the general form:
\[\lambda = \beta_0 + \sum_{i=1,2}^N \beta_i x^i\]
where each $\beta$ are coefficients and $N = 20$ to represent the cluster size, or 10 mass-radius pairs. Polynomial features were created using the machine learning toolbox scikit-learn~\cite{scikit-learn}, and subsequently fit to a linear regression model. This model uses optimization in the form of ordinary least squares, which takes the form:
\[\text{min}_\beta ||X\beta - y||^2_2\]
where $X$ is the $N = 20$ input vector, and $y$ is the target EOS parameter, either $\lambda_1$ or $\lambda_2$.

\subsubsection{Estimation of Uncertainty}

\begin{figure}
    \centering
    \includegraphics[width=\exsize\textwidth]{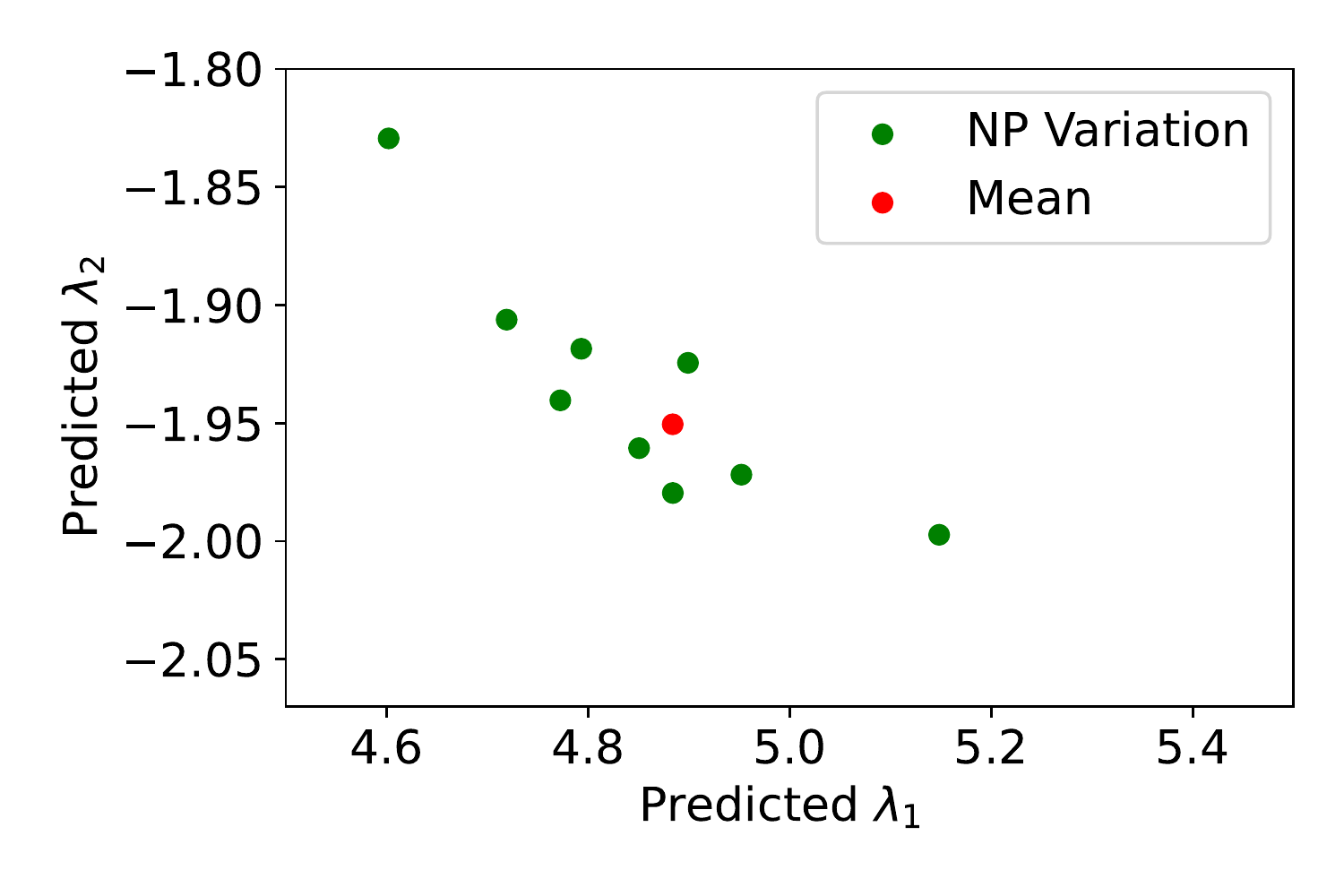}
        \includegraphics[width=\exsize\textwidth]{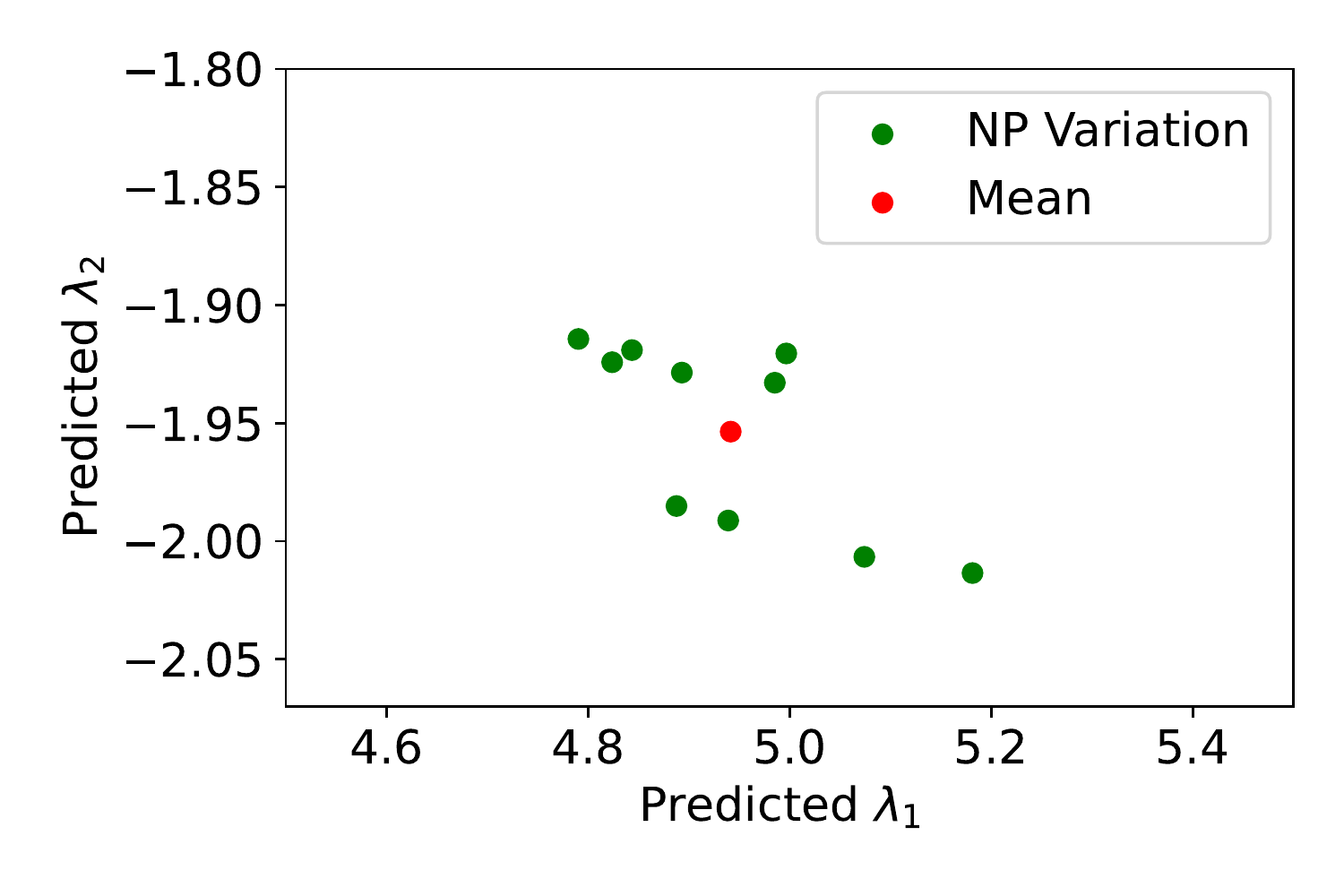}
    \includegraphics[width=\exsize\textwidth]{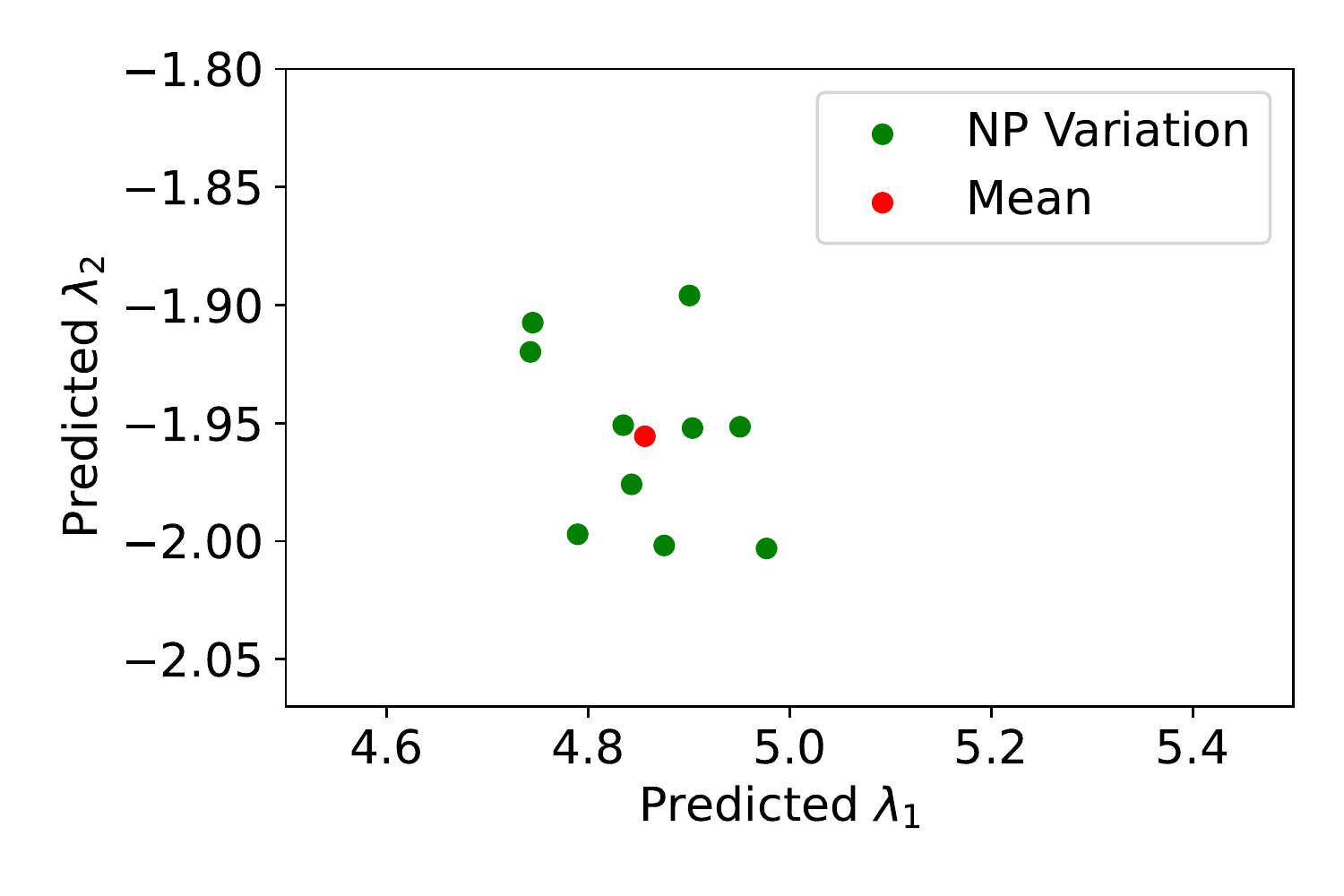}
        \includegraphics[width=\exsize\textwidth]{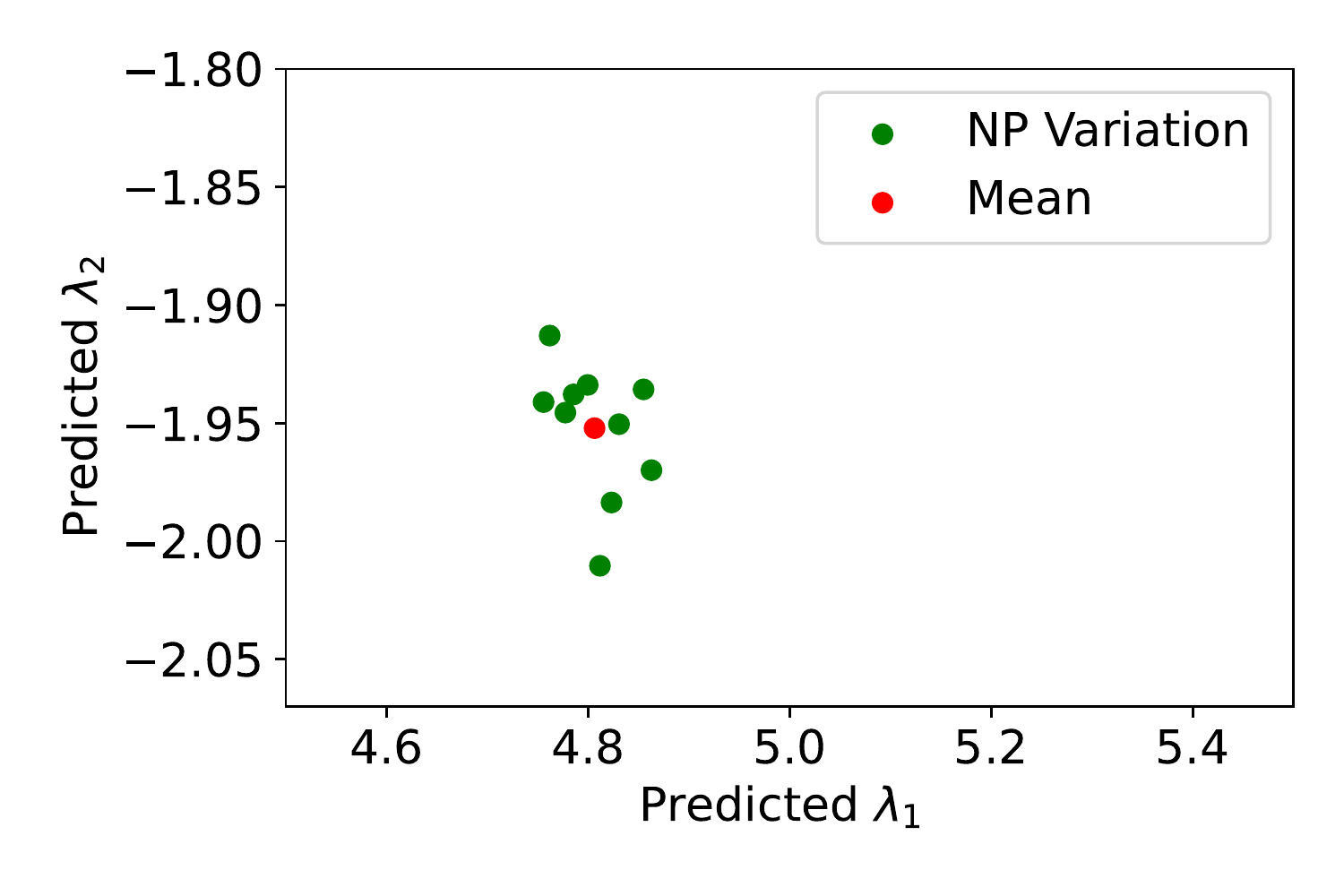}
    \caption{Neural network regression of the EOS parameters $\lambda_1$ and $\lambda_2$ of a set of 10 neutron stars from from their  masses and radii as estimated by \xspec\ from each stars spectrum. Each pane represents an example dataset of 10 simulated stars, and shown (green) are EOS estimates for several independent values of the stellar nuisance parameters drawn from the associated priors, and the mean value (red). Top two cases have loose priors, bottom two have tight.}
    \label{fig:xspec_eos_nps}
\end{figure}

The uncertainty in the underlying NS nuisance parameters has a significant impact on the estimation of EOS parameters. Because the mass and radius estimation is conditioned on the NPs, leading to variations in the mass and radius, (eg see Figure ~\ref{fig:xspec_mr_np}), those variations can be propagated through the EOS estimation. The significant uncertainties in stellar nuisance parameters and the small number of stars observed to date make the treatment of those uncertainties vital. The validity of the final result is only as powerful as the validity of its uncertainties. In Figure ~\ref{fig:xspec_eos_nps}, examples of the variation of the EOS estimates are shown, where the underlying stellar spectra are fixed. Thus, this provides a measure of the uncertainty in the EOS due to the uncertainty in the NPs.

\subsubsection{Performance}

\begin{figure}[hbt!]
    \centering
   \includegraphics[scale=0.3]{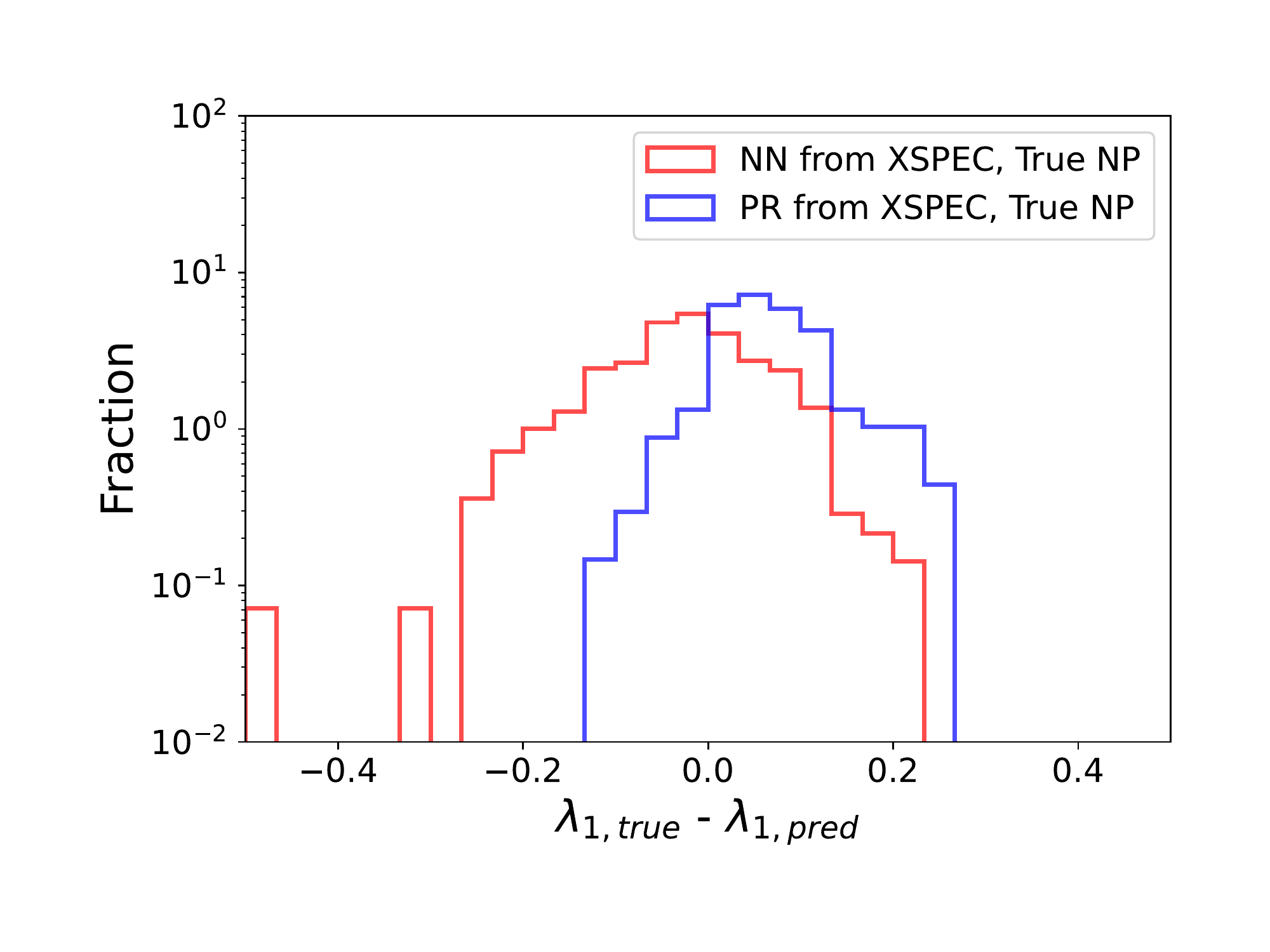}
   \includegraphics[scale=0.3]{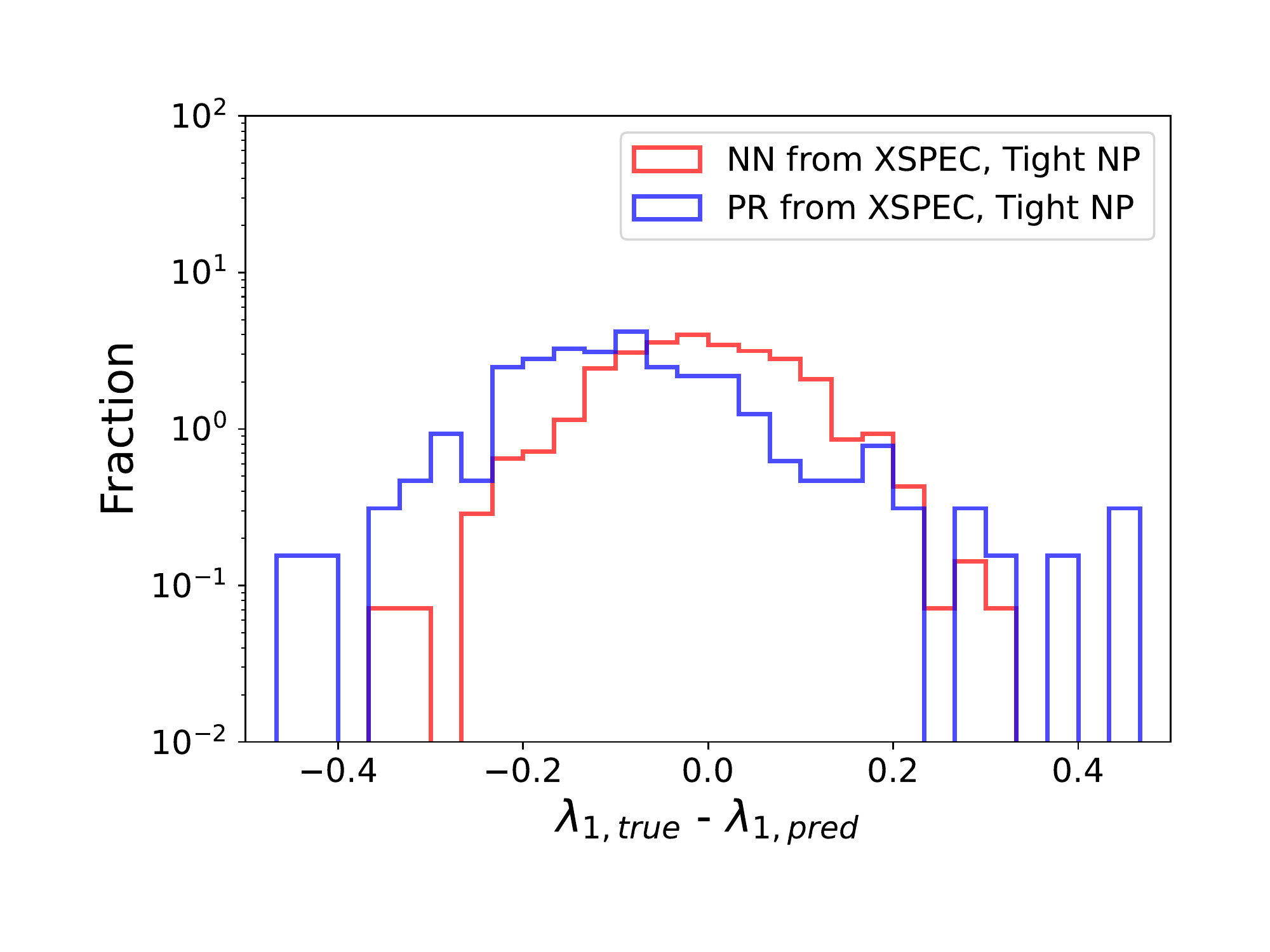}
    \includegraphics[scale=0.3]{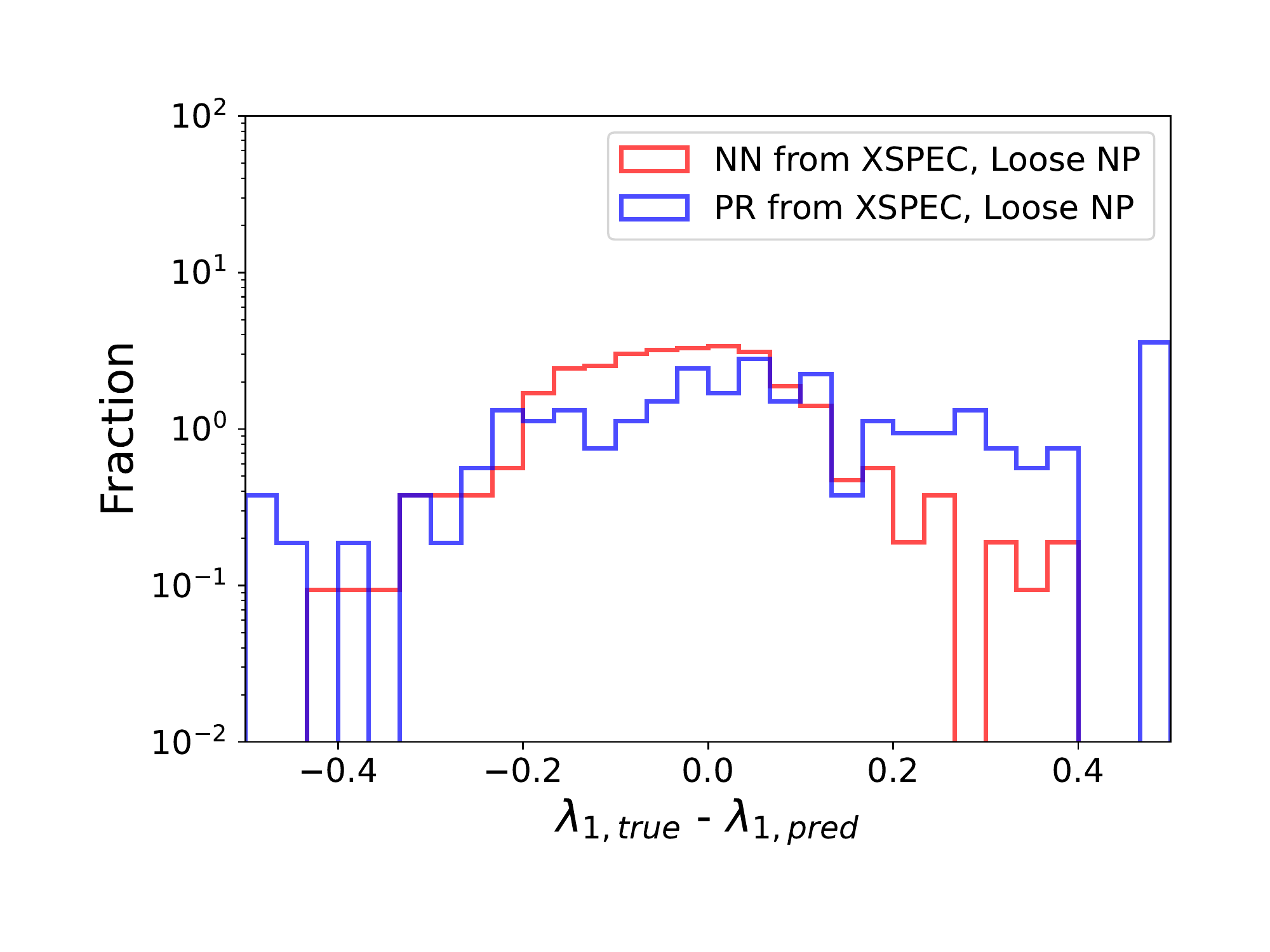}
   
    \caption{ Comparison of the performance of NN regression and polynomial regression of EOS parameter $\lambda_1$ from mass radius pairs inferred by \xspec\  from stellar spectra. Shown is the residual, the difference between the predicted and true values for each of three treatments of the stellar nuisance parameters. In the "true" case, the NPs are fixed to their true values; in the "tight" and "loose" cases, they are drawn from narrow or wide priors, respectively; see text for details. }
    \label{fig:mr_to_eos1}
\end{figure}

\begin{figure}[hbt!]
    \centering
    \includegraphics[scale=0.3]{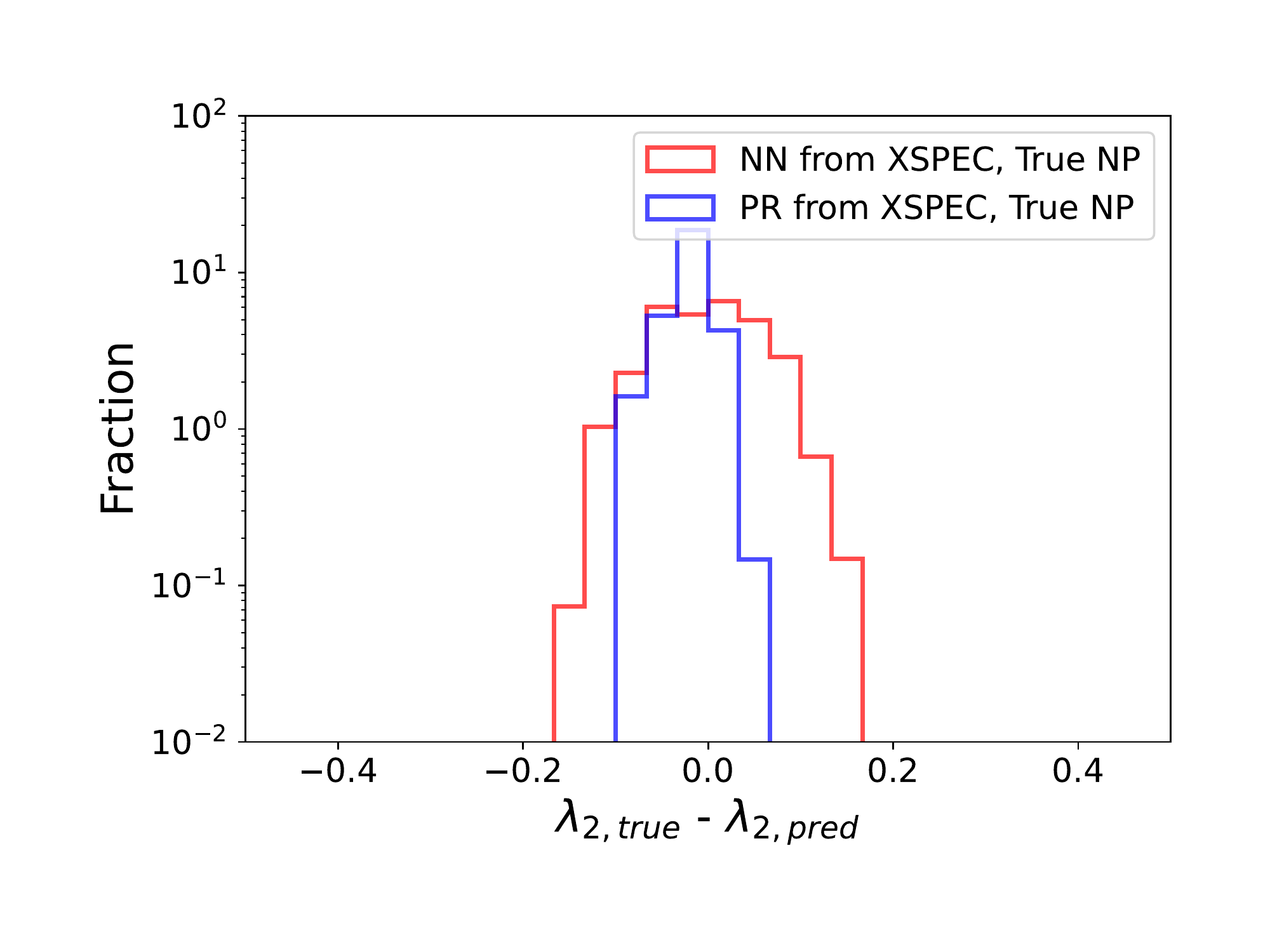}
   \includegraphics[scale=0.3]{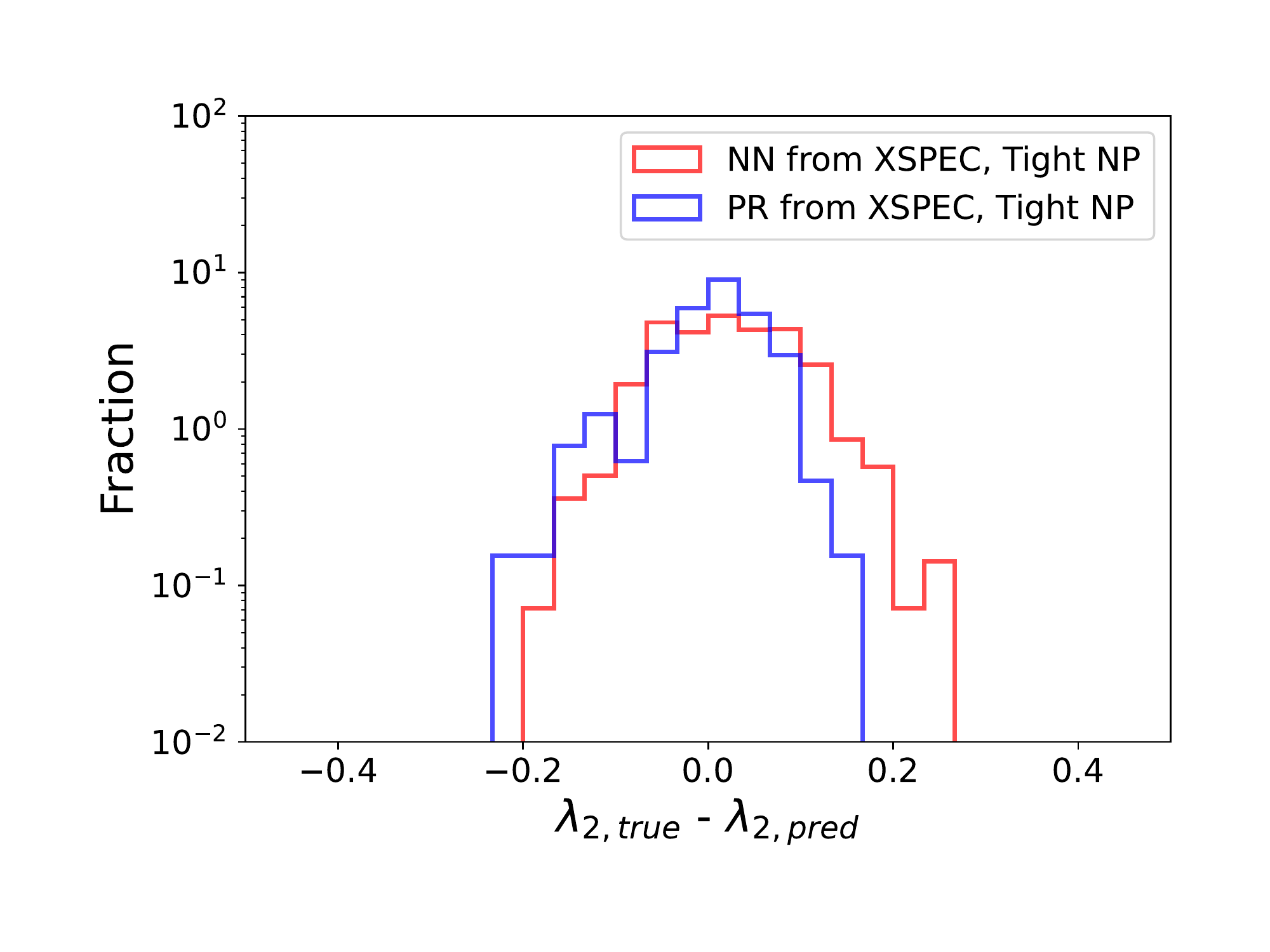}
   \includegraphics[scale=0.3]{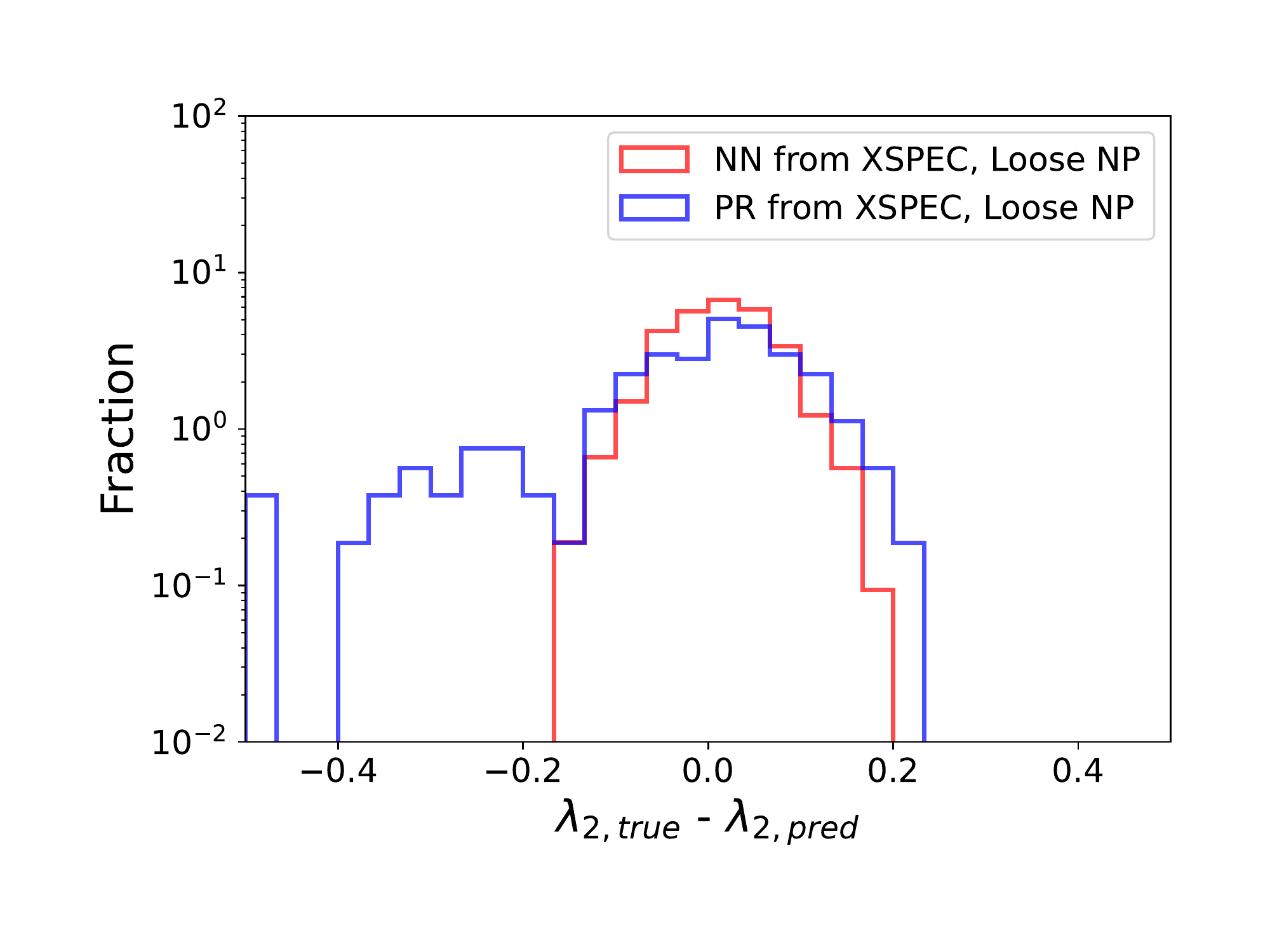}
    \caption{ Comparison of the performance of NN regression and polynomial regression of EOS parameter $\lambda_2$ from mass radius pairs inferred by \xspec\  from stellar spectra. Shown is the residual, the difference between the predicted and true values for each of three treatments of the stellar nuisance parameters. In the "true" case, the NPs are fixed to their true values; in the "tight" and "loose" cases, they are drawn from narrow or wide priors, respectively; see text for details. }
    \label{fig:mr_to_eos2}
\end{figure}

Performance of neural network regression of the EOS parameters $\lambda_1$ and $\lambda_2$ are compared to polynomial regression of the same quantities using identical datasets via comparison of the residuals, the difference between the true and regressed values. As seen in Figures~\ref{fig:mr_to_eos1} and \ref{fig:mr_to_eos2}, while PR is able to achieve narrower residuals in the true case,
the network regression is  more robust in cases with larger uncertainties. 

This result confirms what has been seen in earlier studies of NN regression from mass-radius pairs~\cite{Fujimoto:2019hxv}, but our study extends previous work by using realistic values of the mass and radius inferred from realistic simulated spectra, as well as by demonstrating uncertainty quantification, via full propagation of the underlying uncertainties due to nuisance parameters.

\FloatBarrier

\section{Inference of Mass and Radius from Spectra}
\label{sec:mr}

Previous applications of machine learning to neutron star datasets focus on analysis of mass-radius pairs, as demonstrated above, rather than direct analysis of the stellar spectra by neural networks.

A potential obstacle to direct analysis of spectra by neural networks is that the spectra are high-dimensional, often with $\mathcal{O}(10^3)$ bins of photon energy. 
However deep learning methods combined with GPUs have no trouble analyzing data with similarly high dimensionality~\cite{Baldi:2016fzo}, opening up new opportunities to tackle this important topic.  While the mass and radius are powerful summaries of the information in the lower-level spectra which is relevant to the equation of state, direct ML analysis of the spectra themselves may allow for the extraction of additional information or provide more robust propagation of uncertainties.  As an initial step, we begin by estimating the stellar mass and radius from a single stellar spectrum before moving on to end-to-end inference of EOS parameters directly from a set of spectra in the next section.

In this section, we apply machine learning to the task of extracting the mass and radius from the stellar spectra, training a network we refer to as \mrnet. This serves as a demonstration of the capabilities of ML to grapple with high-dimensional datasets, allows us to harmonize the treatment of nuisance parameters end-to-end from spectra to EOS, and potentially extract more relevant information.

\subsection{\mrnet Method}

We build a network whose inputs correspond to the bins of the stellar X-ray spectrum, and whose outputs are the estimates of the star's mass and radius.  In addition, the mass and radius regressor is parameterized on the stellar nuisance parameters (distance, $N_H$, $\log(T_\textrm{eff})$), which allows the results to be conditioned on the nuisance parameters.

This architecture is composed of two input branches, one to process the star's spectra and another to process the corresponding nuisance parameters. Each branch contains a series of layers that process its inputs in isolation. Following these initial layers, the output from the branches is combined, forming a single vector containing all the information. This vector is then passed to a final series of layers to predict the star's mass and radius. Each segment of the network, both the branches and the main trunk of the network, contain four layers, giving the network eight layers in total. All fully connected layers contain 275 nodes and utilize a dropout probability of 0.25. The network employs skip connections between alternate layers. This stabilizes the training process and adds robustness to the network overall. The network is trained with an MSE loss and an Adam optimizer with an initial learning rate of 0.00017 which is slowly decayed over the course of training.

Mass-radius regression is formulated as a supervised learning problem, where the network learns to minimize the error between the true mass and radius and its predictions. The network weights are updated by stochastic  gradient descent using backpropagation. The Huber loss function is used and the Adam optimizer computes gradients and schedules the backward passes. 

\subsection{\mrnet Performance in Mass, Radius}

\begin{figure}[hbt!]
    \centering
    \includegraphics[scale=0.3]{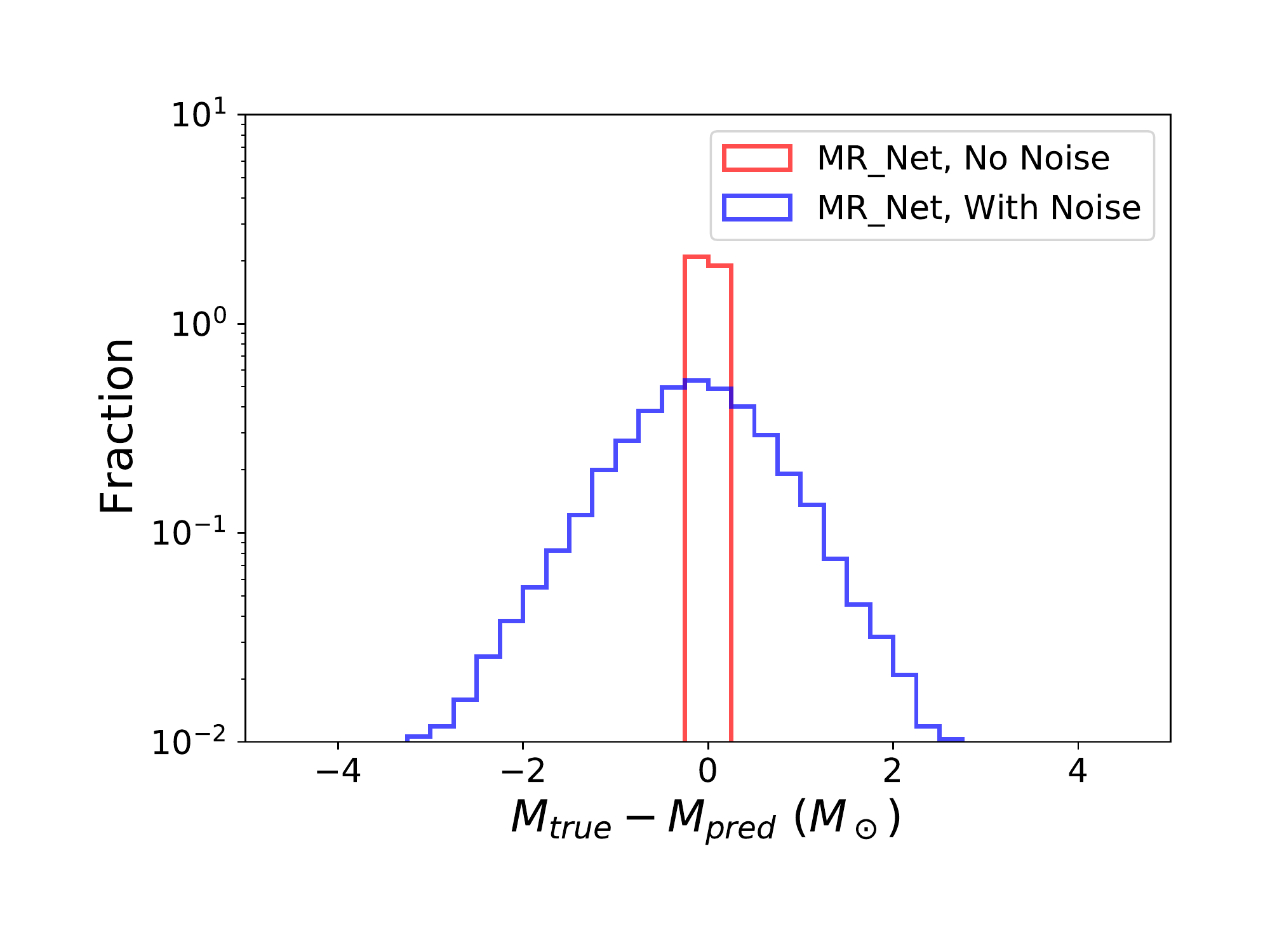}
    \includegraphics[scale=0.3]{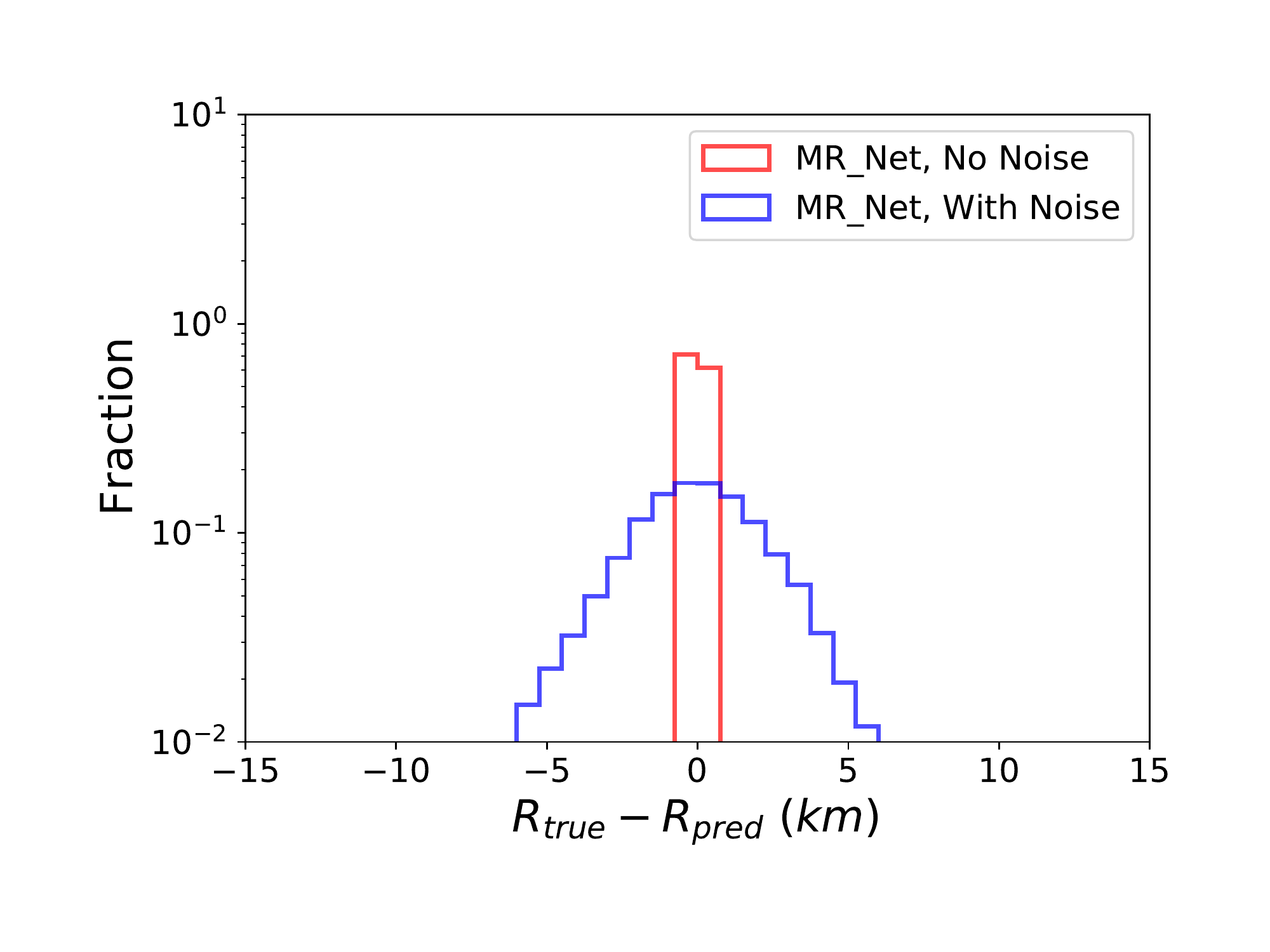}
    \caption{ Performance of the \mrnet regression of a neutron star mass (top) and radius (bottom) from its stellar X-ray spectrum. Shown is the residual, the difference between the true and predicted values, for spectra with statistical noise (blue) corresponding to an observation time of 100ks, and for spectra without statistical noise (red), which demonstrates the capacity of the network. Nuisance parameters are fixed to their true values.}
    \label{fig:regressMR}
\end{figure}

We begin with the best-case scenario in which the nuisance parameters are known with zero uncertainty, referred to as ``true NP" above.  
Figure~\ref{fig:regressMR} shows the performance of \mrnet given neutron star spectra with statistical noise corresponding to 100,000 seconds (100 ks) of observation time, as well as for spectra without statistical noise.  This demonstrates the contribution of statistical noise to the residual and demonstrates the network's ability to digest the spectral information and understand the dependence on mass and radius. 

What is clear is that \mrnet is capable of extracting the mass and radius values of the star directly from the spectrum. We emphasize that the network is trained on examples, but does not benefit from the knowledge of the theoretical model used to generate these stars, while \xspec\  requires precise specification of the theoretical model. This lack of requirement of a theoretical model opens new possibilities, such as training \mrnet to interpolate smoothly between theoretical models by providing a mixed or parameterized training set.

\begin{figure}[hbt!]
    \centering
    \includegraphics[scale=0.3]{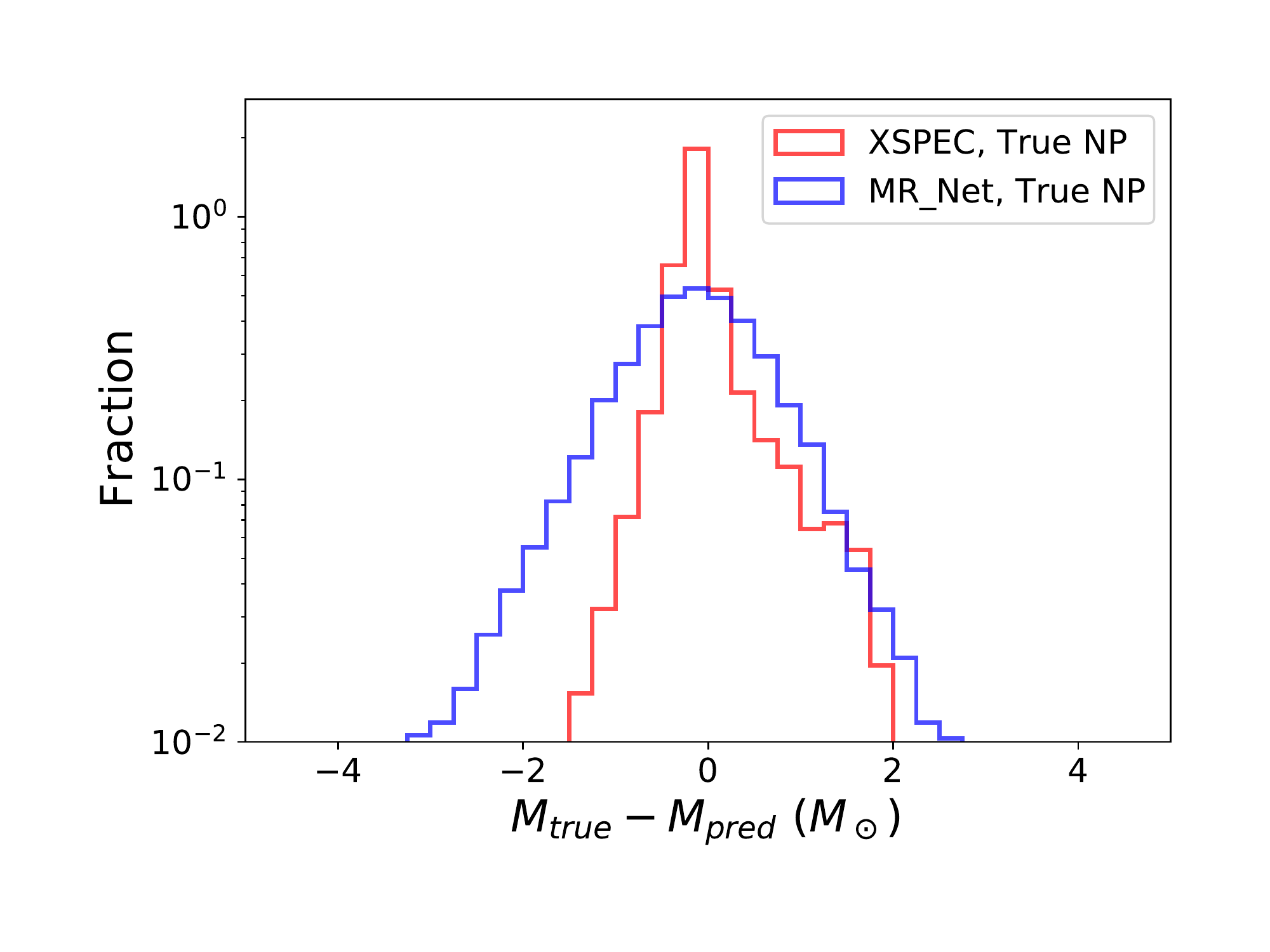}
    \includegraphics[scale=0.3]{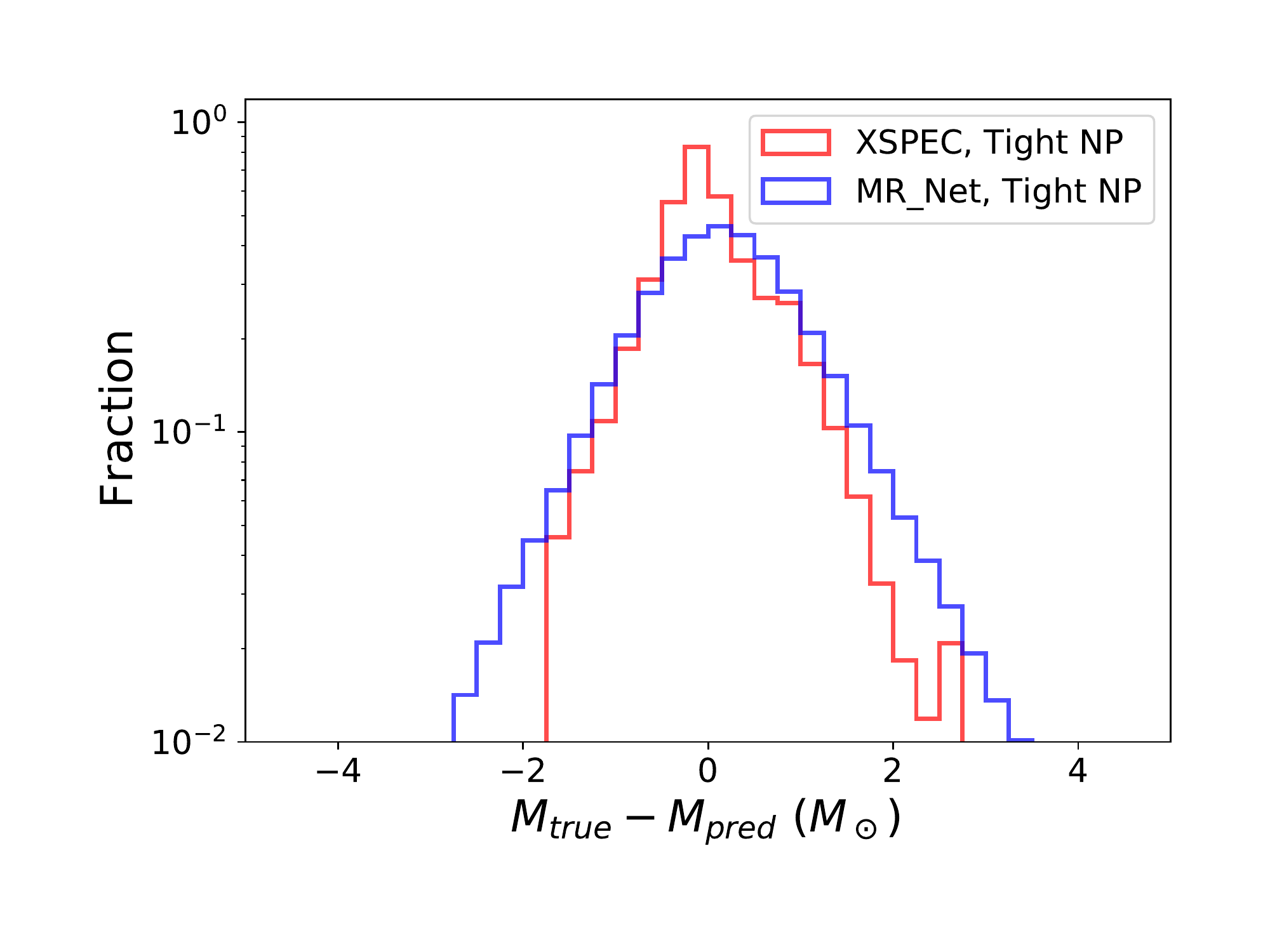}
    \includegraphics[scale=0.3]{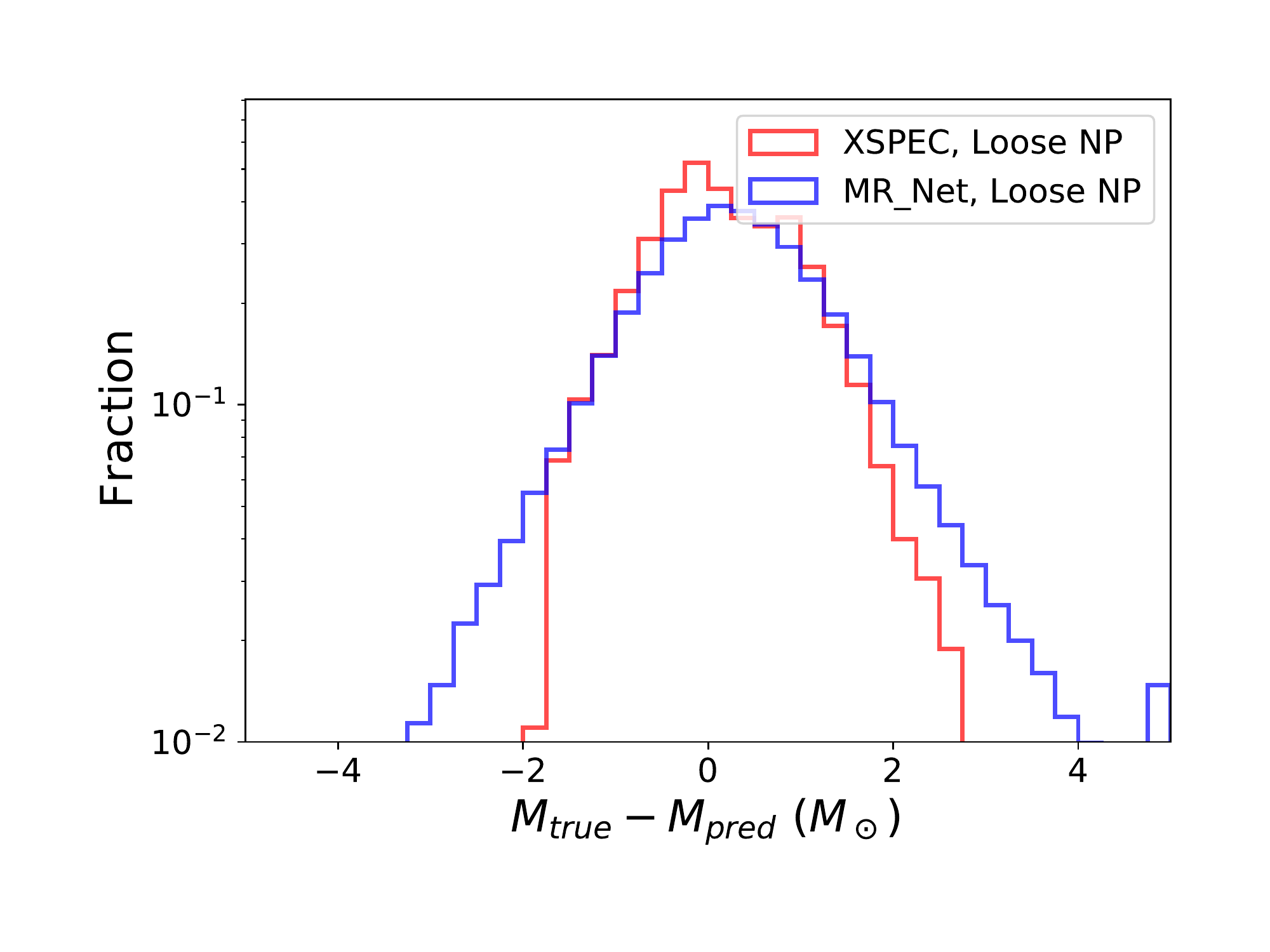}
    \caption{Performance of the \mrnet regression of a neutron star mass  from its stellar X-ray spectrum, compared to regression using \xspec\ . Shown is the residual, the difference between the true and predicted values, for three scenarios of nuisance parameter uncertainties.    In the ``true" case, the NPs are fixed to their true values; in the ``tight" and ``loose" cases, they are drawn from narrow or wide priors, respectively; see text for details.}
    \label{fig:regressMR_np}
\end{figure} 

\begin{figure}[hbt!]
    \centering
    \includegraphics[scale=0.3]{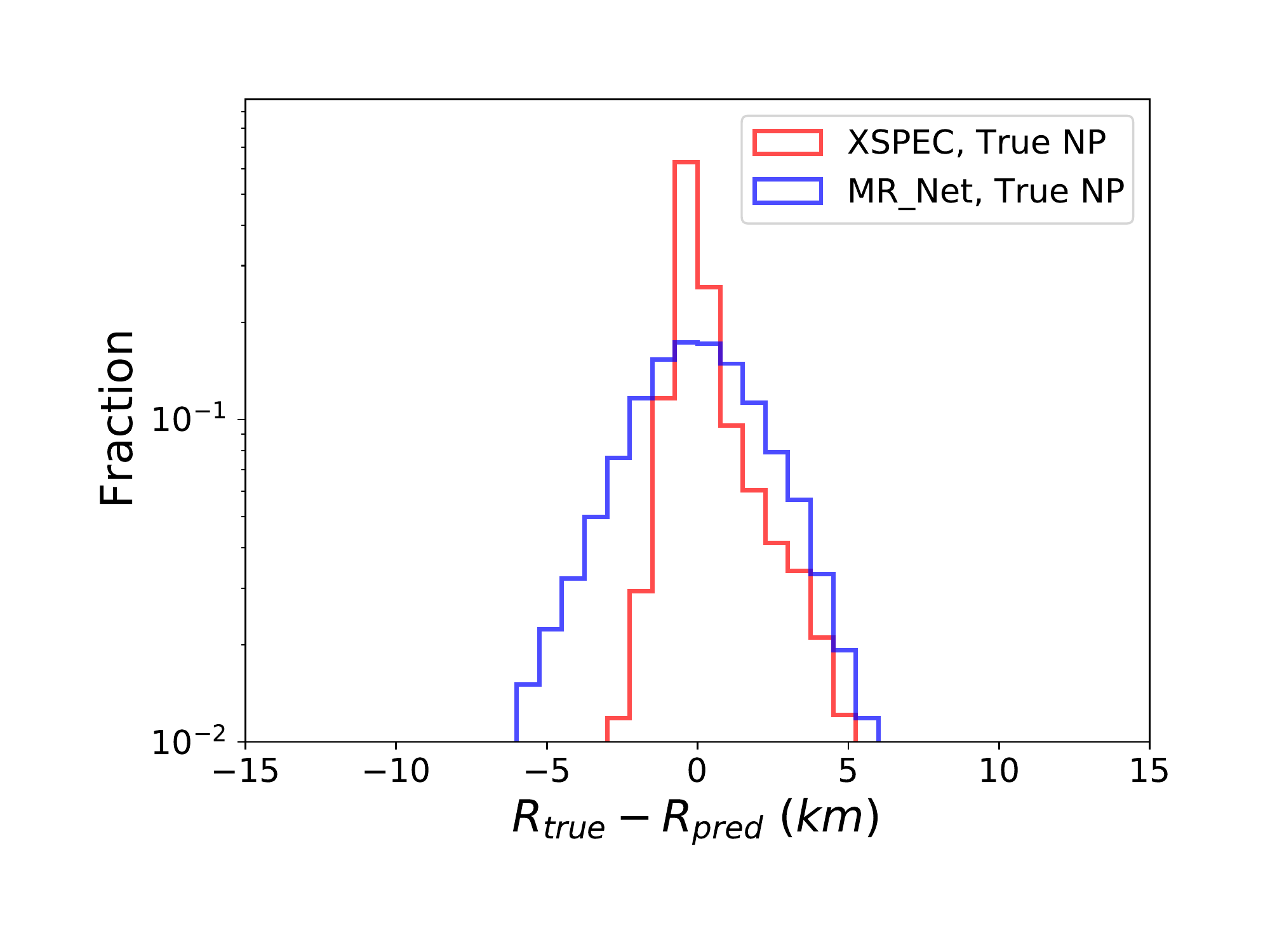}
    \includegraphics[scale=0.3]{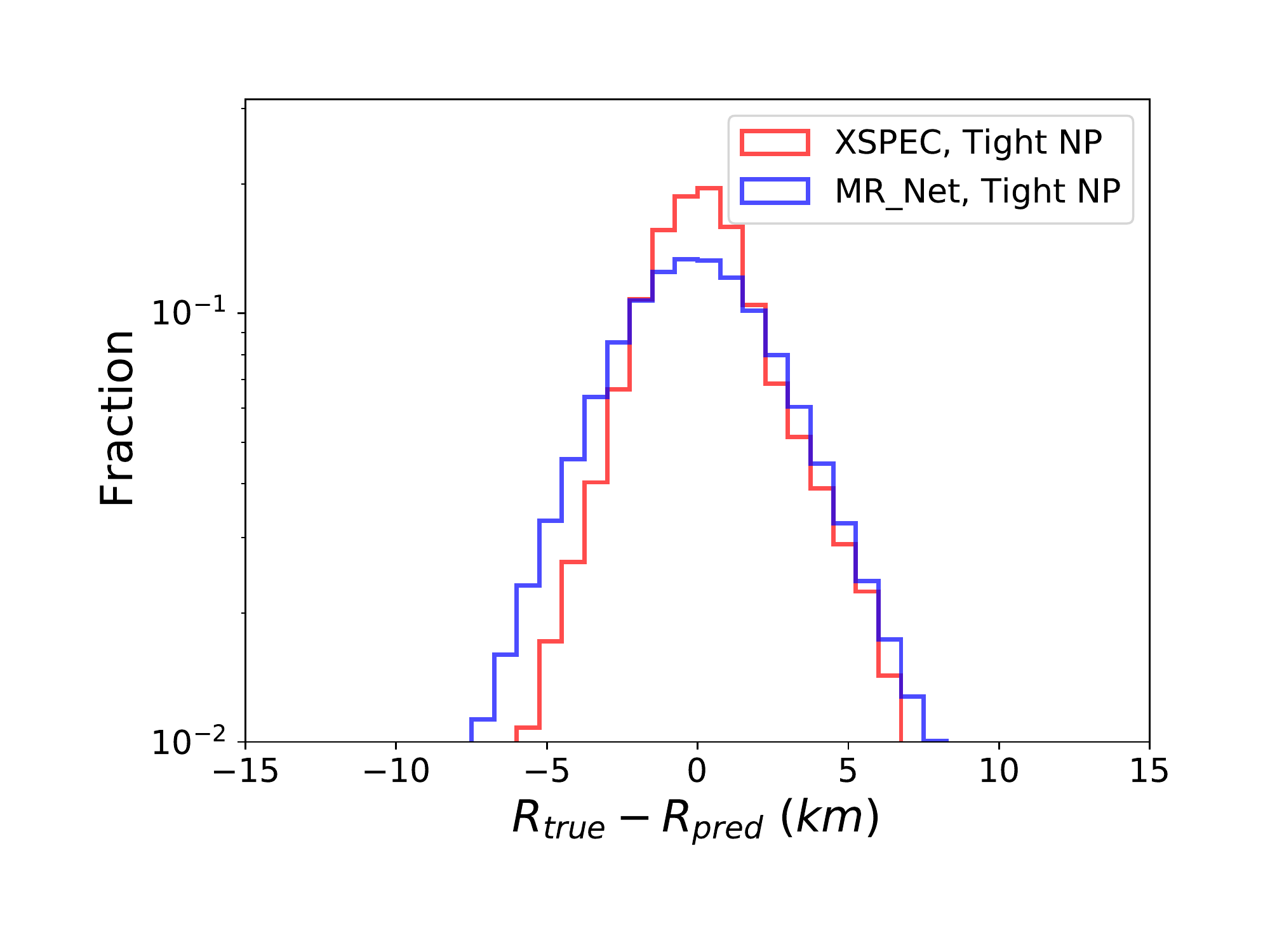}
    \includegraphics[scale=0.3]{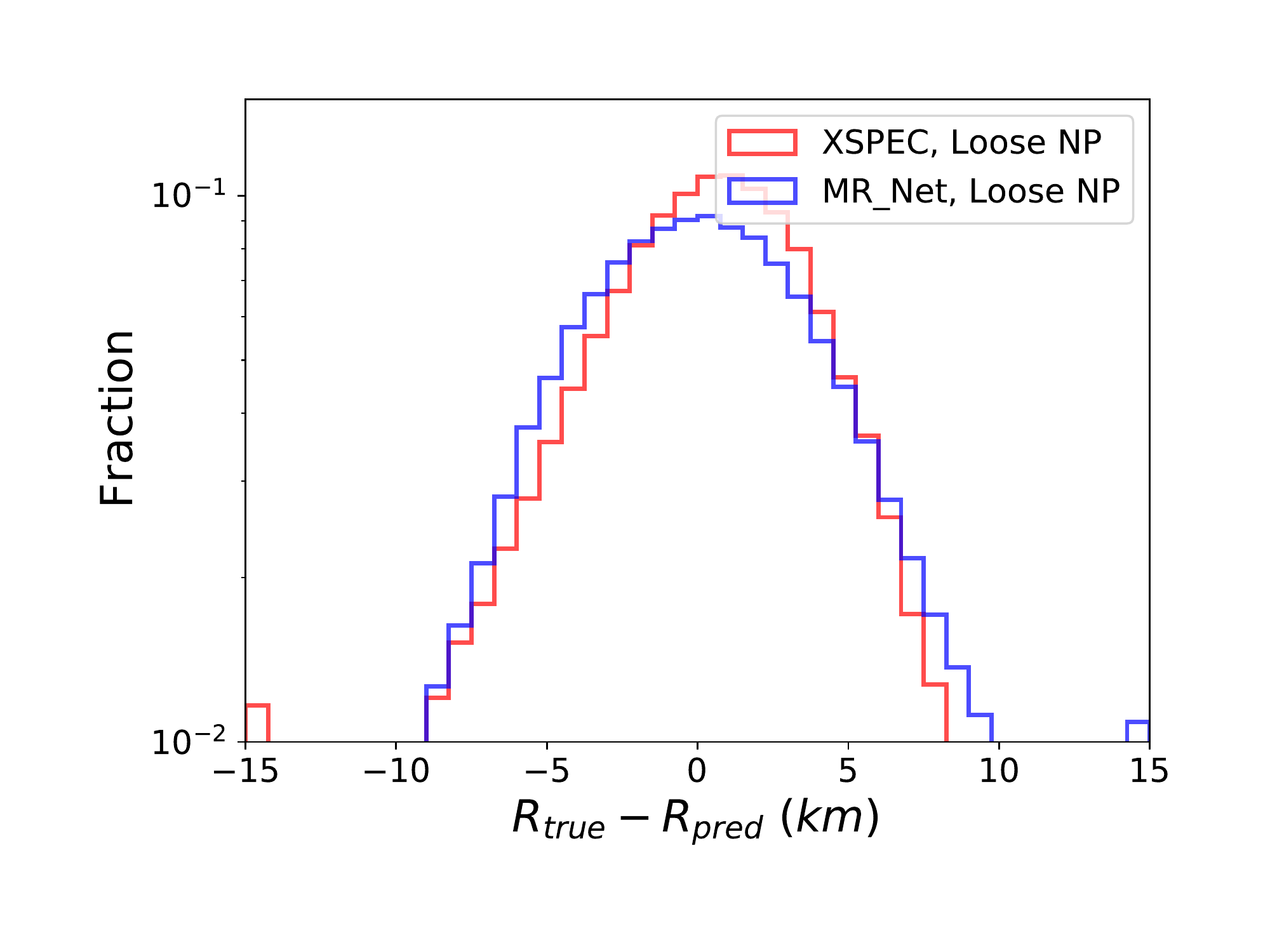}
    \caption{Performance of the \mrnet regression of a neutron star radius from its stellar X-ray spectrum, compared to regression using \xspec\ . Shown is the residual, the difference between the true and predicted values, for three scenarios of nuisance parameter uncertainties.    In the ``true" case, the NPs are fixed to their true values; in the ``tight" and ``loose" cases, they are drawn from narrow or wide priors, respectively; see text for details.}
    \label{fig:regressMR_np2}
\end{figure} 

\begin{figure}[]
    \centering
    \includegraphics[scale=0.35]{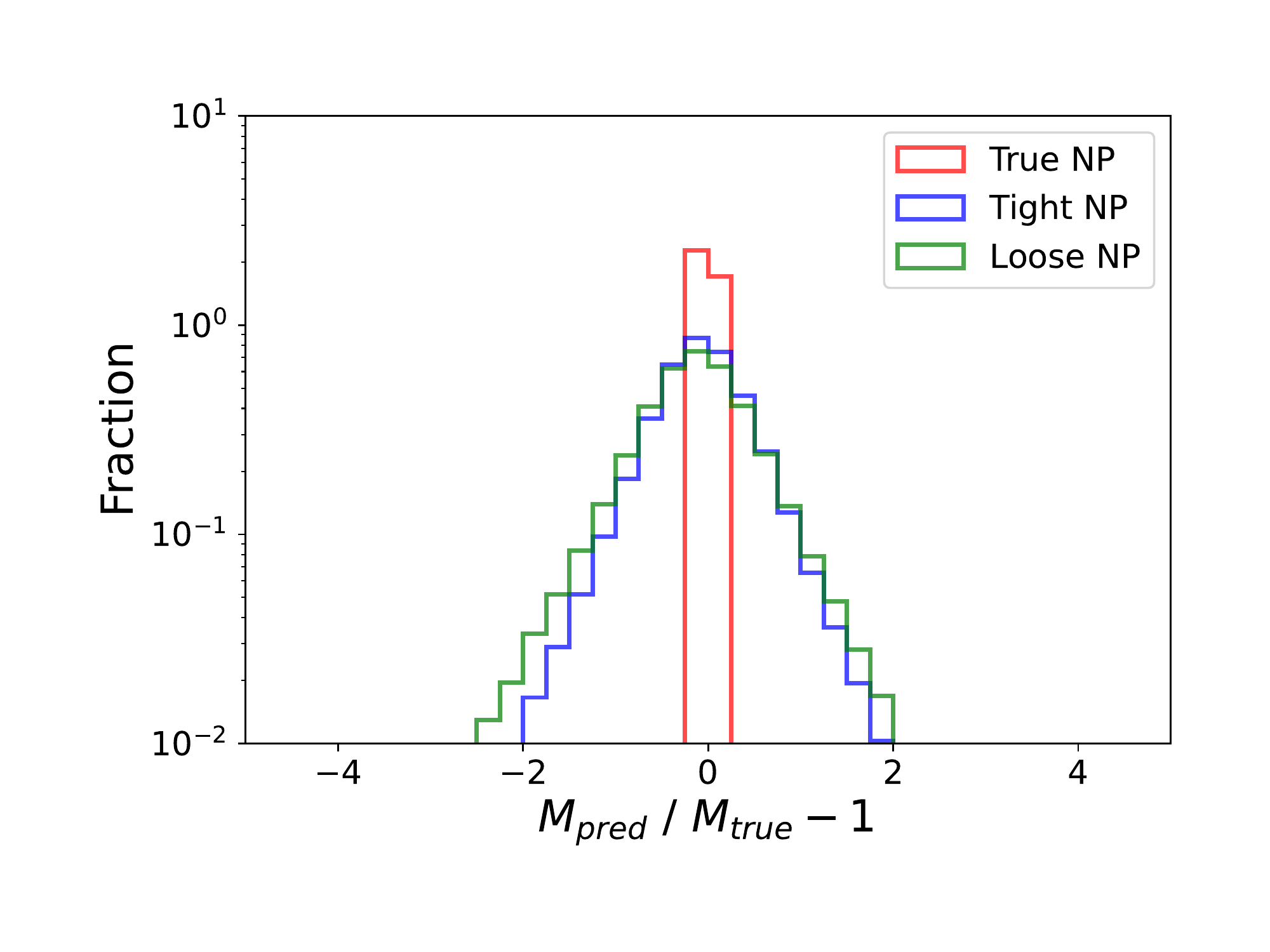}
    \includegraphics[scale=0.35]{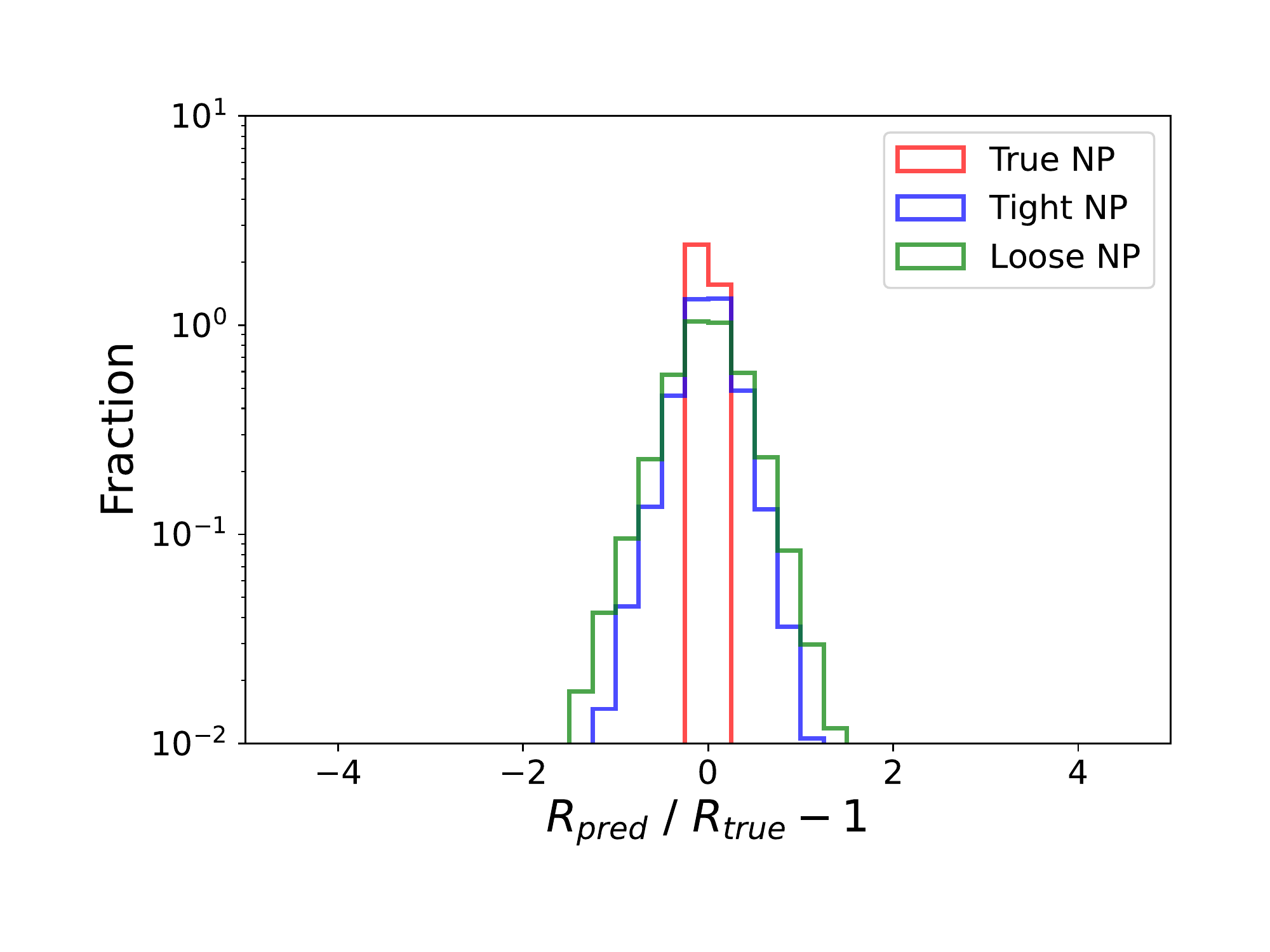}
    \caption{Performance of the \mrnet regression of a neutron star mass (top) and radius (bottom) from its stellar X-ray spectrum. Shown is the ratio of predicted ("pred") to true values minus one for three scenarios of nuisance parameter uncertainties.    In the ``true" case, the NPs are fixed to their true values; in the ``tight" and ``loose" cases, they are drawn from narrow or wide priors, respectively; see text for details.}
    \label{fig:new_mr_compare}
\end{figure}

The \mrnet is conditioned on the nuisance parameters, allowing for propagation of the NP uncertainties through to the regression target as was done with \xspec\ estimates. To assess the impact of NP uncertainty, we draw from priors on the NPs under the ``Tight" and ``Loose" scenarios defined earlier. The residuals widen, as expected. Figures~\ref{fig:regressMR_np} and \ref{fig:regressMR_np2} show the mass and radius residuals, respectively, under each NP scenario for \mrnet and \xspec\. Table~\ref{tab:mr} shows the mean and width of each residual distribution, as well as the combined width. As an additional comparison, Figure~\ref{fig:new_mr_compare} shows the ratio of predicted values from \mrnet to true values subtracted from 1.

\mrnet is capable of analyzing the stellar spectrum directly and extracting stellar parameters in a robust manner that allows for propagation of NP uncertainties.

\begin{table}[]
    \caption{ Performance of the regression of neutron star mass and radius for \xspec\ as well as our neural network regression, \mrnet, which lacks any knowledge of the theoretical model.  Shown are the mean ($\mu$) and standard deviation ($\sigma$) of the residuals under three  scenarios of nuisance parameter uncertainties.    In the ``true" case, the NPs are fixed to their true values; in the ``tight" and ``loose" cases, they are drawn from narrow or wide priors, respectively; see text for details. The combined column is a quadrature sum of the standard deviations of radius and mass.}
    \label{tab:mr}
    \centering
    \begin{tabular}{lcrrrrrr}
             \toprule
         & Nuis. & \multicolumn{2}{c}{Mass} & \ \ & \multicolumn{2}{c}{Radius} & Combined \\
        \cline{3-4} \cline{6-7}\\
         Method & Params & $\mu$ & $\sigma$ && $\mu$ & $\sigma$ & $\sigma$ \\
                  \hline 
        \xspec\  &True          & $-0.01$  & 0.50  && 0.23  & 1.44 &  1.51  \\
        \mrnet   &True          & $-0.14$  & 0.93  && $-0.07$  & 2.80 &  2.99  \\ \hline
        \xspec\  &Tight         & $-0.06$  & 0.73  && 0.24  & 2.61 &  2.69  \\
        \mrnet   &Tight         & 0.17  & 1.06  && 0.06  & 3.52 &   2.76 \\ \hline
        \xspec\  &Loose         & 0.18  & 0.86  && $-0.06$ & 4.32 &  4.40  \\
        \mrnet   &Loose         & 0.28 & 1.29  && 0.14  & 4.93 &   5.10 \\
         \hline \hline
    \end{tabular}
\end{table}

\subsection{Network uncertainty on mass and radius}

\begin{figure} [hbt!]
    \centering
   \includegraphics[width=\exsize\textwidth]{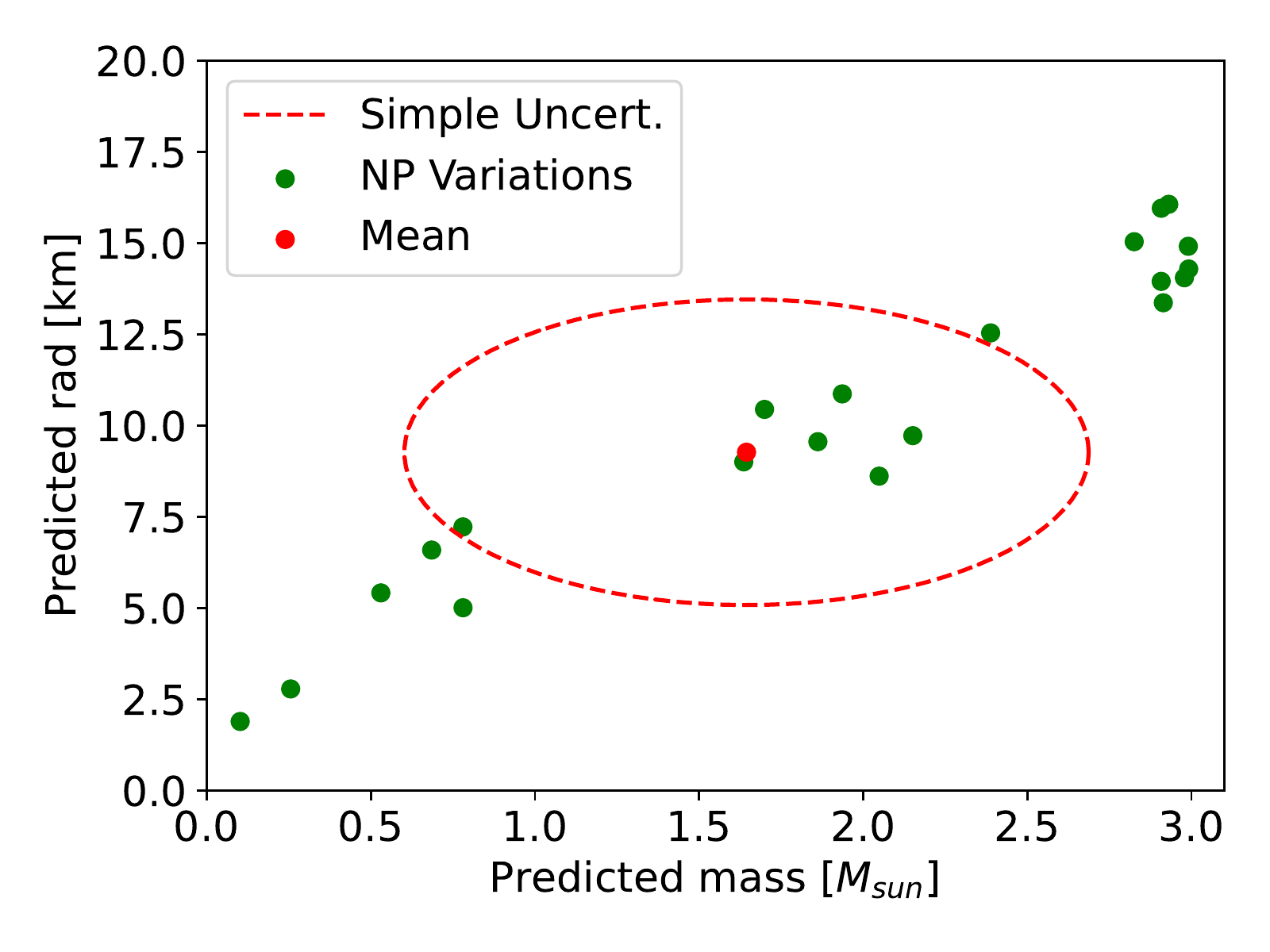}
    \includegraphics[width=\exsize\textwidth]{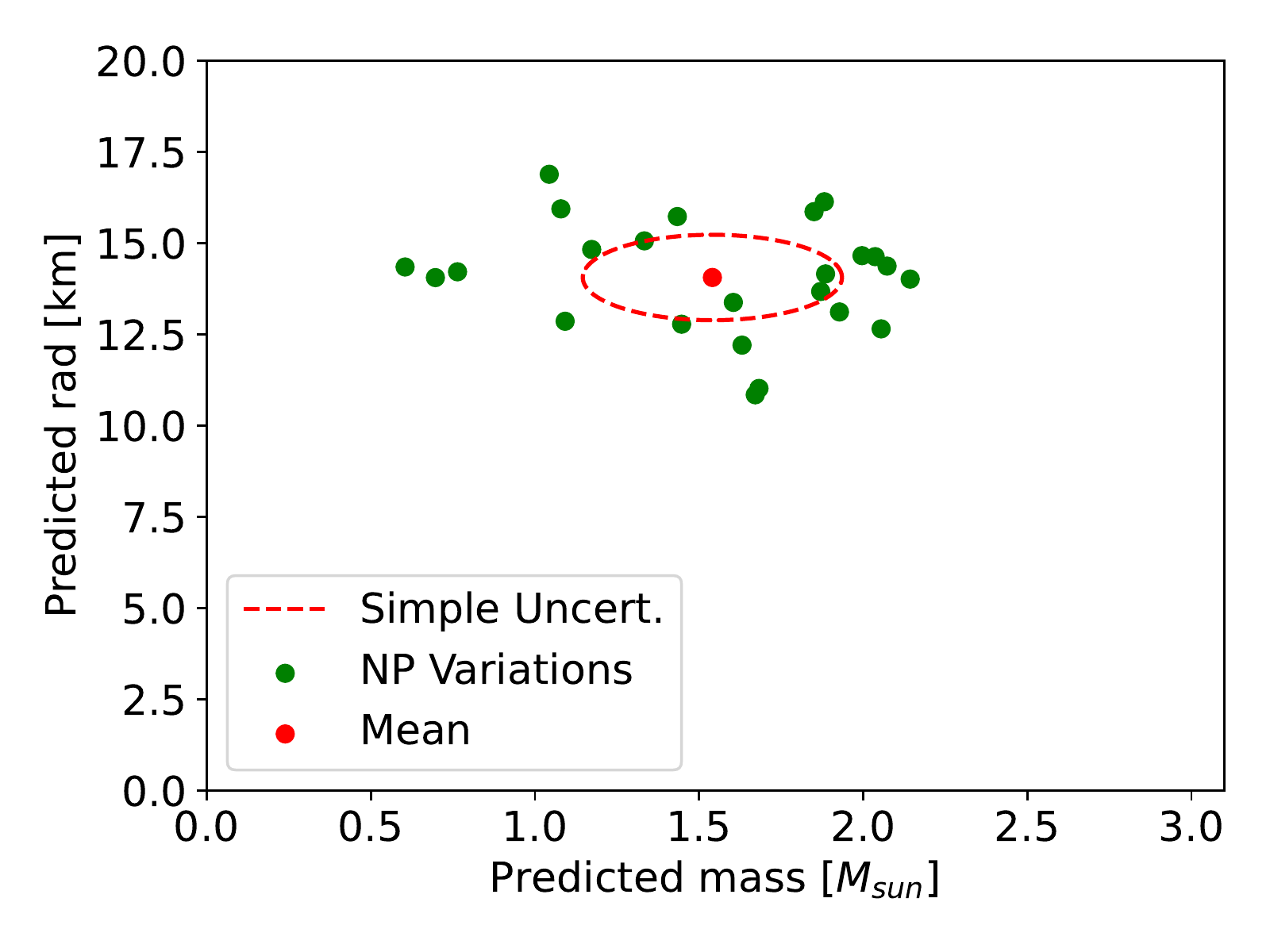}
    \includegraphics[width=\exsize\textwidth]{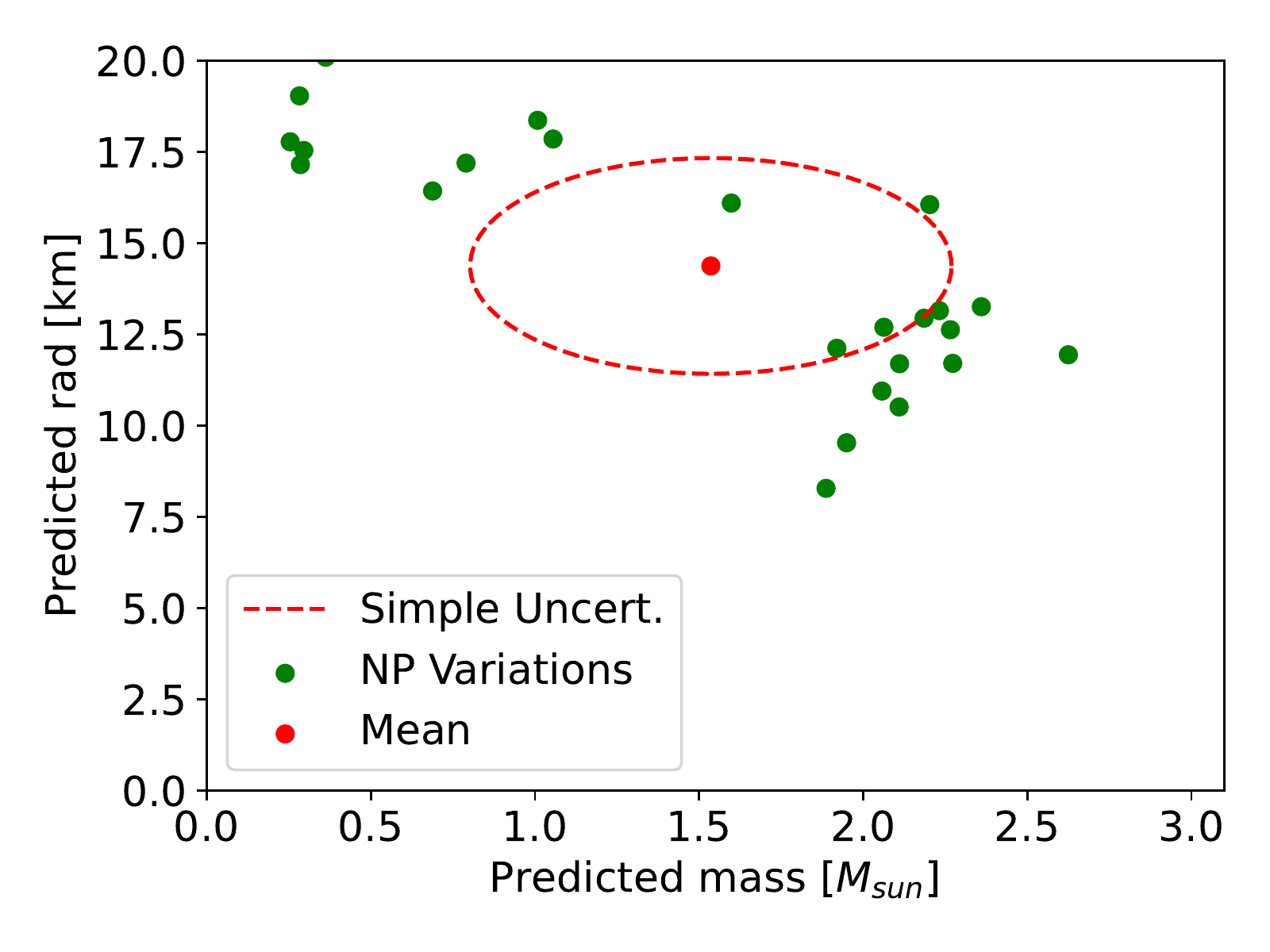}
    \includegraphics[width=\exsize\textwidth]{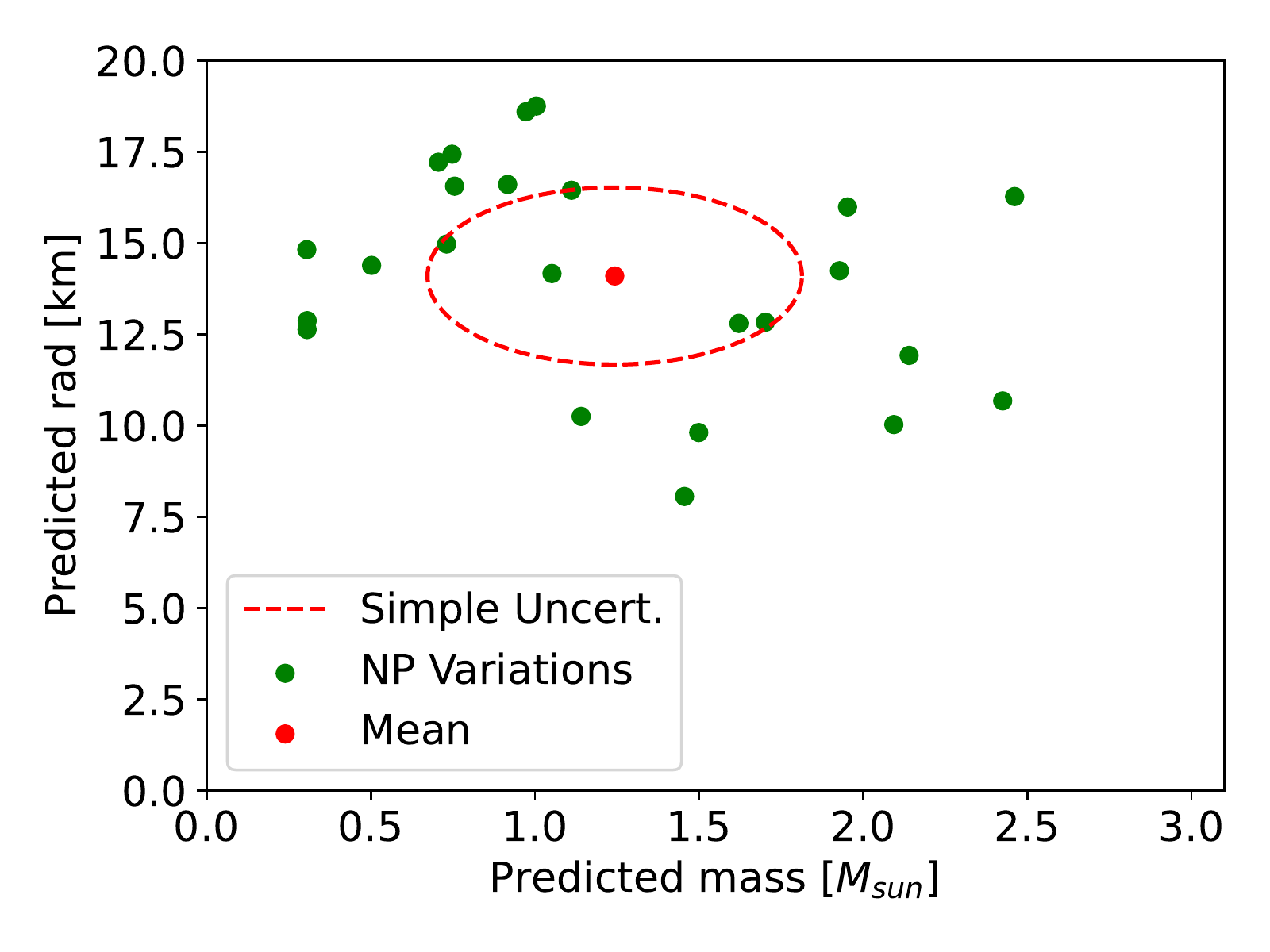}
    \caption{ Estimation of the mass and radius of a neutron star from the underlying stellar spectra, by \mrnet. Each pane represents one star, and shown (green) are estimates for several independent values of the nuisance parameters drawn from the associated priors, and the mean value (red). Top two cases have loose priors, bottom two have tight. The dashed ellipse, whose widths are set to the standard deviation of the mass and radius estimates, is a demonstration of the inadequacy of a simple uncertainty model.}
    \label{fig:mr_examples}
\end{figure}

Conditioning the extraction of $M$ and $R$ on the nuisance parameters allows for the natural propagation of the corresponding uncertainties.  As was done for \xspec\ mass and radius estimates, we propagate the NP uncertainty through to mass and radius uncertainty by sampling from the stellar NP priors several times for a given stellar spectrum, performing the mass and radius regression multiple times.  Figure ~\ref{fig:mr_examples} demonstrates this for several example stars.  
Note that the variation of NP values does not produce variation in $M$ and $R$ which can be accurately summarized by 2D uncorrelated Gaussians, as has been assumed in previous studies~\cite{Fujimoto:2019hxv} with ad-hoc datasets.

\subsection{Revisiting EOS regression}

The question driving the analysis of neutron star spectra is not a desire to measure their masses and radii, but to use those to determine the equation of state parameters. In this section, we push the results of \mrnet through our NN regression of EOS parameters to analyze the performance of spectra$\rightarrow (M,R) \rightarrow$EOS regression. In later sections, we remove the intermediate step and perform direct spectra$\rightarrow$EOS regression.



\begin{figure}
    \centering
\includegraphics[width=\exsize\textwidth]{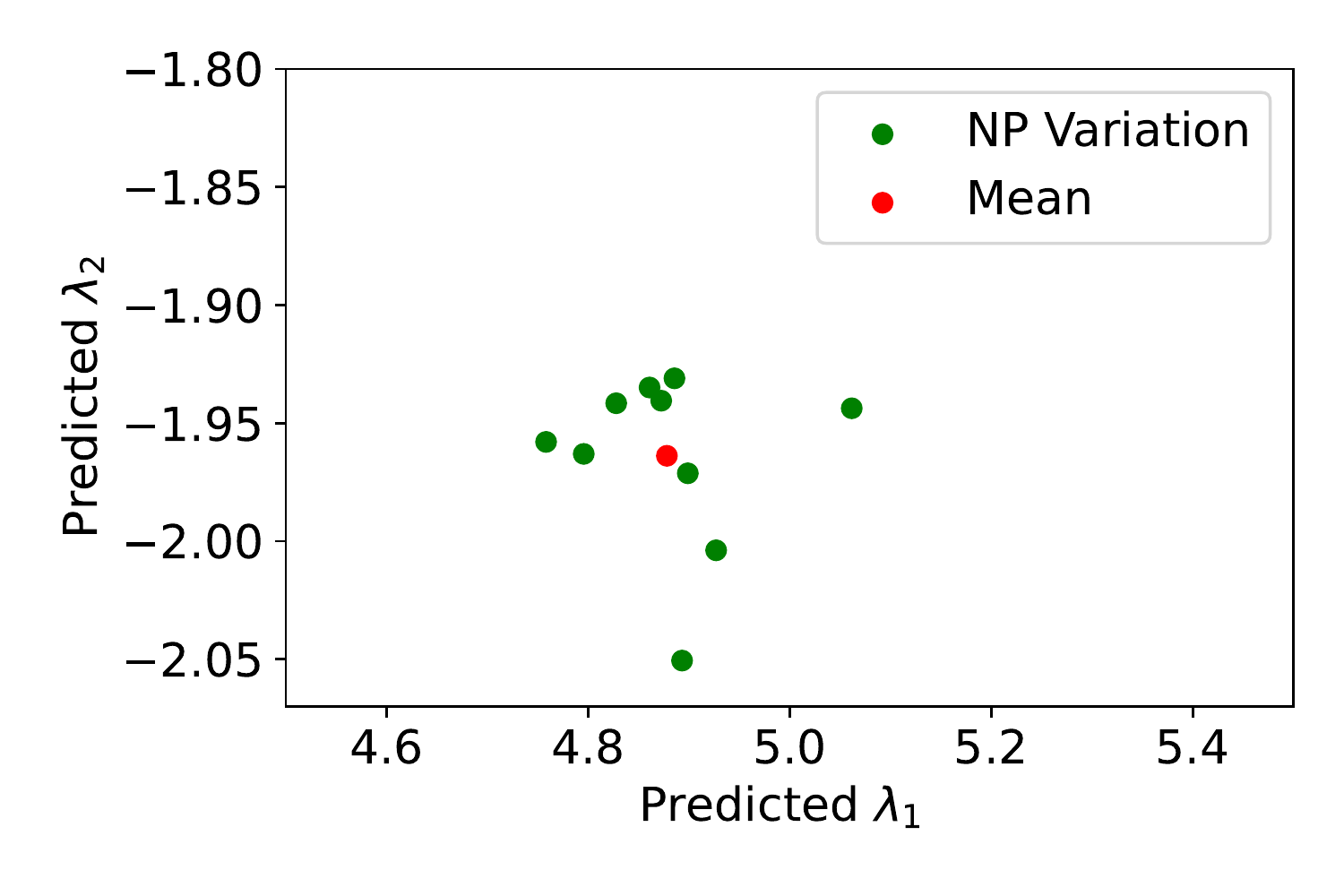}
\includegraphics[width=\exsize\textwidth]{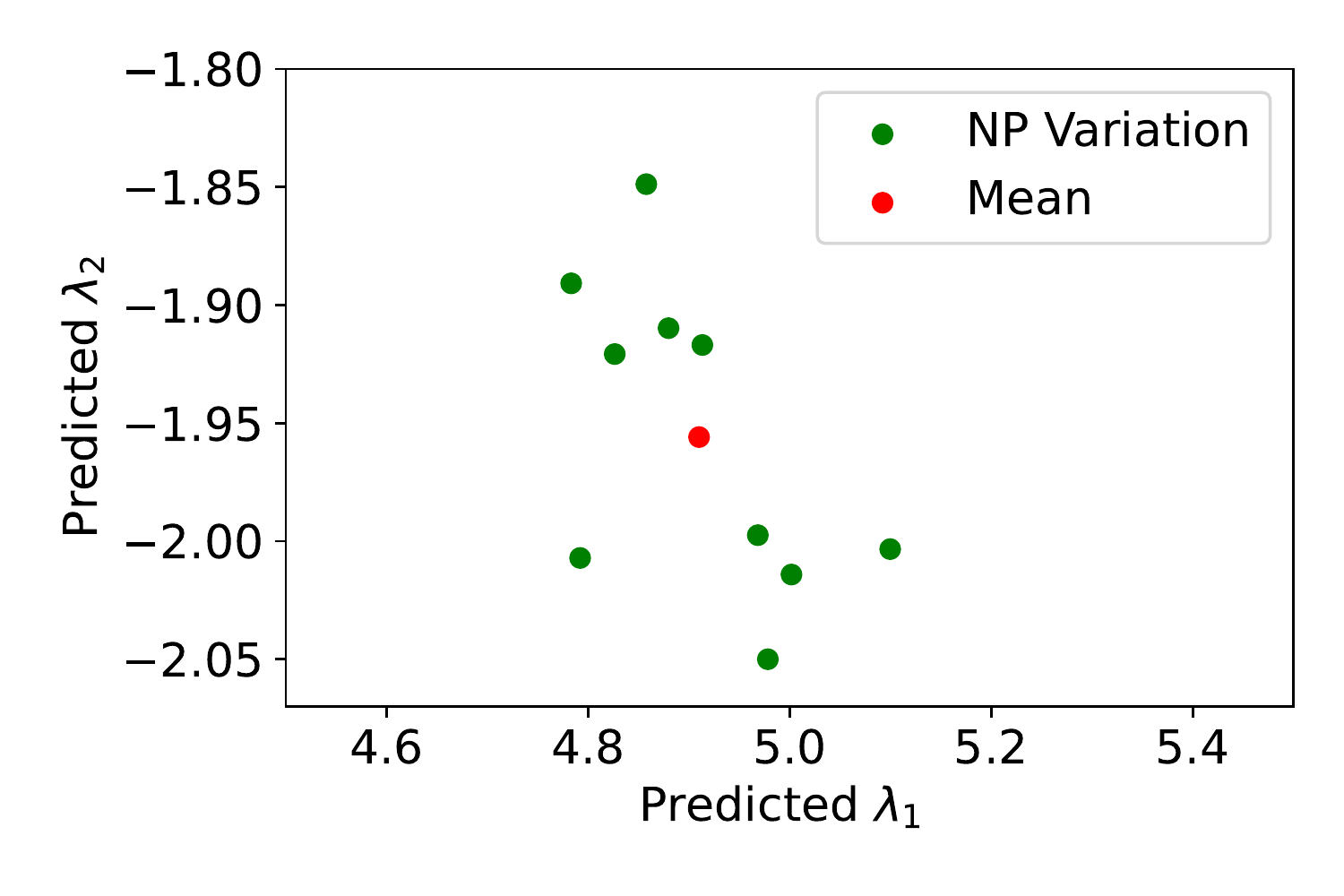}
\includegraphics[width=\exsize\textwidth]{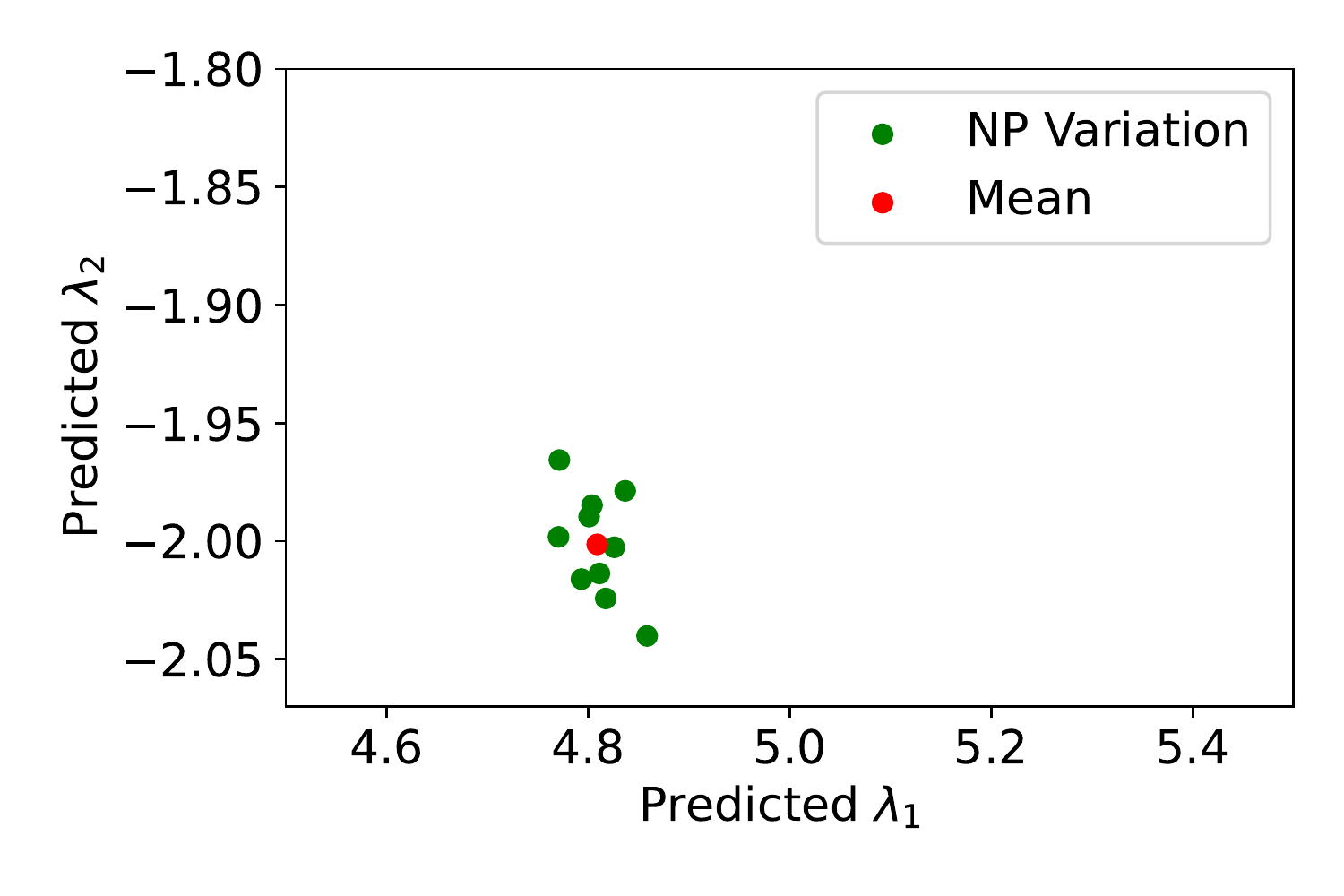}
\includegraphics[width=\exsize\textwidth]{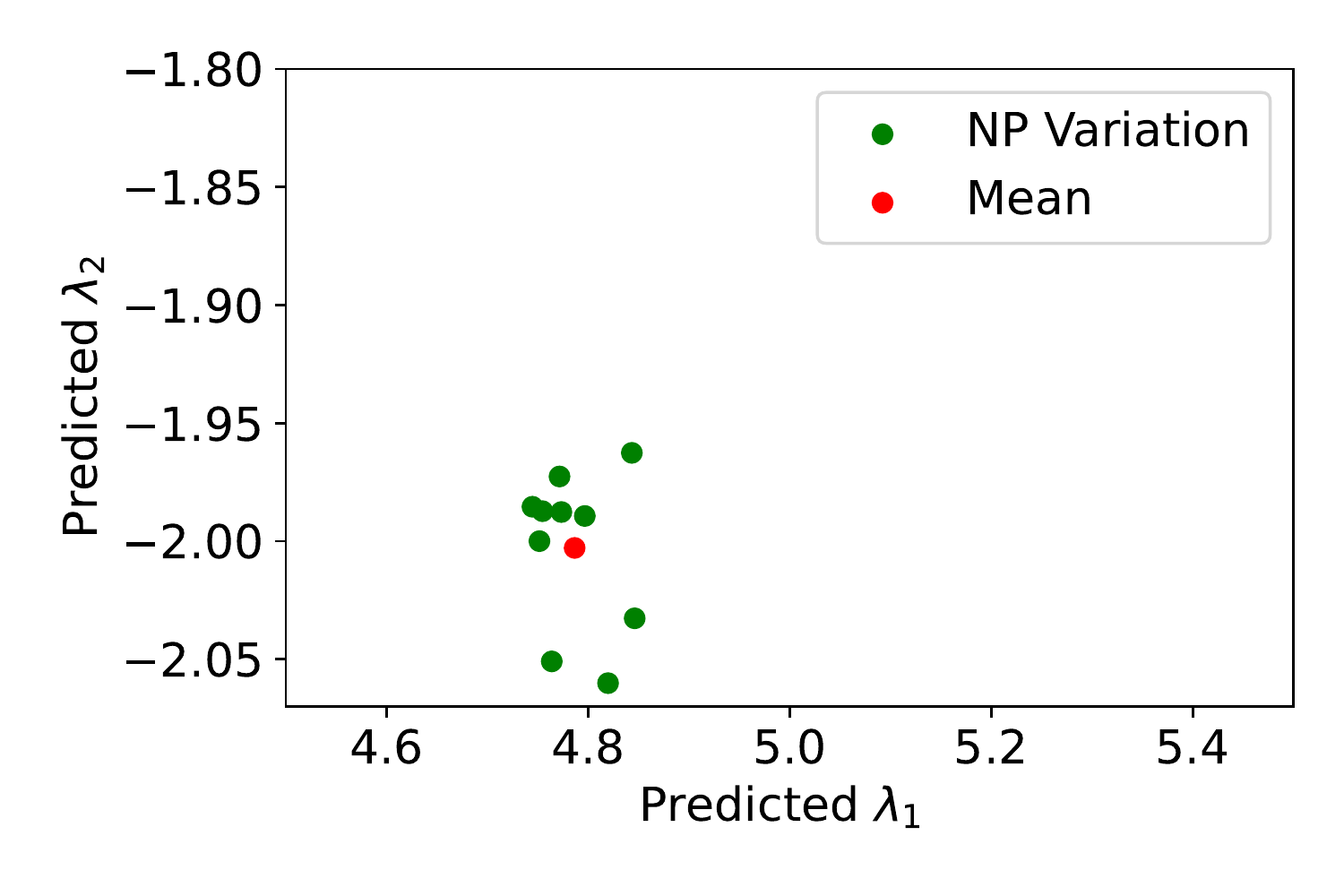}
    \caption{Neural network regression of the EOS parameters $\lambda_1$ and $\lambda_2$ of a set of 10 neutron stars from from their  masses and radii as estimated by \mrnet from each stars spectrum. Each pane represents an example dataset of 10 simulated stars, and shown (green) are EOS estimates for several independent values of the stellar nuisance parameters drawn from the associated priors, and the mean value (red). Top two cases have loose priors, bottom two have tight.} 
    \label{fig:mrnet_eos_nps}
\end{figure}

Performance of the EOS regression using stellar mass and radius information from \mrnet is shown in Figures~\ref{fig:eos_nn_compare1} and \ref{fig:eos_nn_compare2}, and compared to regression from mass and radius information by \xspec. Examples of uncertainty propagation through to EOS estimates are shown in Figure~\ref{fig:mrnet_eos_nps}.

\FloatBarrier

\section{Inference of EOS from Spectra}
\label{sec:e2e}

In the previous section, we connected the two ML models, \mrnet (spectra $\rightarrow  M,R$) and our EOS regression NN ($M,R\rightarrow$ EOS). This required the collapse of the full information and dependence on nuisance parameters into these two physical quantities.  At the same time, \mrnet demonstrated that it is possible to regress physics quantities directly from high-dimensional stellar spectra. In this section, rather than connecting two networks via mass and radius, we use a single network to perform end-to-end regression of EOS parameters from a set of stellar spectra, avoiding the information collapse, keeping the full information, and allowing for robust propagation of the stellar nuisance parameters into uncertainty quantification for the EOS parameters of interest.

In addition to a demonstration of the power of networks to directly analyze low-level data, this allows us to probe the question of whether the mass and radius are sufficient statistics, and whether they contain {\it all} of the information relevant to the problem. There are many examples in the literature in which such well-motived high-level heuristics fail to capture the complete information contained in lower-level data. In this case, while in principle the mass and radius are all that are required to infer the EOS in the context of a fixed theoretical stellar  model, such information is never without uncertainty. Full propagation of the dependence on nuisance parameter uncertainty may allow for more accurate and robust estimates. 

Furthermore, there are properties of neutron stars that can be deciphered from spectra beyond simply mass and radius. Quantities like temperature inhomogeneities~\cite{elshamouty2016impact} may impact a star's equation of state but are not captured by the mass and radius.

\subsection{Architecture}
Many neural network architectures operate on
sequences of vectors, rather than set of vectors. For instance, in natural language processing, the input may be a sentence where each word is converted to a vector and the ordering of the vectors matters. However, in the case of neutron stars and other problems, we need neural networks that operate on sets of vectors, such as the independent spectra observations for multiple stars.

One architectures with invariant properties with respect to permutations of the input vectors is the transformer architecture \cite{vaswani2017attention,quarks2022baldi}. 
Perhaps surprisingly, transformers were originally developed for problems in natural language processing, thus requiring the addition of positional information bits to the vector encoding each word in order to recover the sequential dimension. 
More recently they have been used in other areas, including physics \cite{fenton2022permutationless,10.21468/SciPostPhys.12.5.178} in order to leverage their permutation invariance properties. 
Transformer architectures typically consists of stacks of encoder modules followed by decoder modules. The structure of each encoder module and each decoder module is similar, thus we describe only a typical encoder module. 
A transformer encoder module accepts an unordered set of inputs and produces a set of outputs. The transformer employs a mechanism called {\it self-attention}, which allows it to compare each element in the set against every other. This mechanism allows the network to attend to important features in the set while computing an output prediction. Briefly, self attention (Equation \ref{eq:self_attention}) operates on an input matrix $X$, with $N$ rows and $D$ columns. Three matrices are produced from the projection of $X$ with differing, trainable, weight matrices: $Q=XW_Q, K=XW_K, V=XW_V$ (termed Query, Key, and Value respectively).

\begin{equation}
    S = D(Q, K, V) = \text{softmax}\left(\frac{QK^T}{\sqrt{d_q}}\right) V
    \label{eq:self_attention}
\end{equation}
Thus in short each output corresponds to a different convex combination of the Value vectors, where each convex combination depends on the degree of similarity between the corresponding Query and Key vectors. The similarity is computed by  taking dot products between corresponding Query and Key vectors, and then applying a softmax to yield a convex combination (see \cite{vaswani2017attention,quarks2022baldi} for additional details). 

\begin{figure}
    \centering
    \includegraphics[scale=0.7]{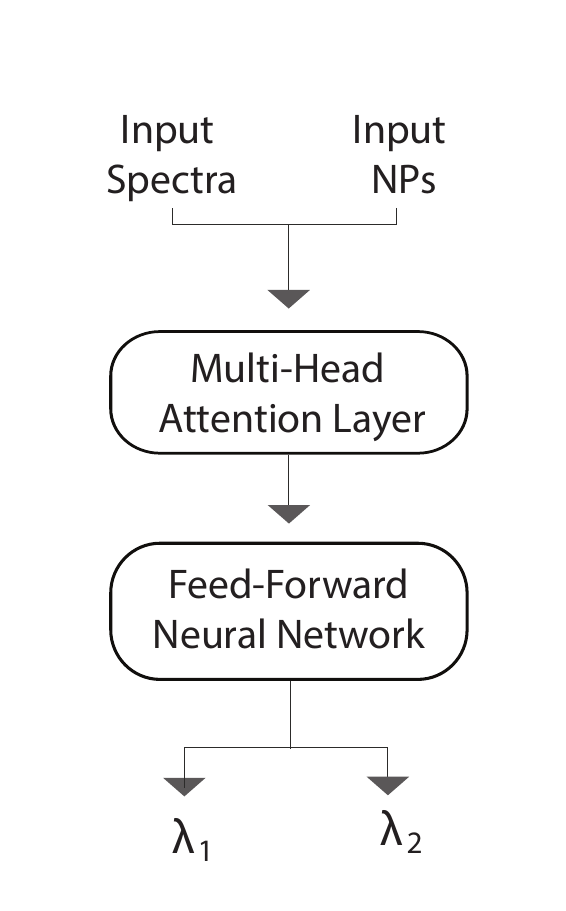}
    \caption{Schematic diagram of the transformer neural network used to determine EOS coefficients $\lambda_1$ and $\lambda_2$ from an input of spectra and NPs. 
    }
    \label{fig:transformer_scheme}
\end{figure}

In the neutron star application, the transformer architecture takes as input spectra and corresponding nuisance parameters for each star in the set. In the results shown below, the network is given a set of 10 stars, though the structure of the network allows it to accept larger or smaller datasets with minimal modification. The final output of the network is the two EOS parameters. This is shown schematically in Figure~\ref{fig:transformer_scheme}.

The network is composed of six consecutive transformer blocks. Each block processes the input through multi-head attention (with eight heads), followed by dropout ($p=0.15$), normalization, and fully connected layers. Following these blocks the output is processed by one final fully connected layer in order to regress the EOS coefficients $\lambda_1$ and $\lambda_2$. All fully connected layers, with the exception of the final one, use the ReLU activation function. The Adam optimizer was used to provide gradient updates with an initial learning rate of 0.000075, which was slowly decade over the course of training. 

\subsection{Training}
The parameters of the network architecture and the learning algorithm, the hyperparameters, were optimized with Sherpa \cite{hertel2020sherpa}, a Python library for hyperparameter tuning. The algorithm used is based on a random search and has the advantage of making no assumptions about the structure of the hyperparameter search problem and thus is well suited for exploring a variety of parameter settings. An initial exploratory search was conducted on a subset of the data to find appropriate hyperparameters. 

Following this exploratory phase, the network is trained for 1,000 epochs on the full dataset. The Adam optimizer~\cite{adam} is used for gradient descent, with early stopping monitoring
of the validation loss to prevent overfitting.

\subsection{Results}

\begin{figure}[hbt!]
    \centering
   \includegraphics[scale=0.4]{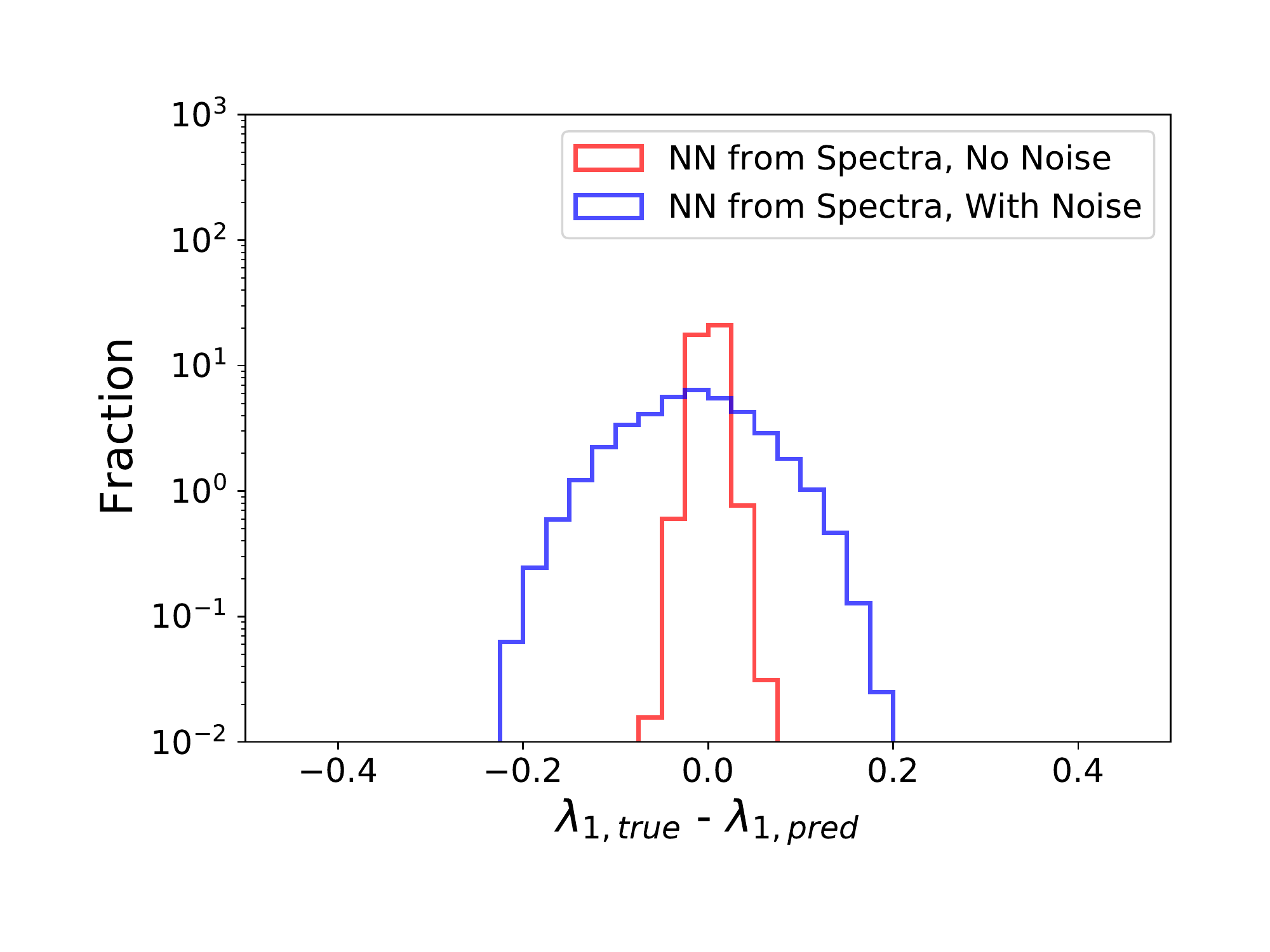}
   \includegraphics[scale=0.4]{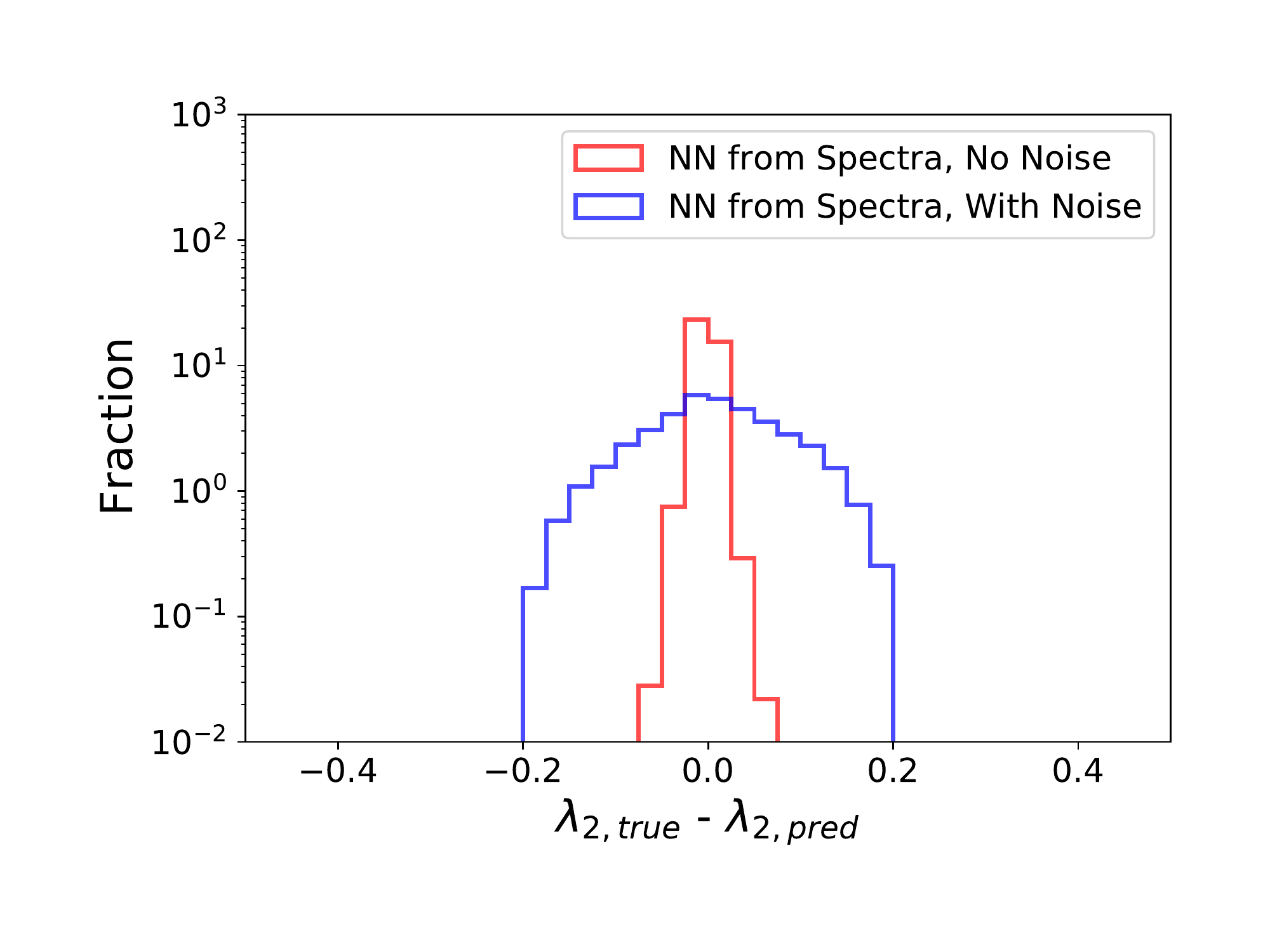}
    \caption{Performance of the neural network regression of the neutron star EOS parameters $\lambda_1$ (top) and $\lambda_2$ (bottom) directly from a set of stellar X-ray spectra, without intermediate prediction of the mass and radius.  Shown is the residual, the difference between the true and predicted values, for spectra with statistical noise (blue) corresponding to an observation time of 100k seconds per star, and for spectra without statistical noise (red), which demonstrates the capacity of the network. Nuisance parameters are fixed to their true values.}
    \label{fig:res_regress_noise}
\end{figure}

Figure~\ref{fig:res_regress_noise} shows the performance of spectra$\rightarrow$EOS regression in the best-case scenario, where the nuisance parameters are perfectly known. Shown is the performance with statistical noise corresponding to 100 ks of observation time per star, as well as for spectra without statistical noise. While there is significant width to the residuals, this is dominated by the statistical uncertainty, not the network's ability to digest the spectral information and understand the dependence on mass and radius. This clearly demonstrates the network's capacity is sufficient for the regression task.

\begin{figure}
    \centering
    \includegraphics[scale=0.35]{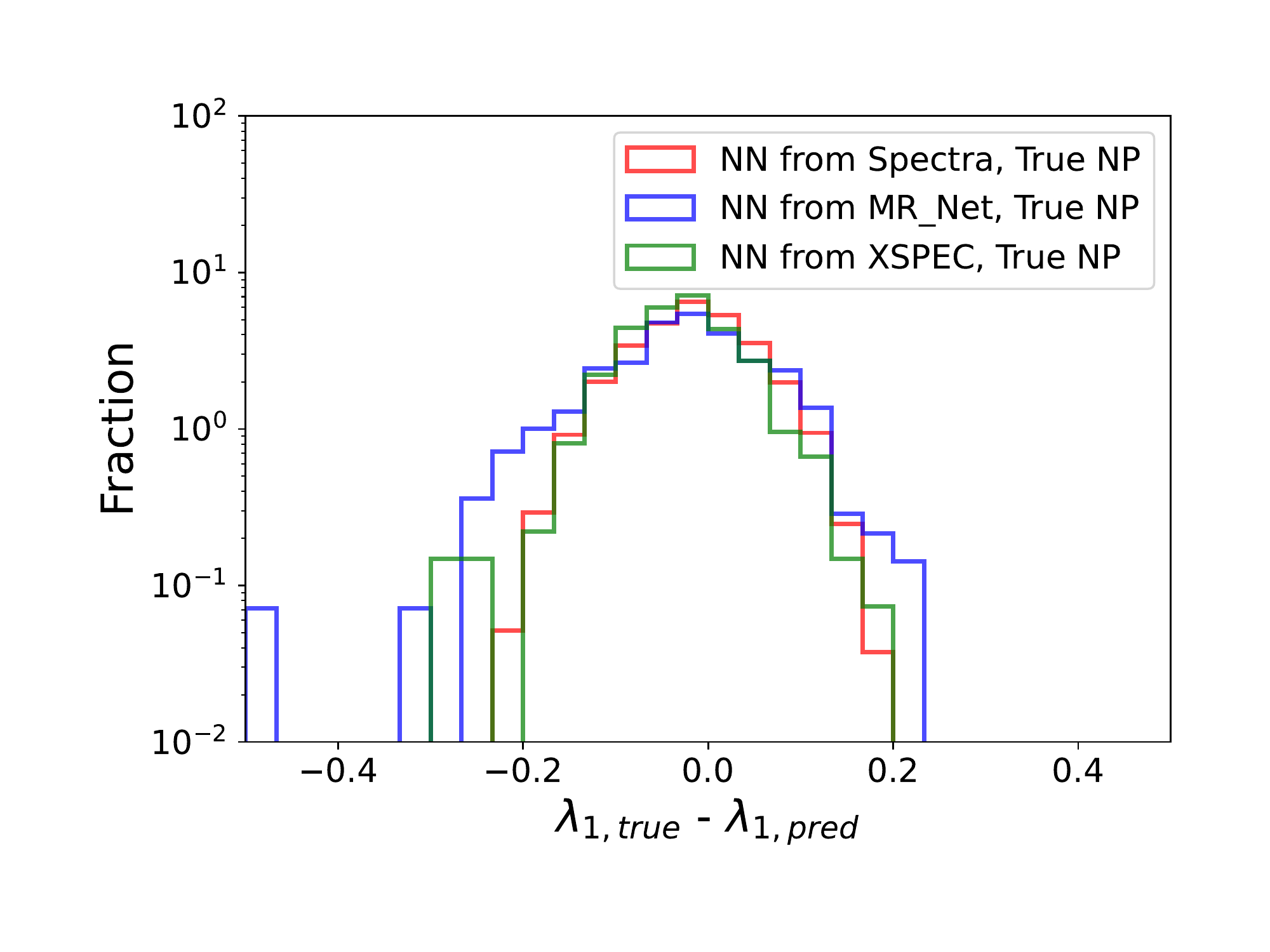}
    \includegraphics[scale=0.35]{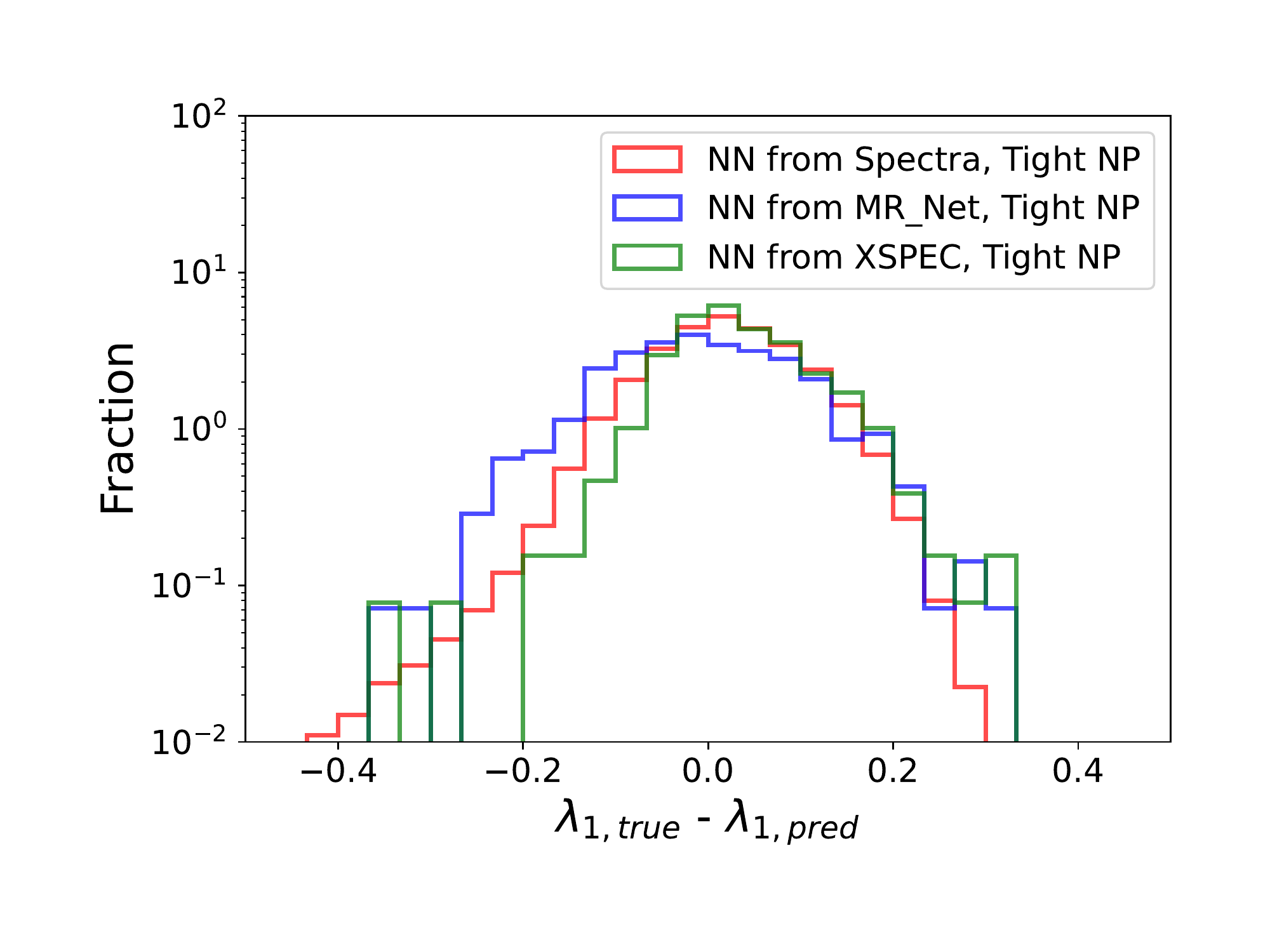}
    \includegraphics[scale=0.35]{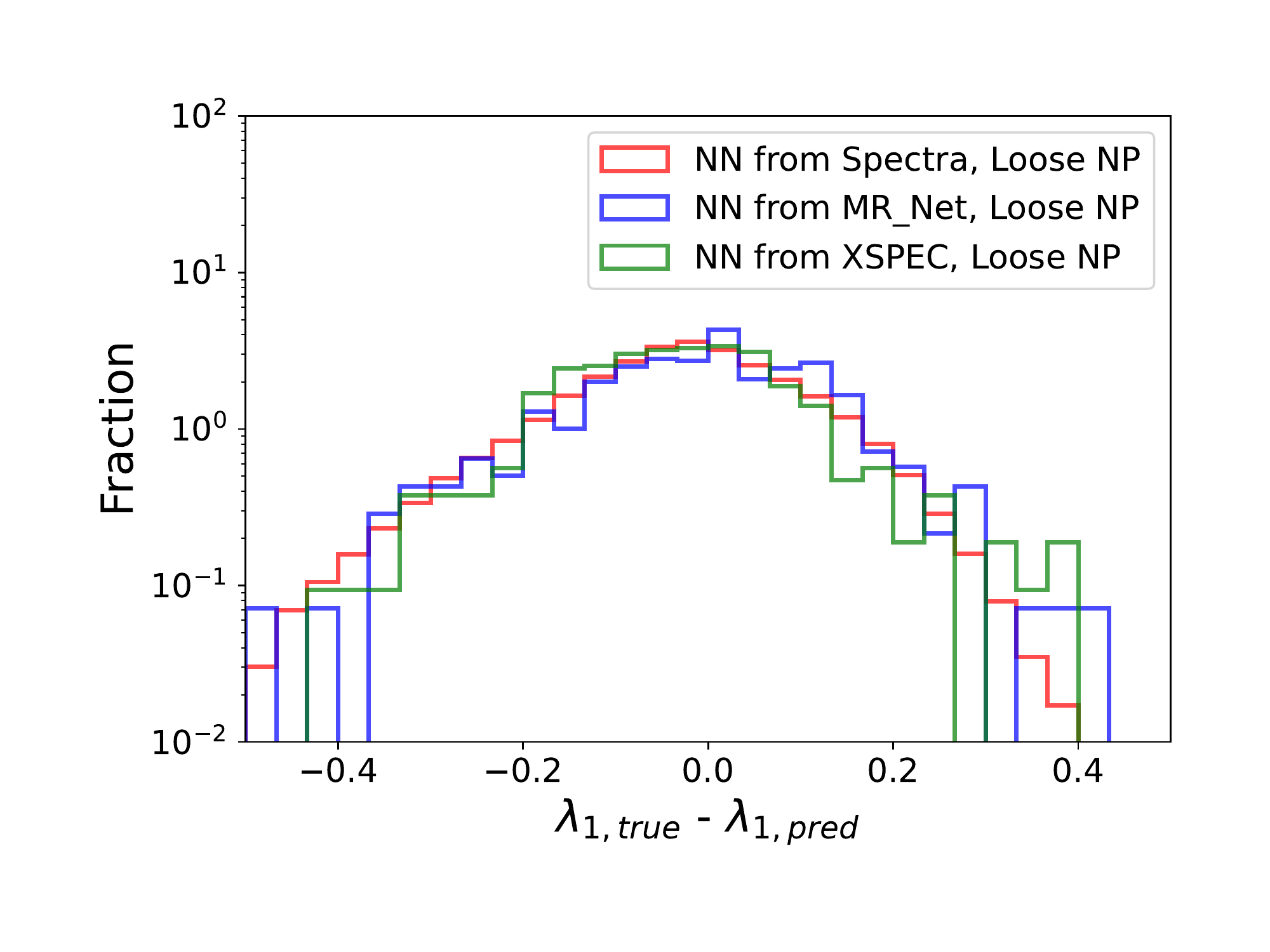}
    \caption{ Performance of the regression of neutron star EOS parameter $\lambda_1$ using direct regression from spectra, as compared to regression from mass and radius information extracted via \mrnet or \xspec. Shown are the residual distributions, the difference between the true and predicted values, under three  scenarios of nuisance parameter uncertainties. See Table~\ref{tab:eos} for quantitative analysis.    In the ``true" case, the NPs are fixed to their true values; in the ``tight" and ``loose" cases, they are drawn from narrow or wide priors, respectively (see text for details).}
    \label{fig:eos_nn_compare1}
\end{figure}

\begin{figure}
    \centering
    \includegraphics[scale=0.35]{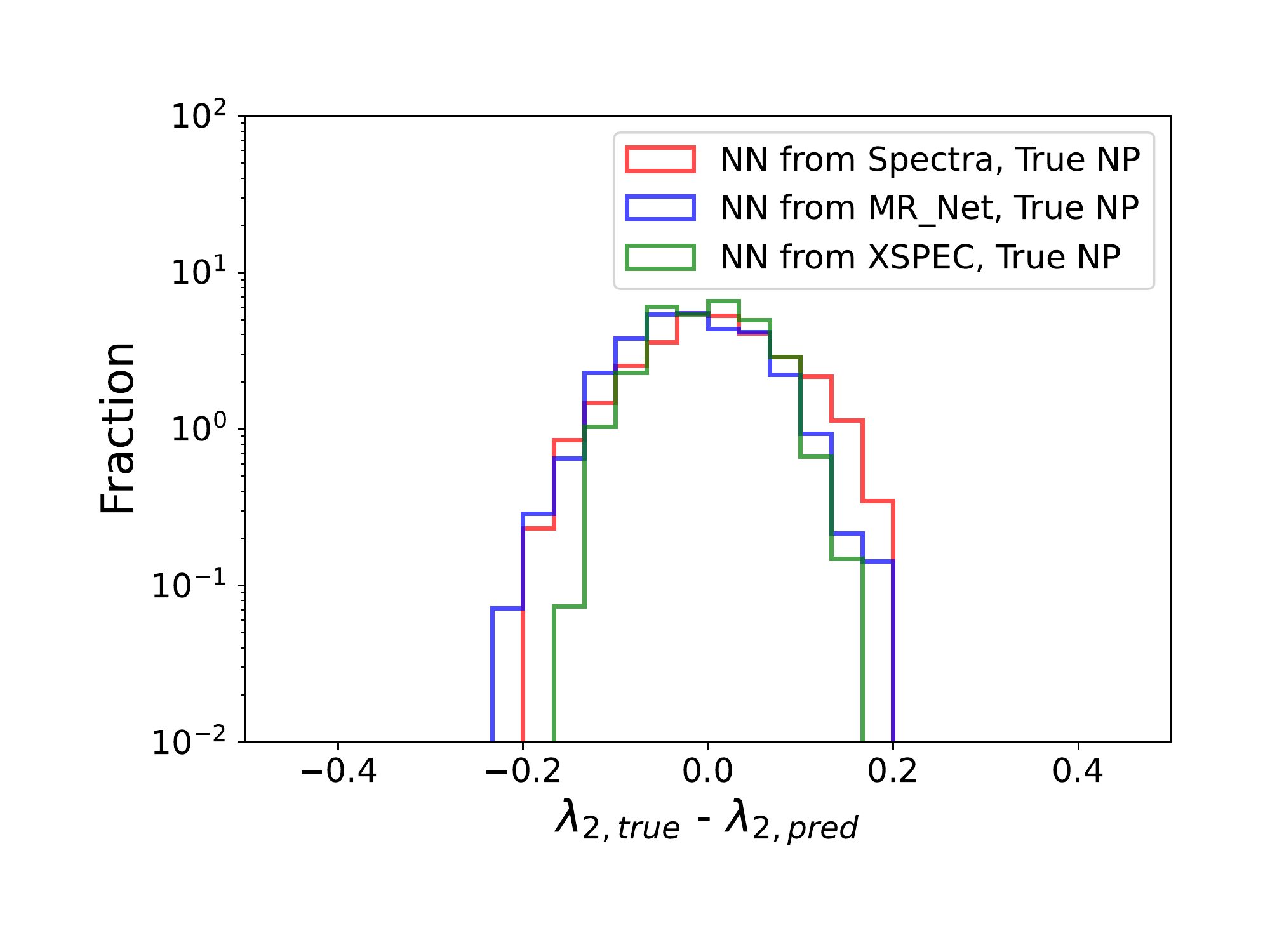}
    \includegraphics[scale=0.35]{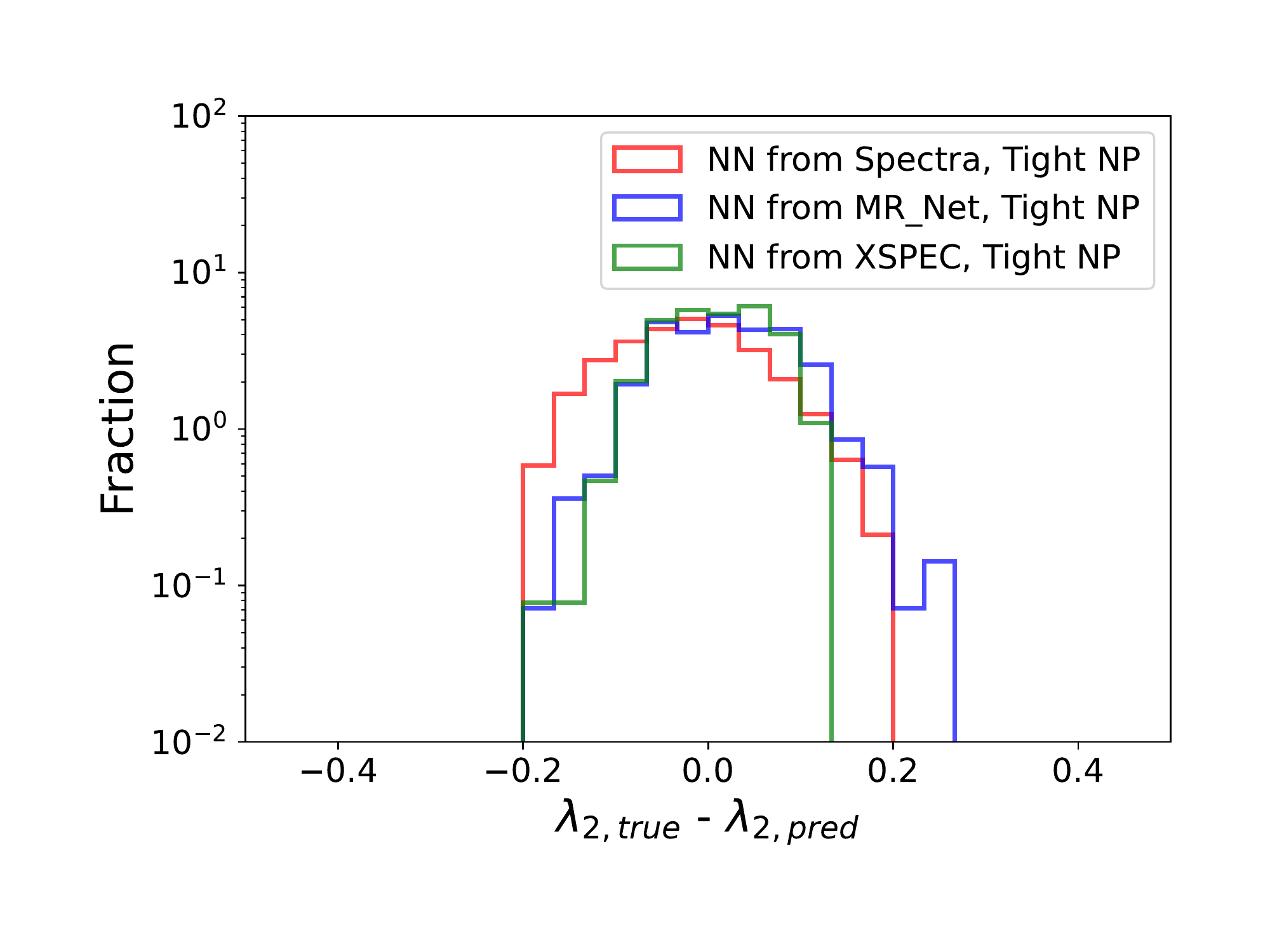}
    \includegraphics[scale=0.35]{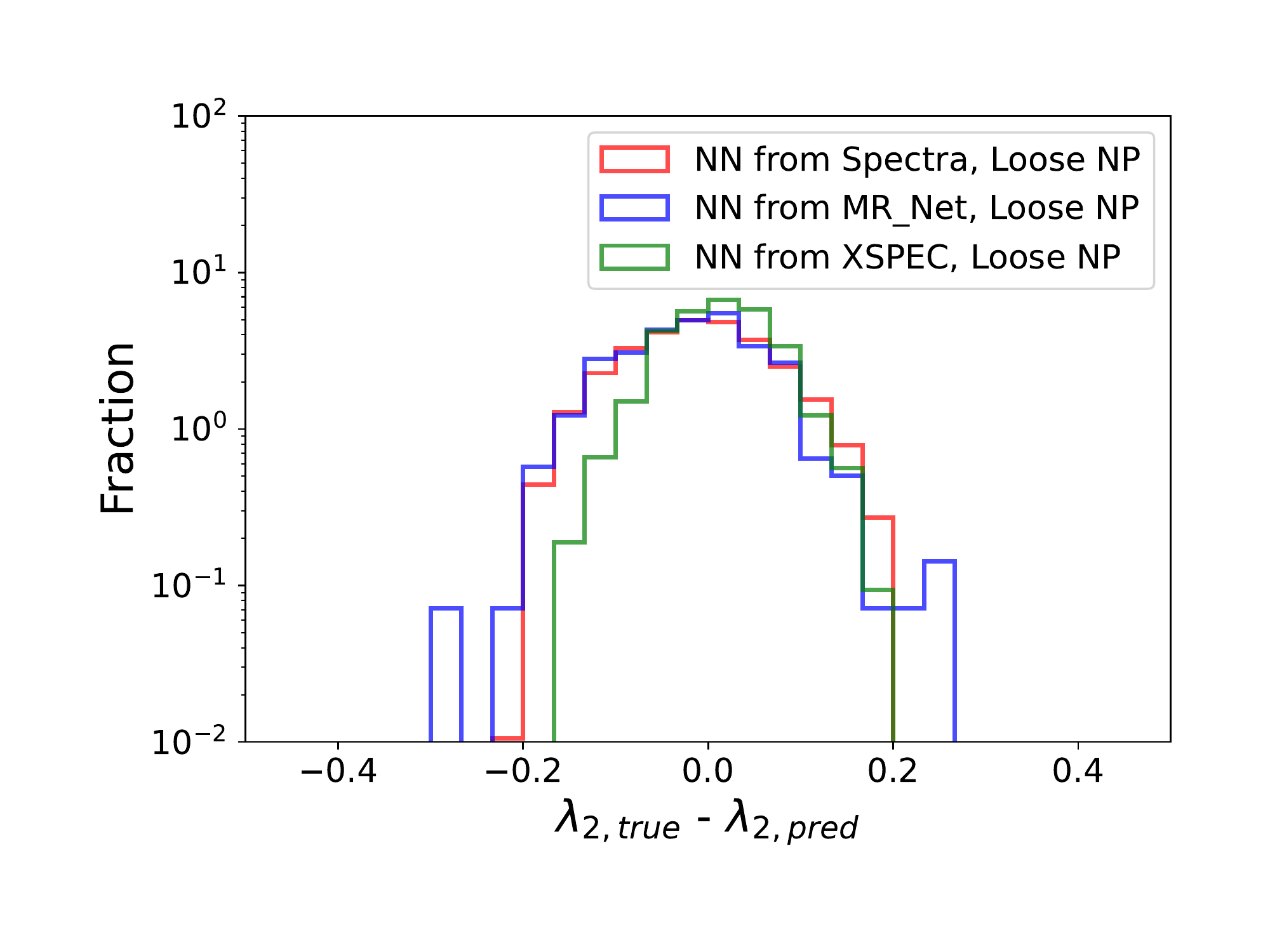}
    \caption{ Performance of the regression of neutron star EOS parameter $\lambda_2$ using direct regression from spectra, as compared to regression from mass and radius information extracted via \mrnet or \xspec. Shown are the residual distributions, the difference between the true and predicted values, under three  scenarios of nuisance parameter uncertainties. See Table~\ref{tab:eos} for quantitative analysis.    In the ``true" case, the NPs are fixed to their true values; in the ``tight" and ``loose" cases, they are drawn from narrow or wide priors, respectively (see text for details).}
    \label{fig:eos_nn_compare2}
\end{figure}

We next analyze the performance of the direct regression in cases where the nuisance parameters are not perfectly determined. Figures ~\ref{fig:eos_nn_compare1} and~\ref{fig:eos_nn_compare2} shows the residuals in the EOS parameters for the end-to-end regression, as compared to regression from mass and radius information provided by \mrnet or \xspec\  from the stellar spectra. Table~\ref{tab:eos} summarizes the performance for each method.

\begin{figure}
    \centering
    \includegraphics[width=\exsize\textwidth]{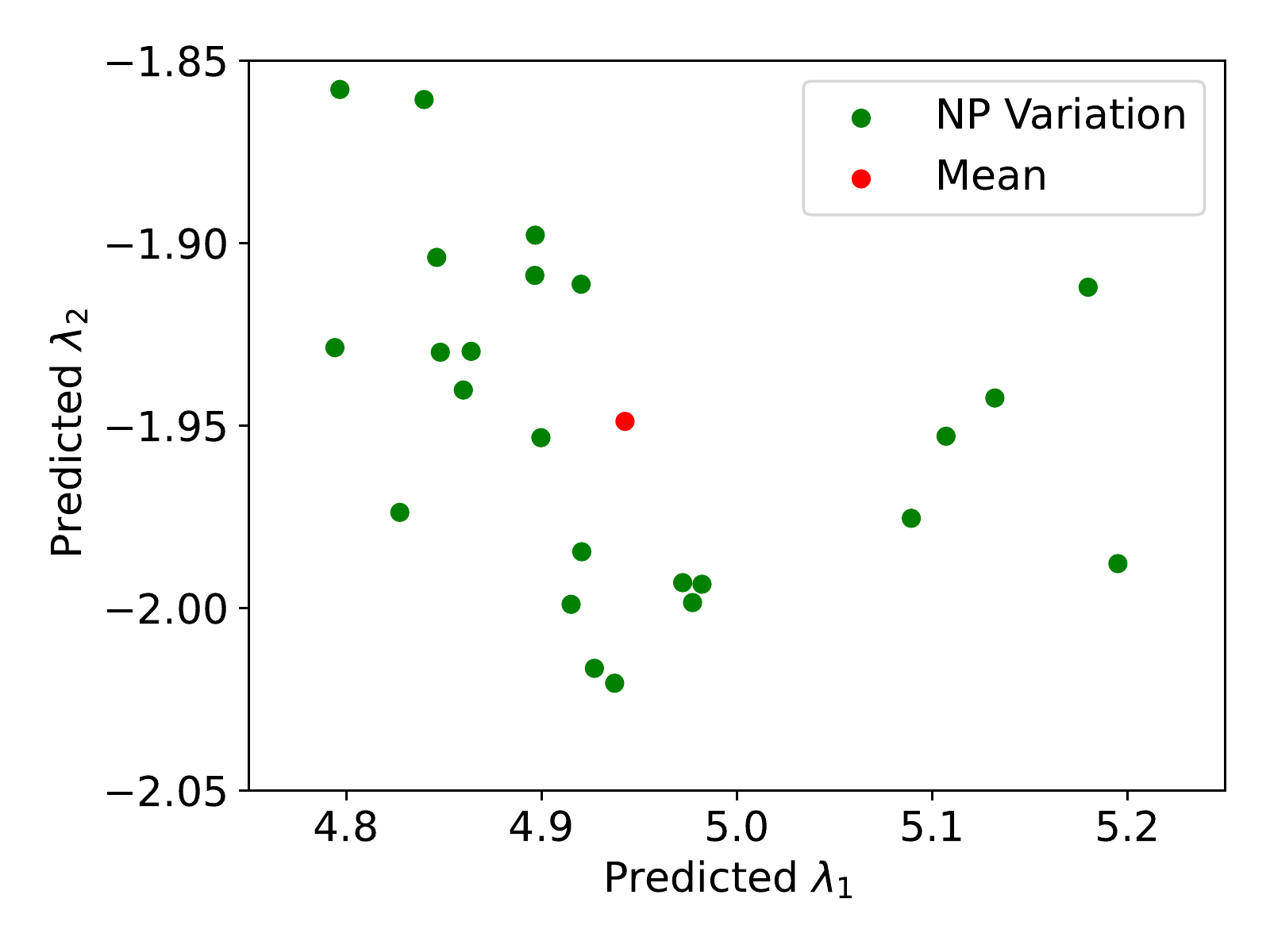}
    \includegraphics[width=\exsize\textwidth]{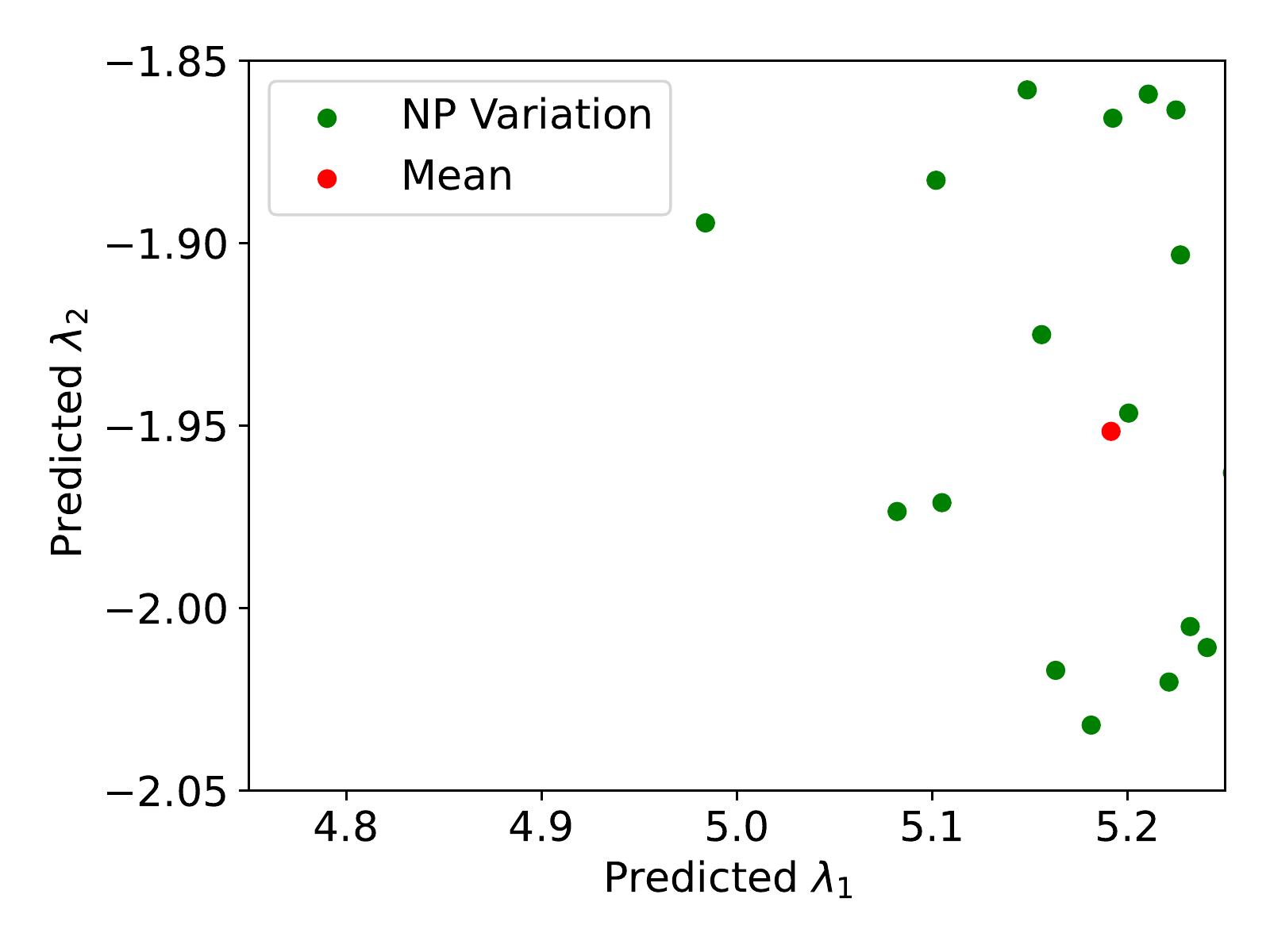}
    \includegraphics[width=\exsize\textwidth]{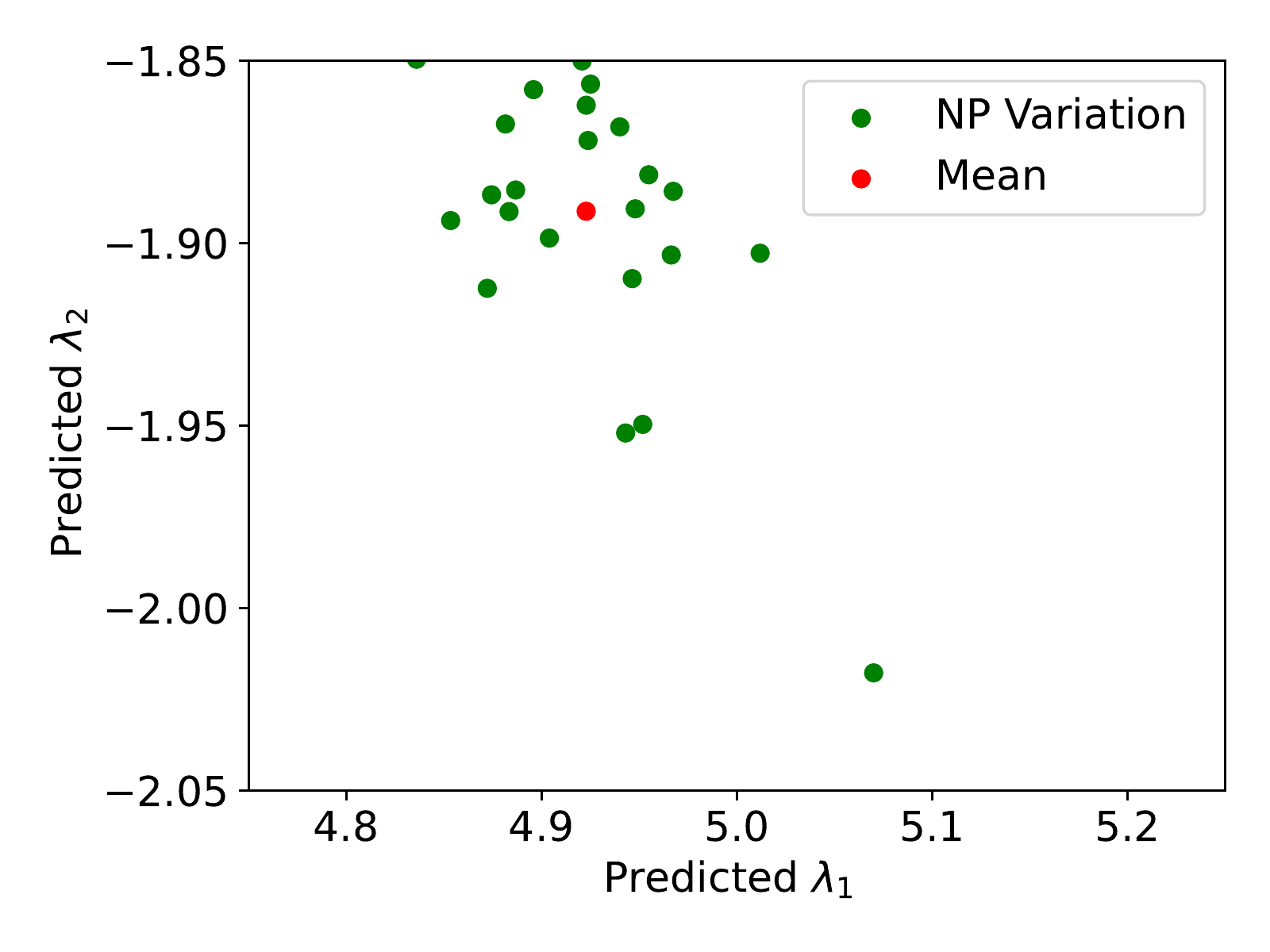}
    \includegraphics[width=\exsize\textwidth]{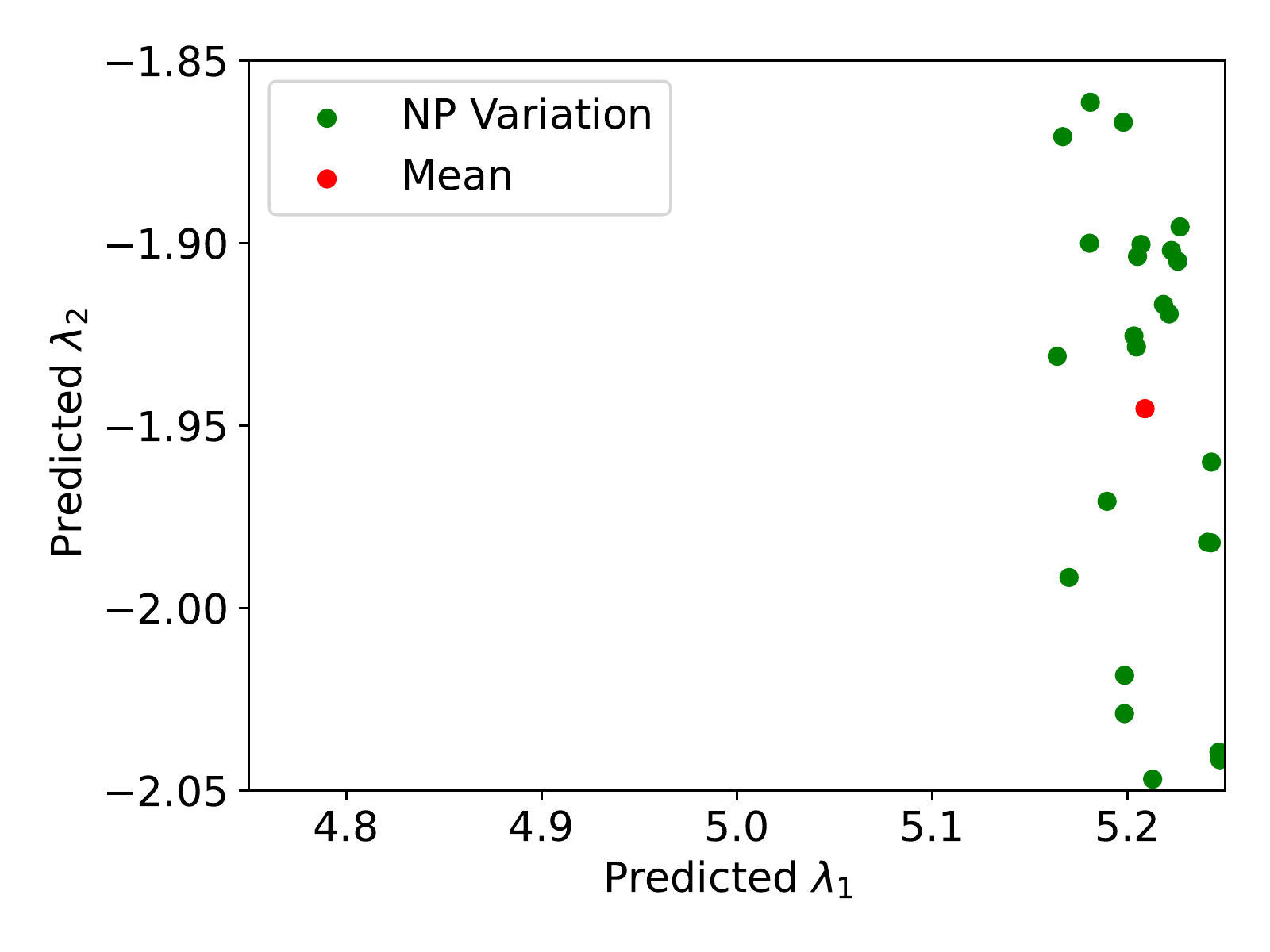}
    \caption{ Neural network regression of the EOS parameters $\lambda_1$ and $\lambda_2$ of a set of 10 neutron stars directly from the set of stellar spectra. Each pane represents an example dataset of 10 simulated stars, and shown (green) are EOS estimates for several independent values of the stellar nuisance parameters drawn from the associated priors, and the mean value (red). Top two cases have loose priors, bottom two have tight.} 
    \label{fig:eos_examples}
\end{figure}

As the full network is again conditioned on the NPs, we can propagate this uncertainty directly through our regression. Figure~\ref{fig:eos_examples} demonstrates how variations of the NPs, drawn from the appropriate priors, provide a measure of the uncertainty on the final result.

\begin{table*}[]
    \caption{Performance of the regression of neutron star EOS parameters $\lambda_1$ and $\lambda_2$ using direct regression from spectra, as compared to NN regression from mass and radius $(M,R)$ information extracted via \mrnet or \xspec. Shown are the mean ($\mu$) and standard deviation ($\sigma$) of the residuals under three  scenarios of nuisance parameter uncertainties; distributions are given in Figures~\ref{fig:eos_nn_compare1} and~\ref{fig:eos_nn_compare2}.    In the ``true" case, the NPs are fixed to their true values; in the ``tight" and ``loose" cases, they are drawn from narrow or wide priors, respectively; see text for details. The combined column is a quadrature sum of the standard deviations of $\lambda_1$ and $\lambda_2$.}
    \label{tab:eos}
    \centering
    \begin{tabular}{lcrrrrrr}
             \hline \hline
         & Nuis. &\multicolumn{2}{c}{$\lambda_1$} & \ \ &\multicolumn{2}{c}{$\lambda_2$} & Combined\\
         \cline{3-4}\cline{6-7}\\
         Method &  Params. &  $\mu$ & $\sigma$ & &$\mu$ & $\sigma$  & $\sigma$  \\
                  \hline 
        NN(Spectra)  & True  & $-0.02$ & 0.066  && 0.01  & 0.075   & 0.099  \\
        NN($M,R$ via \mrnet)   & True  & -0.03 & 0.089  && $-0.02$ & 0.068   &  0.112 \\
        NN($M,R$ via \xspec\ ) & True  & $-0.03$ & 0.065  && 0.01  & 0.055   &  0.085 \\ \hline
        NN(Spectra)  & Tight & 0.02  & 0.085  && $-0.02$ & 0.077   & 0.115  \\
        NN($M,R$ via \mrnet)   & Tight & 0.00  & 0.104  && 0.02  & 0.072   & 0.126  \\ 
        NN($M,R$ via \xspec\ ) & Tight & $-0.03$ & 0.081  && 0.01  & 0.056   &  0.098 \\ \hline
        NN(Spectra)  & Loose & $-0.03$ & 0.131  && $-0.01$ & 0.078   &  0.152 \\
        NN($M,R$ via \mrnet)   & Loose & $-0.01$ & 0.135  && $-0.02$ & 0.078   & 0.156  \\
        NN($M,R$ via \xspec\ ) & Loose & $-0.03$ & 0.123  && 0.01  & 0.058   & 0.136  \\
         \hline \hline
    \end{tabular}
\end{table*}

\begin{figure}
    \centering
    \includegraphics[width=0.4\textwidth]{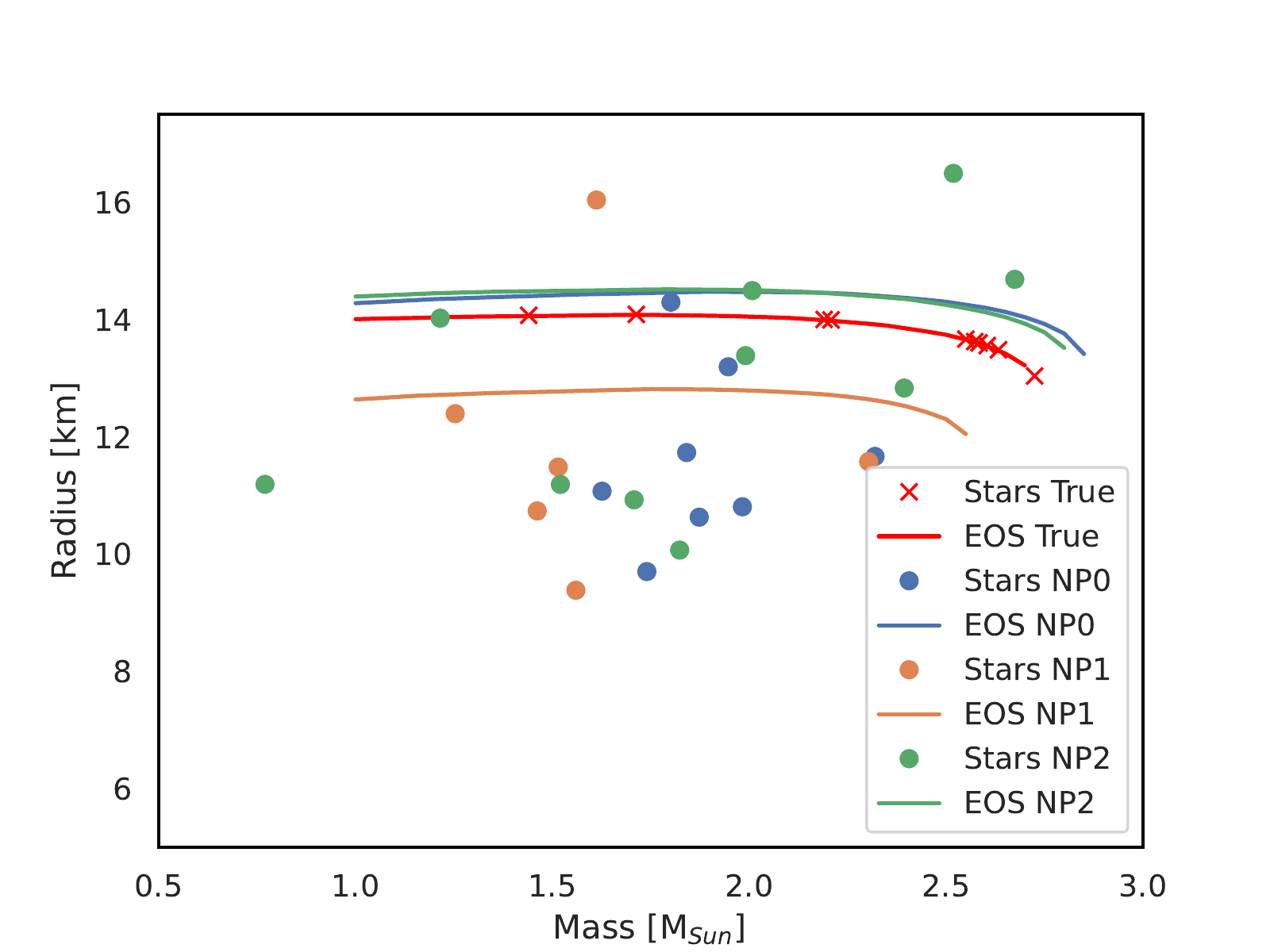}
        \includegraphics[width=0.4\textwidth]{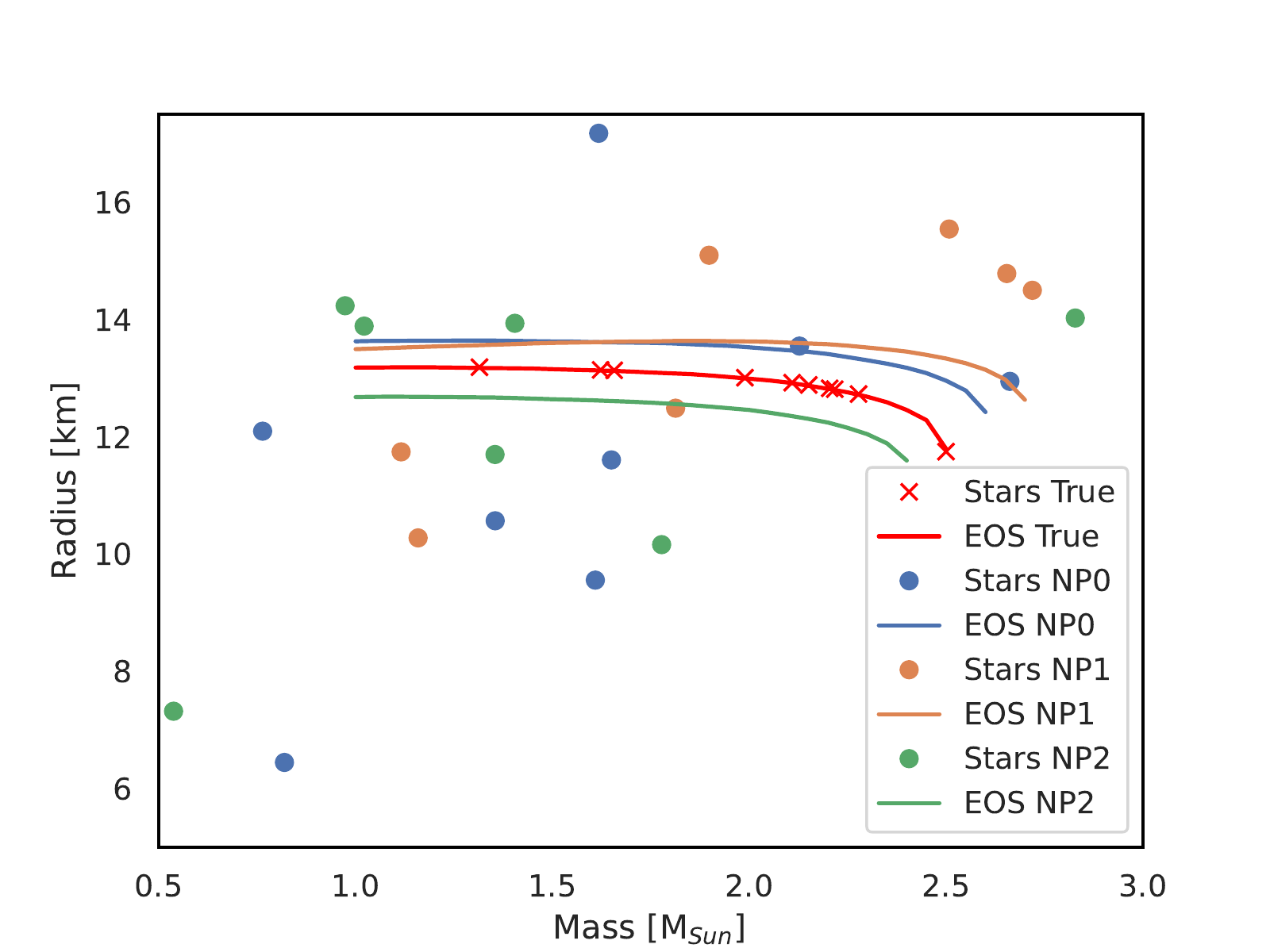}
    \includegraphics[width=0.4\textwidth]{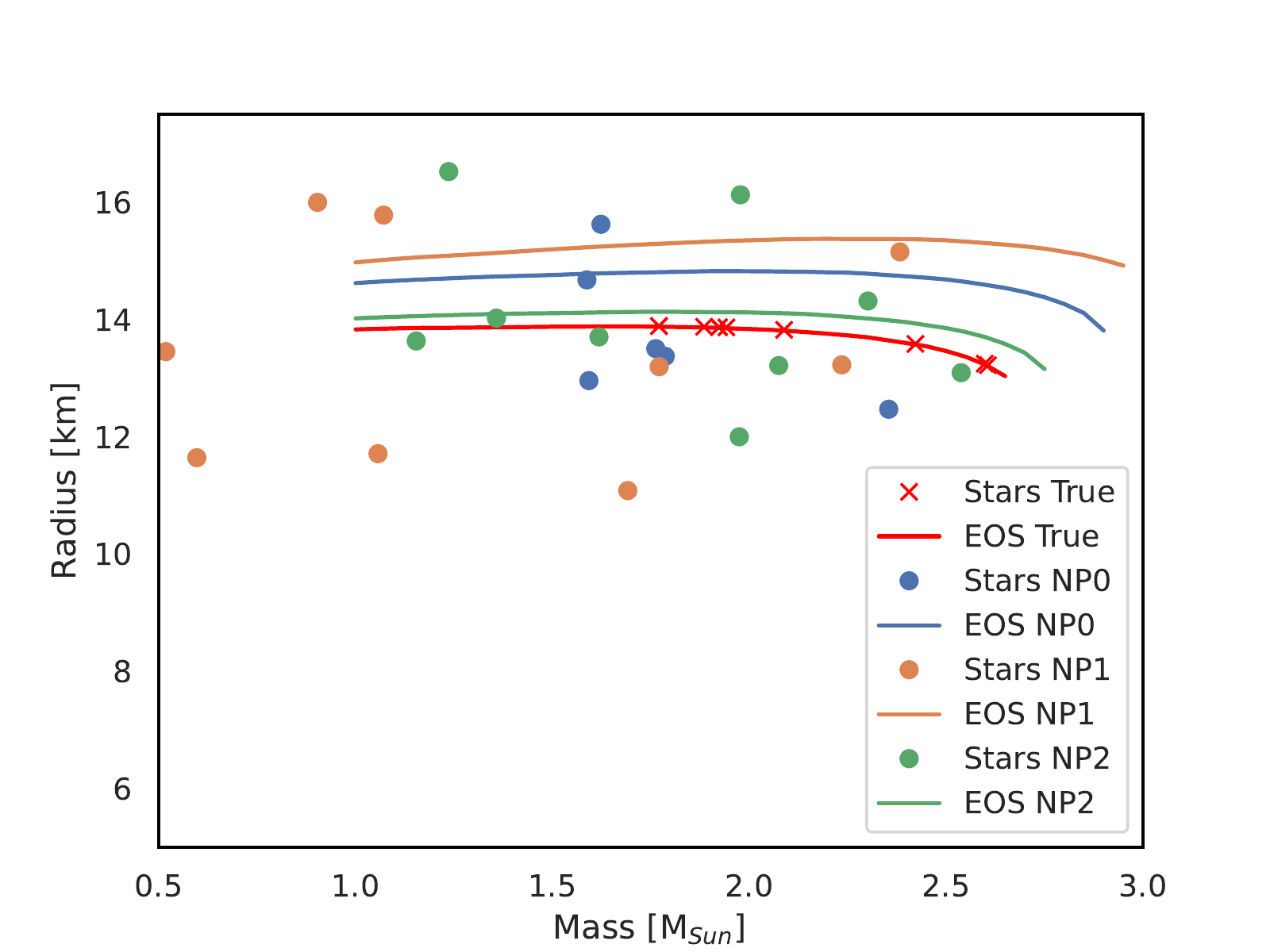}
    \caption{ Demonstration of the impact of nuisance parameters on the regression of EOS parameters for three  sets of observed stellar spectra. In red are the true mass and radius of the stars, drawn from the mass-radius curve determined by the true EOS parameters. In blue dots are the values of the mass and radius deduced by \mrnet for each star given a set of stellar NPs drawn from the priors; the blue line shows the mass-radius curve corresponding to the EOS parameters deduced directly from the stellar spectra and NPs by the proposed end-to-end regression. The results of \mrnet are not used in the EOS regression, and only appear to aid the visualization.  Brown and green are similar to blue, but for independent draws of the NPs from the same stellar priors.}
    \label{fig:eos_vis}
\end{figure}

\section{Discussion} \label{sec:Discussion}

The performance of the three methods are measured in simulated stellar samples generated with the same theoretical stellar model that is assumed by \xspec, which makes it a valuable upper limit for the two fully neural network based methods, which must infer the relationships. The three approaches perform comparably, and the end-to-end method slightly but consistently outperforms the two-step method using \mrnet. Once trained, the end-to-end network can handle any prior on the nuisance parameters, whereas the networks that rely on \xspec\ fits or \mrnet predictions need to first be trained on data with the desired prior.

When the data are simulated and drawn from a known theoretical model, one cannot achieve more statistical power than directly calculating the likelihood. But even powerful theoretical models for spectral fitting still rely on a variety of assumptions about the spectrum's source. The flexibility of these fully neural network based approaches is an important advantage, opening the door to interpolating between theoretical models~\cite{Ghosh:2021roe}, or even learning directly from observational data~\cite{Howard:2021pos,Ghosh:2022zdz}. An inference approach with this flexibility and the capacity for robust propagation of uncertainties is vital.

As a further visualization, Figure ~\ref{fig:eos_vis} shows several example curves in the mass-radius plane fitted to the same stellar spectra with varying nuisance parameters.

\section{Conclusions}

We have demonstrated the network regression of EOS parameters from realistic neutron star mass and radius estimates drawn from simulated stellar spectra. Our approach of conditioning each step on nuisance parameters allows us to fold in the NP uncertainty via multiple sampling from priors and permits full propagation of the uncertainty through to the final regression targets. The full propagation is important because variation in NPs does not produce variations in the mass and radius of neutron stars that can be accurately summarised as two dimensional uncorrelated Gaussians (see Figure ~\ref{fig:mr_examples}), as has been assumed in previous studies. In addition, we have shown that networks can analyze high-dimensional telescope data directly, including sets of multiple stars, and achieve comparable performance to methods that assume perfect knowledge of the theoretical model used to generate the simulated samples. In realistic cases where the nuisance parameter uncertainties are significant, the proposed end-to-end network regression achieves comparable precision in EOS regression to the network using \xspec\ fits or \mrnet predictions.

These results suggest many future directions. Our networks are parameterized in the nuisance parameters, allowing for the propagation of prior uncertainties which are implicity derived from auxiliary data. But the stellar spectra may also offer information that constrains the NP uncertainty. Profiling over the nuisance parameters could reduce this uncertainty, though it may be computationally very expensive without neural likelihood estimation techniques~\cite{Heinrich:2022qlq}.

Alternatively, rather than employing regression to directly produce estimates of the EOS parameters, one might train a generative model to operate as a surrogate of the likelihood~\cite{Cranmer:2019eaq}, allowing for fast evaluation of the likelihood as a function of the EOS parameters and potentially direct profiling.

To reduce the impact on one particular set of theoetical assumptions, such networks may be trained on a collection of theoretical models, and in the future when more telescope data becomes available, even trained directly from observed spectra.

Other future directions for the networks described in this paper would be to test more exotic neutron star equations of state, including those with phase transitions. Including additional parameters will play a key role in conducting similar research using alternative models in \xspec \ that rely on different nuisance parameters, like a Helium atmospheric model. Even more interesting would be extending this type of EOS inference to other compact objects like white dwarfs.

\section{Acknowledgements}

DW and AG are supported by The Department of Energy Office of Science. LL was supported by NSF Grant No. 2012857 to the University of California at San Diego. AWS was supported by NSF AST 19-09490, PHY 21-16686, and the Department of Energy Office of Nuclear Physics. DF and FW are supported by the National Science Foundation (USA) under Grant No. PHY-2012152. The authors are grateful to C. O. Heinke and W. C. G. Ho for their assistance in understanding and navigating the \xspec \ software. 

\clearpage



\bibliography{ns}

\end{document}